\documentclass[final,5p,times,twocolumn,preprint]{elsarticle}

\usepackage{epsfig,float,amssymb,latexsym,amsmath,amsthm,fancyhdr,slashed}
\usepackage{graphics,psfrag,longtable,color}
\usepackage{bm}

\usepackage{mathtools}
\allowdisplaybreaks
\newcommand{\be}{\begin{eqnarray}}
\newcommand{\ee}{\end{eqnarray}}
\newcommand{\beq}{\begin{equation}}
\newcommand{\eeq}{\end{equation}}

\begin{document}

\begin{frontmatter}



\title{Peripheral transverse densities of the baryon octet \\
from chiral effective field theory and dispersion analysis}

\author[HISKP,JLab]{J. M. Alarc\'on}
\author[IFIC]{A. N. Hiller Blin}
\author[IFIC]{M. J. Vicente Vacas}
\author[JLab]{C. Weiss}
\address[HISKP]{Helmholtz-Institut f\"ur Strahlen- und Kernphysik \& Bethe Center for Theoretical Physics,
Universit\"at Bonn, 53115 Bonn, Germany}
\address[JLab]{Theory Center, Jefferson Lab, Newport News, VA 23606, USA}
\address[IFIC]{Departamento de F\'isica Te\'orica and IFIC, Centro Mixto Universidad de Valencia-CSIC,\\
Institutos de Investigaci\'on de Paterna, E-46071 Valencia, Spain}

\begin{abstract}
The baryon electromagnetic form factors are expressed in terms of two-dimensional densities describing 
the distribution of charge and magnetization in transverse space at fixed light-front time.
We calculate the transverse densities of the spin-$1/2$ flavor-octet baryons at peripheral
distances $b = \mathcal{O}(M_\pi^{-1})$ using methods of relativistic chiral effective field theory ($\chi$EFT) 
and dispersion analysis. The densities are represented as dispersive integrals over the imaginary 
parts of the form factors in the timelike region (spectral functions). The isovector spectral functions 
on the two-pion cut $t > 4 \, M_\pi^2$ are calculated using relativistic $\chi$EFT including octet and 
decuplet baryons. The $\chi$EFT calculations are extended into the $\rho$ meson mass region using an 
$N/D$ method that incorporates the pion electromagnetic form factor data. The isoscalar spectral functions 
are modeled by vector meson poles. We compute the 
peripheral charge and magnetization densities in the octet baryon states, estimate the uncertainties, 
and determine the quark flavor decomposition. The approach can be extended to baryon form factors of 
other operators and the moments of generalized parton distributions.
\end{abstract}


\end{frontmatter}

\section{Introduction}
\label{Sec:Introduction}
Exploring the spatial structure of hadrons and their interactions is a central objective of 
modern strong interaction physics. The description of hadrons as extended objects in relativistic 
space-time enables an intuitive understanding of their structure and permits the formulation 
of novel approximation methods based on physical distance scales.
The most basic information about spatial structure comes from the transition matrix elements
(or form factors) of electroweak current operators, which reveal the distributions of charge
and current in the system. In non-relativistic systems, such as nuclei in nuclear many-body 
theory, the form factors are represented as the Fourier transform of 3-dimensional spatial distributions 
of charge and current at a fixed instant of time $x^0 = $ const. These distributions can be 
related to probability 
densities of the $N$-particle Schr\"odinger wave function. In relativistic systems such as hadrons
the formulation of a density requires separating the structure of the ``system'' from that of the 
``probe'' in the presence of vacuum fluctuations that change the number of constituents. This is 
accomplished in a frame-independent manner by viewing the system at fixed light-front time 
$x^+ = x^0 + x^3$ (light-front quantization) \cite{Dirac:1949cp,Leutwyler:1977vy,Lepage:1980fj,Brodsky:1997de}. 
In this framework the form factors are represented in terms 
of 2-dimensional spatial distributions of charge and current at fixed light-front time 
$x^+ = $ const.\ \cite{Soper:1976jc,Burkardt:2000za,Burkardt:2002hr,Miller:2007uy}. 
The transverse densities are frame-independent (they remain invariant under longitudinal boosts and 
transform kinematically under transverse boosts) and provide an objective spatial representation 
of the hadron as a relativistic system. The properties of the transverse densities of mesons and 
baryons, their extraction from experimental data, and their explanation in dynamical models 
have been discussed extensively in the 
literature \cite{Miller:2007uy,Carlson:2007xd,Vanderhaeghen:2010nd,Venkat:2010by,Miller:2009sg}; 
see Ref.~\cite{Miller:2010nz} for a review. An important 
aspect is that the transverse densities are naturally related to the partonic structure of hadrons probed in 
high-energy short-distance processes (generalized parton distributions, or GPDs). The densities
describe the cumulative charge and current carried by the quarks/antiquarks in the hadron at 
a given transverse position and represent the integral of the GPDs over the quark/antiquark
light-front momentum fraction \cite{Burkardt:2000za,Burkardt:2002hr,Diehl:2002he}; see
Refs.~\cite{Belitsky:2005qn,Boffi:2007yc} for a review. This relation places the study of elastic 
form factors in the context of exploring the 3-dimensional quark-gluon structure of hadrons in QCD.

Of particular interest are the transverse densities at distances of the order
$b = \mathcal{O}(M_\pi^{-1})$, where the pion mass is regarded as parametrically small on the 
hadronic mass scale (as represented e.g.\ by the vector meson mass $M_V$)
\cite{Strikman:2010pu,Granados:2013moa}. At such ``peripheral'' distances the densities are 
governed by the effective dynamics resulting from the spontaneous breaking of chiral symmetry 
and can be computed model-independently using methods of chiral effective field theory ($\chi$EFT)
\cite{Gasser:1983yg,Gasser:1984gg}; see Refs.~\cite{Bernard:1995dp,Scherer:2002tk,Scherer:2012zzd} for a review.
The peripheral densities are generated by chiral processes in which the hadron couples to the 
current operator through soft-pion exchange in the $t$-channel. The nucleon's peripheral charge 
and magnetization densities (related to the electromagnetic form factors $F_1$ and $F_2$) were studied 
in Ref.~\cite{Granados:2013moa} using relativistic $\chi$EFT in the leading-order approximation. 
The densities decay exponentially at large $b$ with a range of the order $1/(2 M_\pi)$, and exhibit a 
rich structure due to the coupling of the two-pion exchange to the nucleon. It was found that 
the peripheral charge and current densities are similar in magnitude and related by an approximate
inequality. These properties are a consequence of the essentially relativistic nature of chiral dynamics 
[the typical pion momenta are $\mathcal{O}(M_\pi)$] and can be understood in a simple quantum-mechanical 
picture of peripheral nucleon structure \cite{Granados:2015rra,Granados:2015lxa,Granados:2016jjl}.
It was also shown that the $\chi$EFT densities exhibit the correct $N_c$-scaling behavior
if contributions from $\Delta$ isobar intermediate states
are included \cite{Granados:2013moa,Granados:2016jjl}.

The study of peripheral transverse densities is closely connected with the dispersive representation of 
the nucleon form factors. The form factors are analytic functions of the invariant momentum transfer $t$ 
and can be represented as dispersive integrals of their imaginary parts (spectral functions) on the 
two-pion cut at timelike $t > 4 \, M_\pi^2$. The integration extends over the unphysical region
below the nucleon-antinucleon threshold, where the spectral functions cannot be measured 
directly, but can be calculated theoretically using 
$\chi$EFT \cite{Gasser:1987rb,Bernard:1996cc,Becher:1999he,Kubis:2000zd,Kaiser:2003qp}
and dispersion theory \cite{Frazer:1960zza,Frazer:1960zzb,Hohler:1974ht,Belushkin:2005ds,Hoferichter:2016duk}, 
or determined by fits to the spacelike form factor data \cite{Hohler:1976ax,Belushkin:2006qa,Lorenz:2012tm}.
The transverse densities are given by similar dispersive integrals, in which the spectral functions 
are integrated with an exponential weight factor $\sim \exp (-b\sqrt{t})$, suppressing the 
contribution from large masses \cite{Strikman:2010pu,Miller:2011du}. At distances $b = \mathcal{O}(M_\pi^{-1})$
the integration is limited to the near-threshold region $t - 4 \, M_\pi^2 = \mathcal{O}(M_\pi^2)$,
where the spectral functions are governed by chiral dynamics. The peripheral densities thus 
represent ``clean'' chiral observables and are ideally suited for $\chi$EFT calculations.
Quantitative studies in Ref.~\cite{Miller:2011du} found that at distances $b > 3\, \textrm{fm}$ the 
dispersive integrals for the densities are dominated by the region $4 \, M_\pi^2 < t \lesssim 10 \, M_\pi^2$, 
where the isovector spectral functions can be computed directly using fixed-order $\chi$EFT. 
To obtain the densities at smaller distances $b \sim 1\, \textrm{fm}$ one needs to construct the
spectral functions at larger values of $t$, where they are affected by strong $\pi\pi$ rescattering in the
$t$-channel and the $\rho$ meson resonance at $t \sim M_\rho^2 = 30 \, M_\pi^2$. 
This can be accomplished in an approach that combines $\chi$EFT with a dispersive technique
based on elastic unitarity in the $t$-channel.

In this article we study the peripheral transverse densities of the spin-1/2 flavor-octet baryons 
using methods of $\chi$EFT and dispersive analysis. We calculate the spectral functions of the isovector
electromagnetic form factors on the two-pion cut at $t > 4 \, M_\pi^2$, using relativistic $\chi$EFT 
with $SU(3)$ flavor group and explicit spin-3/2 decuplet degrees of freedom at $\mathcal{O}(\epsilon^3)$
in the small-scale expansion. The calculations
are extended from the near-threshold region to masses $t \sim 16 \, M_\pi^2$ and higher, using a dispersive 
technique that incorporates $\pi\pi$ rescattering through elastic unitarity and the pion electromagnetic 
form factor data. The method allows us to compute 
the isovector transverse densities at distances $b \gtrsim 1 \, \textrm{fm}$ with controlled 
uncertainties. For the isoscalar densities we construct an empirical parametrization of the spectral 
function in terms of vector meson poles ($\omega, \phi$) using $SU(3)$ symmetry. We present
the isovector and isoscalar peripheral transverse densities of the individual octet baryons and
discuss their properties. We also calculate the flavor decomposition of the peripheral densities
and compare it to the baryons' valence quark content.

The treatment of spectral functions and transverse densities 
presented here goes beyond that of previous studies in 
several aspects. First, the $\chi$EFT calculations are performed with the $SU(3)$ flavor group, which allows 
us to study for the first time the peripheral densities of all octet baryons resulting from two-pion exchange in the $t$-channel
(contributions from two-kaon exchange are included as well but turn out to be negligible at peripheral 
distances). Second, the spin-3/2 decuplet baryons are included in the $\chi$EFT calculations using a consistent
Lagrangian, which eliminates off-shell effects in physical observables. This approach, together 
with a fully relativistic formulation of $\chi$EFT with baryons via the extended-on-mass-shell (or EOMS)
scheme \cite{Fuchs:2003qc}, overcomes certain specific difficulties that arise in $\chi$EFT 
calculations of fundamental 
processes \cite{Becher:2001hv,Alarcon:2011kh, Alarcon:2012kn,Lensky:2014dda,Blin:2016itn} 
and has been applied successfully to the form factors of the nucleon and the $\Delta$ isobar \cite{Ledwig:2011cx}, 
the magnetic moments of the octet and decuplet baryons \cite{Geng:2008mf},
the axial-vector charge of the octet \cite{Ledwig:2014rfa},
and to the calculation of quantities like the nucleon $\sigma$-terms and two photon exchange corrections to the muonic hydrogen Lamb shift \cite{Alarcon:2011zs,Alarcon:2012nr,Alarcon:2013cba}, which are used in searches of physics beyond the standard model. Earlier $\chi$EFT calculations of 
peripheral densities \cite{Granados:2013moa,Granados:2016jjl} included the $\Delta$ isobar through a 
``minimal'' $\pi N \Delta$ coupling with off-shell dependence; this scheme was sufficient to achieve 
proper $N_c$-scaling of the spectral functions and densities, but could not fix the contact terms 
associated with the isobar contribution; see Ref.~\cite{Granados:2013moa,Granados:2016jjl} for details. 

Third, we extend the range of the $\chi$EFT calculation of the spectral function beyond the near-threshold 
region by combining it with a dispersive method based on elastic unitarity. It uses the representation of 
the spectral functions on the two-pion cut as the product of the $\pi\pi \rightarrow B\bar B$ $t$-channel 
partial-wave amplitudes in the $I = J = 1$ channel,
$\Gamma^B_i (t) \; (i = 1, 2)$, and the complex-conjugate pion form factor, $F_\pi^\ast(t)$. 
Chiral EFT is used to calculate the real function $\Gamma^B_i (t) / F_\pi(t) \; (i = 1, 2)$, in which the 
common phase due to $\pi\pi$ rescattering 
cancels by virtue of the Watson theorem \cite{Frazer:1960zza,Frazer:1960zzb,Hohler:1974ht,Watson:1954uc};
the result is then multiplied by the empirical $|F_\pi(t)|^2$ measured in $e^+e^-$ annihilation 
experiments, which contains the $\pi\pi$ rescattering effects and the $\rho$ resonance. 
The ``improved'' $\chi$EFT spectral functions thus obtained agree well with the 
empirical spectral functions (obtained by analytic continuation of the partial 
wave amplitudes) \cite{Hohler:1976ax,Belushkin:2005ds} up to $t \sim 16 \, M_\pi^2$ and have 
qualitatively correct behavior even in the $\rho$ meson mass region. The agreement is achieved 
without adding free parameters and represents a genuine prediction of $\chi$EFT at 
$\mathcal{O}(\epsilon^3)$; the inclusion of higher-order chiral corrections might further improve 
the accuracy of the results in the $\rho$ meson mass region.
The method allows us to compute the isovector transverse densities down to distances 
$b \sim 1 \, \textrm{fm}$ with controlled accuracy and 
results in a major overall improvement in the description of peripheral 
nucleon structure. The method could be extended to the spectral functions of other form 
factors with two-pion cut and their transverse densities \cite{InPreparation}.
The technique described here is a variant of the $N/D$ method of amplitude 
analysis \cite{Chew:1960iv}, which was employed 
in applications of $\chi$EFT to meson-meson, meson-baryon, and baryon-baryon 
scattering \cite{Alarcon:2011kh,Oller:1998zr,Oller:2000ma,Meissner:1999vr,Oller:2000fj,%
Albaladejo:2011bu,Albaladejo:2012sa}.  
A similar approach was used in a recent study of the $\Sigma$-$\Lambda$ 
transition form factors \cite{Granados:2017cib}.

The article is structured as follows. In Sec.~\ref{Sec:Transverse_densities} we summarize
the basic properties of the transverse densities and discuss their dispersive representation.
In particular, we analyze the distribution of strength in the dispersion integral for the
densities at different distances and identify the regions of $t$ where we need to calculate
the spectral functions. In Sec.~\ref{Sec:Spectral_Functions} we describe the $\chi$EFT calculation
of the isovector spectral functions, the improvement using dispersion theory, and the 
modeling of isoscalar spectral functions. In Sec.~\ref{Sec:Transverse_densities_octet} we
present the results for the charge and magnetization densities of the octet baryons and
discuss their properties. We also perform the flavor separation
of the transverse densities and discuss the connection with partonic structure.
A summary and outlook on further studies are given in Sec.~\ref{Sec:Summary}.
Technical material is collected in the appendices. \ref{App:Validation} 
describes the validation of the form factor calculations. \ref{App:Delta} 
discusses the role of the off-shell behavior of the $\chi$EFT with $\Delta$ isobars and
compares our results with previous calculations. \ref{App:Vector} describes
our model for the coupling of isoscalar vector mesons to the $SU(3)$ octet baryons.
Preliminary results of the present study were reported in Ref.~\cite{Alarcon:2017lkk}.
\section{Transverse densities}
\label{Sec:Transverse_densities}
\subsection{Form factors and transverse densities}
\label{SubSec:emFFs}
The transition matrix element of the electromagnetic current between states of a spin-1/2 
baryon $B$ with 4-momenta $p$ and $p'$ is of the general form
\begin{align}
\label{Eq:EM-current-matrix-element}
\langle B(p')| J^\mu | B(p) \rangle = 
\bar{u}(p')\left[\gamma^\mu F_1^B(t)+\frac{i\sigma^{\mu\nu}\Delta_\nu}{2m_B} F_2^B(t)\right]u(p),
\end{align}
where $u(p)$ and $\bar u(p')$ are the Dirac spinors of the baryon states, $m_B$ is the baryon mass, 
$\Delta=p'-p$ is the 4-momentum transfer, and 
$\sigma^{\mu\nu} = (i/2)\left[\gamma^\mu,\gamma^\nu\right]$. The form factors $F_1^B(t)$ 
and $F_2^B(t)$ (Dirac and Pauli form factor) are functions of the Lorentz-invariant momentum transfer 
$t = \Delta^2$, 
with $t < 0$ 
in the physical region of elastic scattering. They are invariant functions and can be measured and 
interpreted without specifying the form of relativistic dynamics or choosing a reference frame.
The isospin decomposition of Eq.~(\ref{Eq:EM-current-matrix-element}) in the case of $SU(3)$ octet baryons 
($B = N, \Lambda, \Sigma, \Xi$) will be described below. 

%
%
\begin{figure}
\begin{center}
\epsfig{file=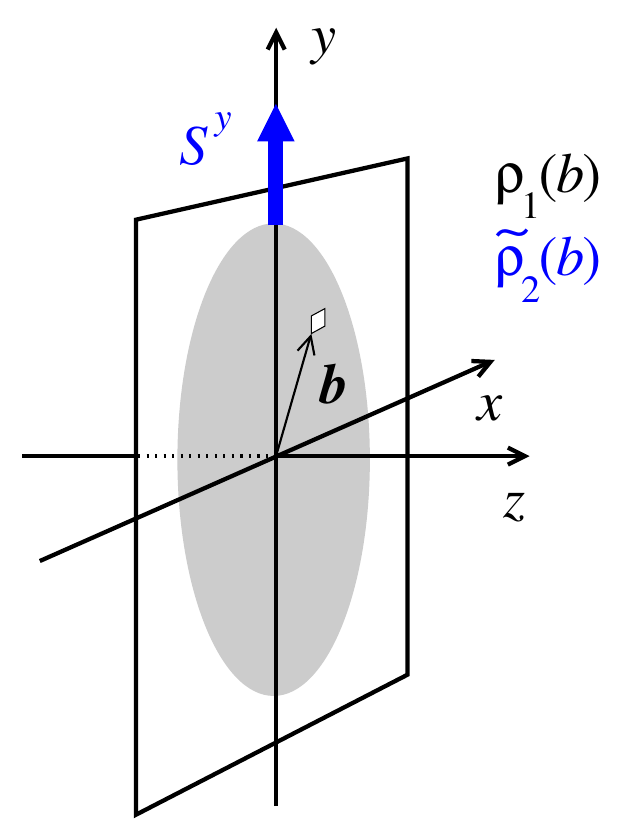,width=.26\textwidth,angle=0} 
\caption[]{Interpretation of the baryon
transverse densities, cf.~Eq.~(\ref{Eq:EM-current-matrix-element}). 
The function $\rho_1(b)$ describes the spin-independent part of the expectation 
value of the $J^+$ current in a baryon state localized at the transverse origin.
The function $\cos\phi\ \widetilde\rho_2(b)$ describes the spin-dependent part 
of the current in a state polarized in the positive $y$-direction.}
\label{fig:interpretation}
\end{center}
\end{figure}
In the light-front form of relativistic dynamics one considers the evolution of strong interactions 
in light-front time $x^+ \equiv x^0 + x^3 = x^0 + z$. The baryon states are parametrized by their 
light-front momentum $p^+ \equiv p^0 + p^z$ and transverse momentum $\bm{p}_T \equiv (p^x, p^y)$, 
whereas $p^- \equiv p^0 - p^z$ plays the role of the energy, with $p^- = (p^2_T + m_B^2)/p^+$
on the baryon mass shell $p^2 = m_B^2$. 
In this context one naturally considers the current matrix element
in a class of reference frames where the momentum transfer has only transverse components, 
$\Delta^+ = \Delta^- =0, \bm{\Delta}_T \neq 0, |\bm{\Delta}_T|^2 = -t$. 
The form factors are then represented as Fourier transforms of two-dimensional 
spatial densities (transverse densities)
\begin{align}
\label{Eq:rho_def}
F_i^B (t = -|\bm{\Delta}_T|^2) \;\; = \;\; \int d^2 b \; 
e^{i \bm{\Delta}_T \cdot \bm{b}} \; \rho_i^B (b) 
\hspace{2em} (i=1,2),
\end{align}
where $\bm{b} \equiv (b^x, b^y)$ is a transverse coordinate variable and $b \equiv |\bm{b}|$;
we follow the conventions of Ref.~\cite{Granados:2013moa}.
The spatial integral of the densities reproduces the total charge and anomalous magnetic moment 
of the baryons,
\begin{align}
\label{Eq:rho_normalization}
\int d^2 b \; \rho_1^B (b) \;\; = \;\; F_1^B(0) \;\; = \;\; Q^B ,
\\
\int d^2 b \; \rho_2^B (b) \;\; = \;\; F_2^B(0) \;\; = \;\; \kappa^B .
\end{align}
The functions thus describe the transverse spatial distribution of charge and magnetization in the baryon.
The interpretation of $\rho_1^B (b)$ and $\rho_2^B(b)$ as spatial densities has been discussed extensively 
in the literature \cite{Burkardt:2000za,Burkardt:2002hr,Miller:2010nz} and is summarized in Ref.~\cite{Granados:2013moa}.
In a state where the baryon is localized in transverse space at the origin, and its spin (or light-front 
helicity) quantized in the $y$-direction, the expectation value of the current $J^+ (x)$ at 
$x^+ = x^- = 0$ and transverse position $\bm{x}_T = \bm{b}$ is given by
\begin{align}
\langle J^+ (\bm{b}) \rangle_{\text{\scriptsize localized}}
& = (...) \left[\rho_1^B (b) + (2 S^y) \cos\phi \, \widetilde\rho_2^B (b) \right] , 
\label{j_plus_rho}
\\[2ex]
\widetilde\rho_2^B (b) & \equiv \frac{\partial}{\partial b} 
\left[ \frac{\rho_2^B (b)}{2 m_B} \right] .
\label{rho_2_tilde_def}
\end{align}
Here $(...)$ represents a factor resulting from the normalization of states 
(see Ref.~\cite{Granados:2013moa} for details), $\phi$ is the angle of the vector 
$\bm{b}$ relative to the $x$-axis, and $S^y = \pm 1/2$ is the spin projection 
in the $y$-direction in the baryon rest frame (see Fig.~\ref{fig:interpretation}).
$\rho_1^B(b)$ describes the spin-independent part of the plus current in a localized
baryon state, while $\cos\phi \, \widetilde\rho_2^B (b)$ describes the spin-dependent part of 
the current in a transversely polarized baryon. Other aspects of the transverse densities, such as 
their connection with the partonic structure and GPDs, 
are discussed in the literature and reviewed in Ref.~\cite{Miller:2010nz}.

In the present work the concepts of light-front quantization are used only for the 
interpretation of the transverse densities but will not be employed in the actual calculations.
Dynamical calculations (chiral EFT, dispersion theory) will be performed at the level 
of the invariant form factors, from which the densities can be obtained by inverting
the Fourier transform of Eq.~(\ref{Eq:rho_def}). The advantage of transverse densities
is precisely that they connect light-front structure with the invariant form factors, 
for which well-tested theoretical methods are available. An alternative approach, in which 
dynamical calculations in chiral EFT are performed directly in light-front quantization,
is described in Refs.~\cite{Granados:2015rra,Granados:2015lxa,Granados:2016jjl}.
\subsection{Dispersive representation}
\label{Subsec:dispersive_representation}
\begin{figure}[t]
\begin{center}
\epsfig{file=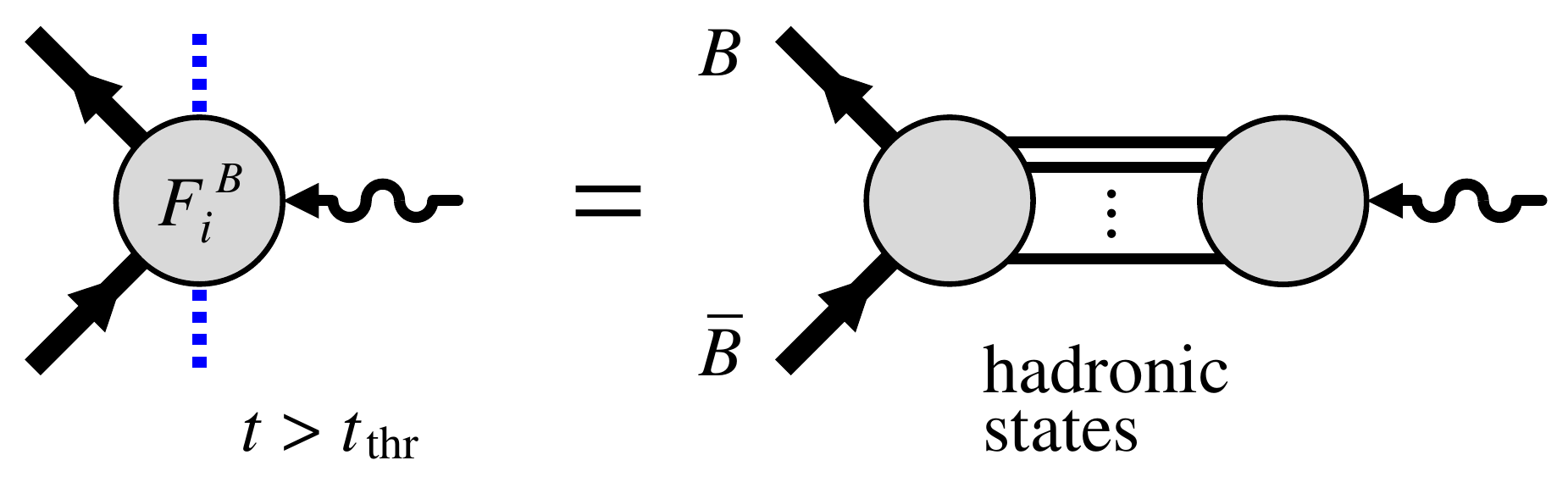,width=.45\textwidth,angle=0} 
\end{center}
\caption[]{Spectral function of the baryon form factors.}
\label{fig:spectral_hadron}
\end{figure}

The baryon form factors are analytic functions of $t$, with singularities (branch points, poles)
on the positive real axis. Assuming power-like asymptotic behavior $F_i^B(t) \sim t^{-(i+1)} \; 
(i = 1, 2)$, as expected in perturbative QCD and supported by present experimental data for 
the nucleon, the form factors satisfy unsubtracted dispersion relations of the form
\begin{align}
F_i^B (t) \;\; = \;\; 
\int_{t_{\rm thr}}^\infty \frac{dt'}{t' - t - i0} 
\; \frac{\textrm{Im}\, F_i^B (t')}{\pi} 
\hspace{2em} (i=1,2) .
\label{Eq:ff-spectral-rep}
\end{align}
Here the form factors are expressed as integrals of their imaginary parts (or spectral functions) 
on the principal cut on the physical sheet. The singularities at $t > t_{\rm thr}$ 
correspond to processes
in which the current produces a hadronic state that couples to the baryon-antibaryon system,
current $\rightarrow$ hadronic state $\rightarrow$ $B\bar B$ (see Fig.~\ref{fig:spectral_hadron}). 
The hadronic state with the lowest mass is the two-pion state with threshold at 
$t_{\rm thr} = 4 \, M_\pi^2$, 
produced by the isovector current 
(anomalous thresholds with $t_{\rm thr} < 4 M_\pi^2$ 
in the strange octet baryon form factors will be discussed in Sec.~\ref{Subsec:Improvement}).
At larger values $t \lesssim 1\, \textrm{GeV}^2$ 
the dominant states are the vector mesons appearing as resonances 
in the two-pion channel ($\rho$, isovector) and the three-pion channel ($\omega$, isoscalar). 
At even higher $t$ the spectral functions receive contributions from $K\bar K$ states and the 
$\phi$ resonance, and from other multi-hadron states whose composition is not known in detail.
Most of the relevant states lie below the $B\bar B$ threshold ($t = 4 \, m_N^2 = 3.5$ GeV$^2$ for the nucleon),
in the unphysical region of the current $\rightarrow$ $B\bar B$ process, where the spectral functions 
cannot be measured directly in conversion experiments, but can only be calculated theoretically.
Methods for constructing the spectral functions in different regions of $t$ ($\chi$EFT, dispersion
analysis) will be described below.
%
%
\begin{figure*}[t]
\begin{tabular}{ll}
\epsfig{file=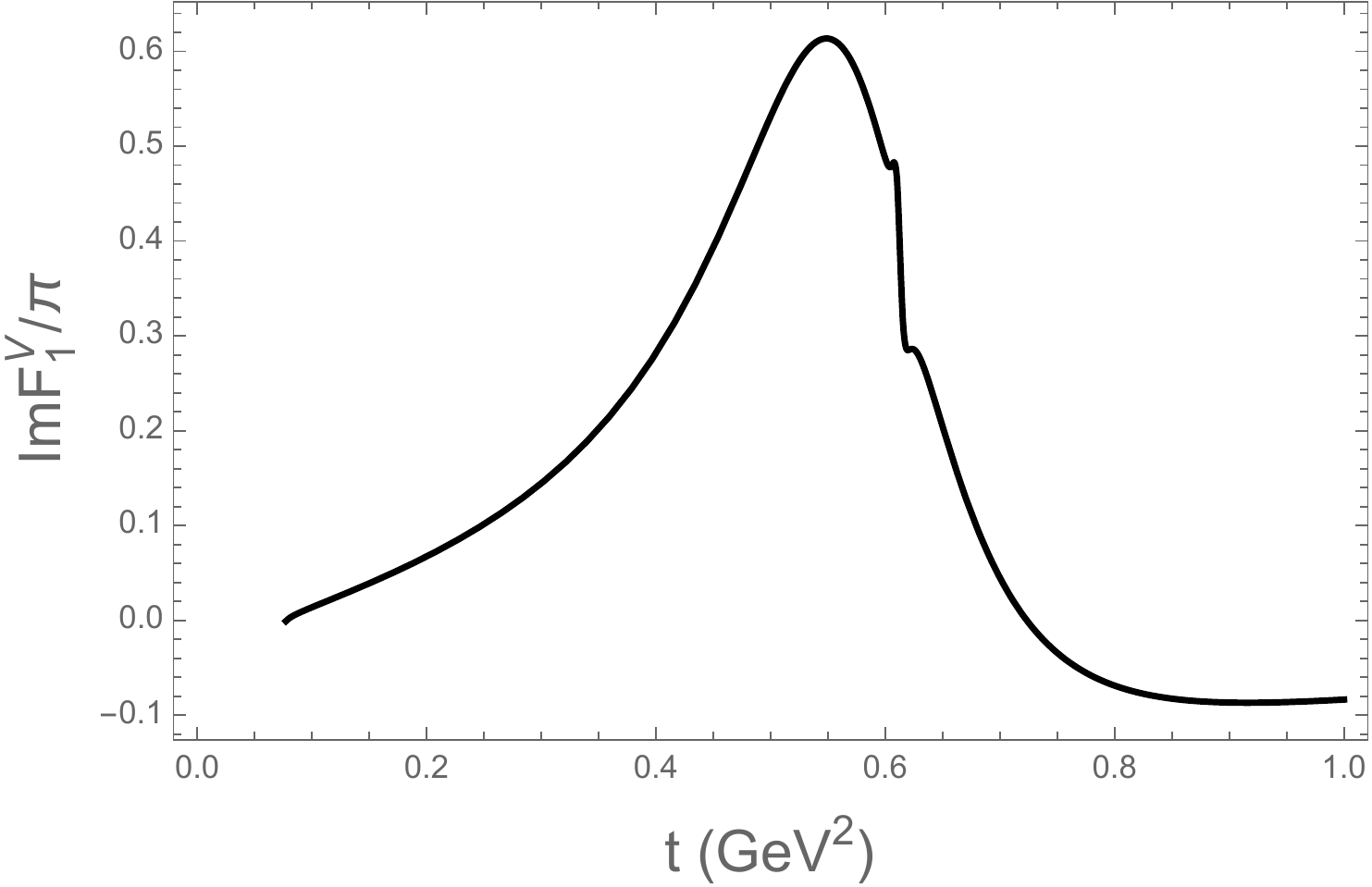,width=.45\textwidth,angle=0} &
\epsfig{file=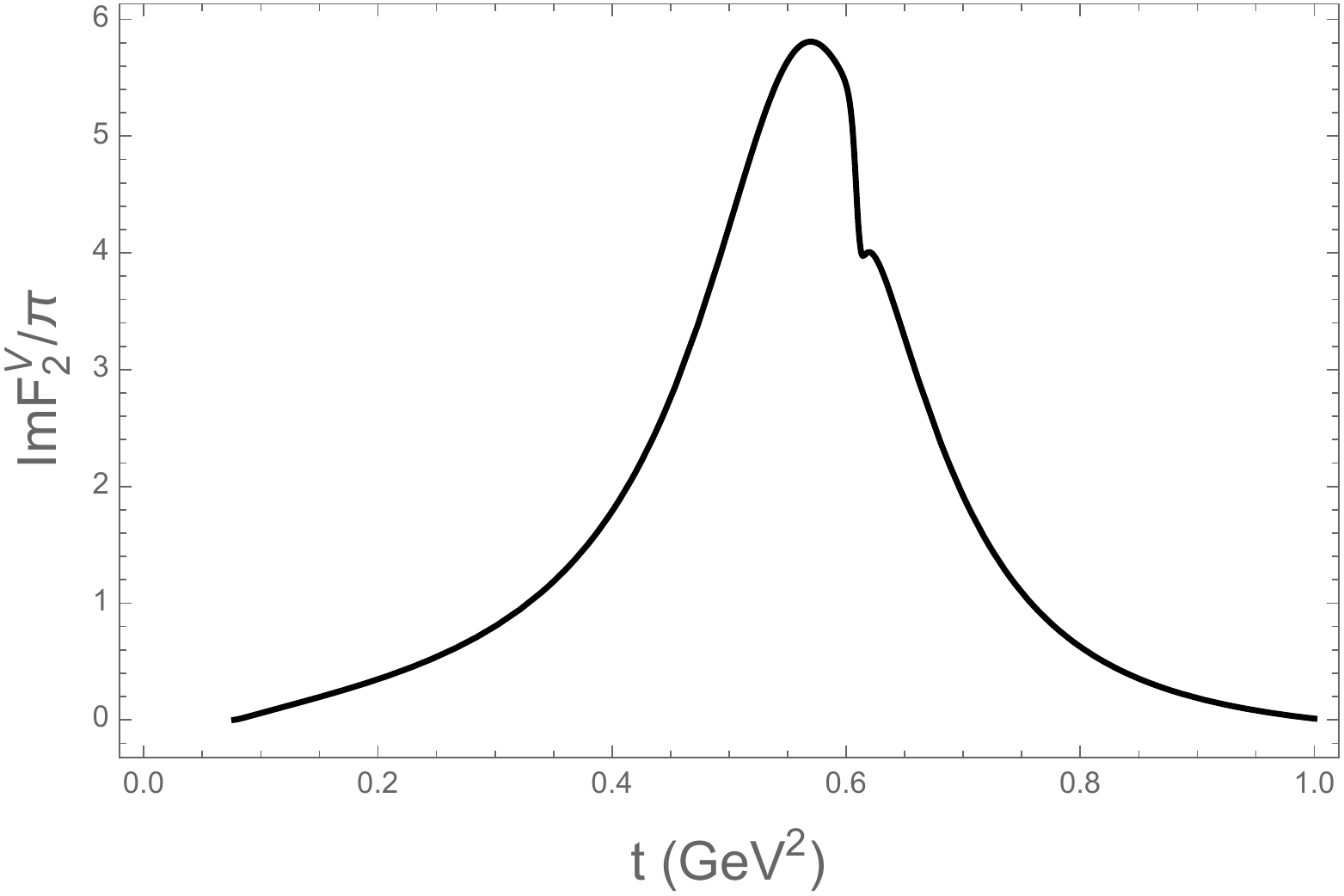,width=.45\textwidth,angle=0}
\\[-2ex]
{\footnotesize (a)} & {\footnotesize (b)}
\\[2ex]
\epsfig{file=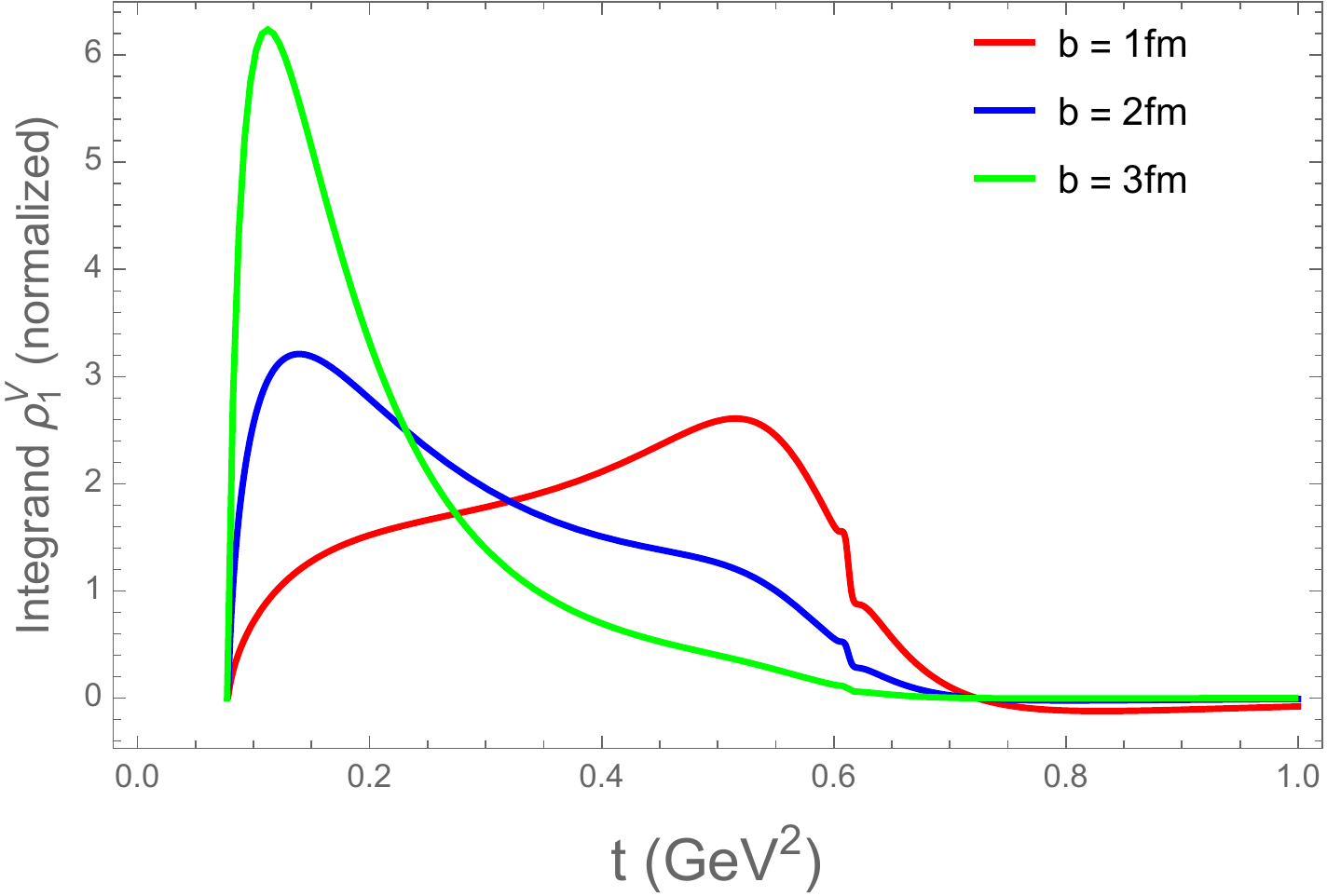,width=.45\textwidth,angle=0} &
\epsfig{file=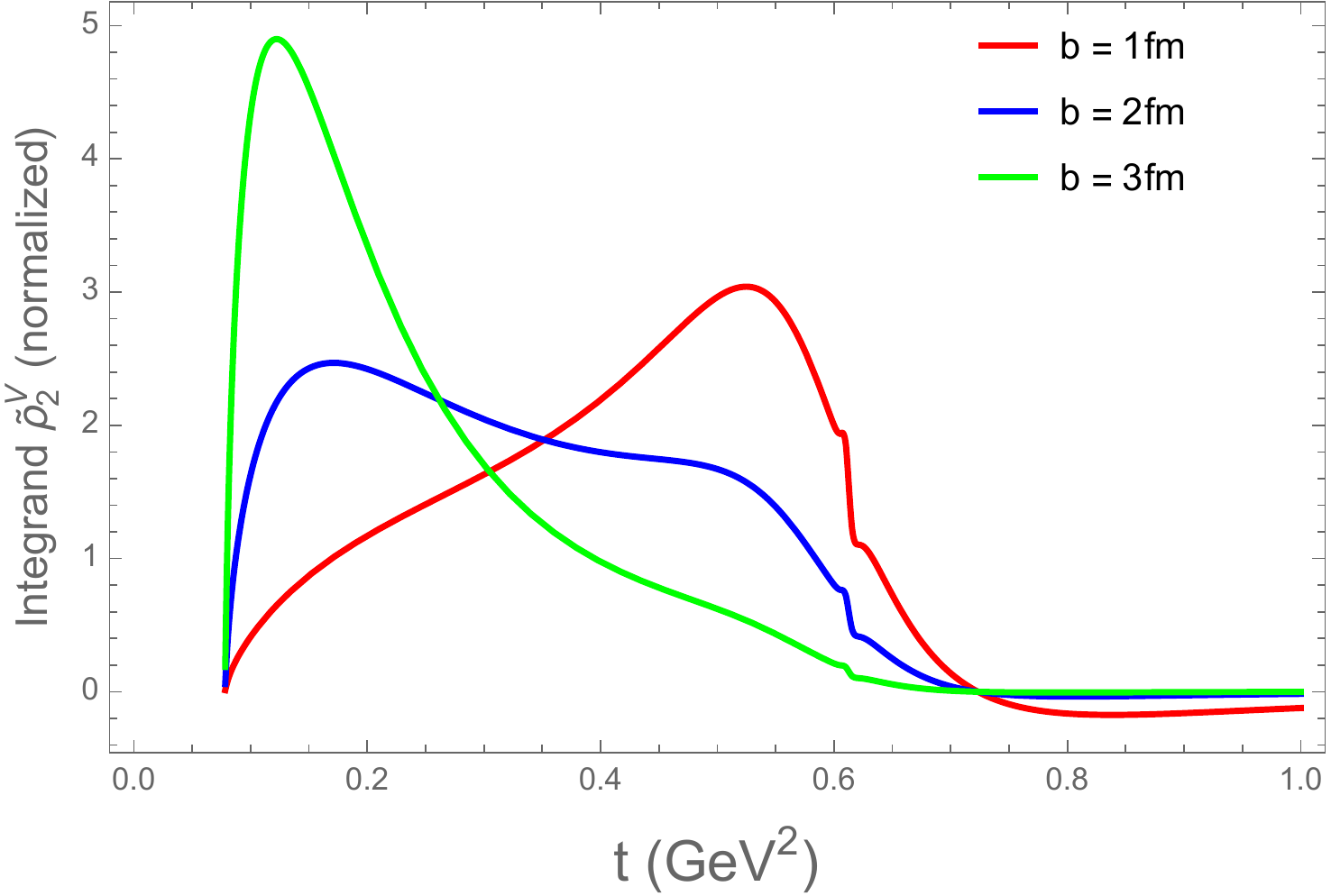,width=.45\textwidth,angle=0}
\\[-2ex]
{\footnotesize (c)} & {\footnotesize (d)}
\end{tabular}
\caption[]{\small (a, b) Spectral functions of the nucleon isovector form factors, 
$\textrm{Im} \, F_1^V(t)$ and $\textrm{Im} \, F_2^V(t)$, as determined in Ref.~\cite{Hoferichter:2016duk}.
(c, d) Distribution of strength in $t$ in the dispersive integrals for the transverse charge
density $\rho_1^V(b)$, Eq.~(\ref{Eq:rho1-spectral-rep}), and the transverse magnetization density 
$\widetilde\rho_2^V(b)$, Eq.~(\ref{Eq:rho2tilde-spectral-rep}), at different distances $b$.
The plots show the integrands as functions of $t$, divided by the value of the integral, i.e., 
normalized to unit area under the curves.
\label{Fig:rho_integrand}}
\end{figure*} 
A similar dispersive representation can be derived for the transverse densities associated
with the form factors, Eq.~(\ref{Eq:rho_def}). Calculating the transverse Fourier transform 
of the form factors as given by Eq.~(\ref{Eq:ff-spectral-rep}), and using Eq.~(\ref{rho_2_tilde_def}), 
one obtains \cite{Strikman:2010pu,Miller:2011du} 
\begin{align}
& \rho_1^B (b) = \phantom{-} \int_{t_{\rm thr}}^{\infty} \! dt\ \frac{K_0(\sqrt{t}b)}{ 2\pi}
\frac{\text{Im}F_1^B(t)}{\pi} ,
\label{Eq:rho1-spectral-rep} \\
& \widetilde{\rho}_2^B (b) = - \int_{t_{\rm thr}}^{\infty} \! dt\ \frac{\sqrt{t}
K_1(\sqrt{t}b)}{ 2\pi} \frac{\text{Im} F_2^B(t)}{\pi},
\label{Eq:rho2tilde-spectral-rep}
\end{align}
where $K_n(z) \; (n = 0, 1)$ denotes the modified Bessel functions of the second kind.
This representation has several interesting properties. (a) The modified Bessel functions
decay exponentially at large arguments,
\begin{align}
K_n (\sqrt{t} b) \;\; \sim \;\; [\pi / (2 \sqrt{t} b)]^{1/2}
\; e^{-\sqrt{t} b} 
\hspace{2em} (\sqrt{t} b \; \gg \; 1).
\label{K0_asymptotic}
\end{align}
The dispersive integrals for the densities therefore converge exponentially at large $t$,
while the original integrals for the form factors converge only like a power. This greatly 
reduces the contribution from the high-mass region where the spectral functions are poorly known.
(b) The density at a certain $b$ is given by the integral of the spectral function with the
``exponential filter'' $K_n (\sqrt{t} b)$. The distance $b$ therefore represents an external 
parameter that allows one to effectively select (emphasize or de-emphasize) certain mass regions 
in the spectral function. The large-distance behavior of the densities is governed by the
lowest-mass states in a given channel. We shall use this property to identify the region
of $b$ in which the densities are dominated by values of $t$ close to the two-pion threshold, 
where the spectral functions can be computed using $\chi$EFT and dispersion theory. 
(c) The representations Eqs.~(\ref{Eq:rho1-spectral-rep}) and 
(\ref{Eq:rho2tilde-spectral-rep})
incorporate the correct analytic behavior of the form factors and can be used to study the
large-distance asymptotics of the densities.\footnote{Form factor representations with incorrect
analytic properties in $t$ (e.g., with artificial singularities at complex $t$) are principally 
not adequate for studying the large-distance asymptotics of the densities, even if they provide
good fits to the spacelike form factor data at small $t < 0$; see Ref.~\cite{Miller:2011du}
for a discussion.}

It is instructive to study the distribution of strength in $t$ in the dispersive integrals for 
the baryon densities, Eqs.~(\ref{Eq:rho1-spectral-rep}) and 
(\ref{Eq:rho2tilde-spectral-rep}), at different $b$ \cite{Miller:2011du}. 
For this purpose we consider the nucleon form factors, whose isovector spectral functions at
$t < 1\,\textrm{GeV}^2$ have been calculated using amplitude analysis techniques
(analytic continuation of the $\pi\pi \rightarrow N\bar N$ partial-wave amplitudes
\cite{Belushkin:2005ds,Hohler:1976ax}, Roy-Steiner equations \cite{Hoferichter:2016duk}). 
Figures~\ref{Fig:rho_integrand}a and b show the spectral functions of the nucleon's isovector 
form factors $F_1^V$ and $F_2^V$ obtained in Ref.~\cite{Hoferichter:2016duk}.
One observes the rise of the spectral functions in the region above the 
two-pion threshold $t > 4 \, M_\pi^2$, and the prominent role of the $\rho$ resonance at
$t \sim 30 \, M_\pi^2 = 0.6 \, \textrm{GeV}^2$ (the rapid variation on the right shoulder of the $\rho$ is due to $\rho$-$\omega$ 
mixing caused by isospin breaking). 
Figures~\ref{Fig:rho_integrand}c and d show 
the distribution of strength in the dispersive integrals Eqs.~(\ref{Eq:rho1-spectral-rep})
and (\ref{Eq:rho2tilde-spectral-rep}),
i.e., the spectral functions multiplied by $K_0(\sqrt{t} b)$ and $\sqrt{t} K_1(\sqrt{t} b)$, 
for $b = 1, 2$ and 3 fm. (The plots show the distributions in $t$ normalized to unit area under 
the curves, to facilitate comparison of the relative distribution of strength at different $b$.) 
One sees that for $b = 3$ fm the dominant contribution to the integrals arises from the near-threshold
region $t < 0.3$ GeV$^2$, while the vector meson region $t \sim 0.5 \, \textrm{GeV}^2$ is very strongly 
suppressed. For $b = 2$ fm the near-threshold region still gives the main contribution to the integrals,
but the vector meson region becomes noticeable, especially in $\widetilde\rho_2^V$.
For $b = 1$ fm, finally, the vector meson mass region gives the main contribution to the integrals. 
These estimates illustrate in what regions of $t$ we need to calculate (or model) the spectral 
functions in order to compute the densities at a given $b$, and how the uncertainties of the 
spectral function in the different regions affect the overall uncertainty of the densities. 
Similar estimates can be performed for the isoscalar densities of the nucleon and the densities 
of the other flavor-octet baryons, using the spectral functions calculated below.

In this work we use relativistic $\chi$EFT to calculate the isovector spectral functions of
the flavor-octet baryons on the two-pion cut. The relativistic $\chi$EFT framework correctly
implements the subthreshold singularities of the form factors on the unphysical sheet,
which result from the baryon Born terms and account for the steep rise of the isovector spectral 
function above threshold \cite{Gasser:1987rb,Bernard:1996cc,Becher:1999he,Kubis:2000zd,Kaiser:2003qp}. 
The inclusion of the decuplet baryons as dynamical degrees of freedom implements also the Born term 
singularities further removed from the physical region, which become important at larger $t$. 
To extend the $\chi$EFT calculations of the spectral functions into the vector meson mass region we combine 
them with a dispersive technique that incorporates the pion form factor data from $e^+e^-$ annihilation 
experiments \cite{Frazer:1960zza,Frazer:1960zzb,Hohler:1976ax}. In this way we construct the isovector 
spectral functions of the octet baryons up to $t \sim 0.3$ GeV$^2$ with an estimated 
uncertainty $< 30\%$, and up to $t \sim 1$ GeV$^2$ with somewhat larger uncertainty. 
The isoscalar spectral functions we model by phenomenological vector meson exchange ($\omega, \phi$).
Non-resonant isoscalar contributions from kaon loops in the $SU(3)$ EFT turn out to be negligible.
Altogether, these methods allow us to compute the peripheral densities of the octet baryons at
distances $b \gtrsim$ 1 fm with controlled accuracy.
\subsection{Isospin structure}
In the present study we consider the form factors and densities of baryons in the octet representation 
of the $SU(3)$ flavor group. The electromagnetic current operator in QCD with 3 flavors is given by
$J^\mu = e \bar\psi Q \gamma^\mu \psi$, where $\psi$ is the quark field and $Q$ the quark 
charge matrix,
\beq 
Q = {\textstyle \textrm{diag}(\frac{2}{3}, -\frac{1}{3}, -\frac{1}{3})} 
= {\textstyle\frac{1}{2}} \lambda^3 + {\textstyle\frac{1}{2\sqrt{3}}} \lambda^8,
\label{quark_charge}
\eeq
where $\lambda^a$ are the Gell-Mann matrices.
The term $\propto \lambda^3$ in the current transforms as an isovector under $SU(2)$ isospin
rotations ($V$); the term $\propto \lambda^8$ transforms as an isoscalar ($S$), and the 
current can be represented as the sum of an isovector and an isoscalar current,
$J^\mu = J^{\mu, V} + J^{\mu, S}$. The matrix elements of the isovector and isoscalar currents 
can be specified for each baryon state.

The baryons in the octet representation of $SU(3)$ form four isospin multiplets
\beq
\left. 
\begin{array}{ccc}
\textrm{multiplet} & \textrm{baryons} & \textrm{isospin} \\
N & p, n & I = {\textstyle\frac{1}{2}} 
\\
\Lambda & \Lambda & I = 0
\\
\Sigma & \Sigma^+, \Sigma^-, \Sigma^0 & I = 1
\\
\Xi & \Xi^0, \Xi^- & I = {\textstyle\frac{1}{2}}. 
\end{array}
\hspace{1em} \right\}
\eeq
Within each multiplet we write the form factors as the sum/difference of
an isoscalar and isovector component,
\beq
\left.
\begin{array}{rcl}
\{ F_i^p, F_i^n \} &=& F_i^{N, S} \pm F_i^{N, V} ,
\\
F_i^{\Lambda} &=& F_i^{\Lambda, S} ,
\\
\{ F_i^{\Sigma^+}, F_i^{\Sigma^-} \} &=& F_i^{\Sigma, S} \pm F_i^{\Sigma, V} ,
\\
F_i^{\Sigma^0} &=& F_i^{\Sigma, S} ,
\\
F_i^{\Lambda - \Sigma} &=& F_i^{\Lambda-\Sigma, V} ,
\\
\{ F_i^{\Xi^0}, F_i^{\Xi^-} \} &=& F_i^{\Xi, S} \pm F_i^{\Xi, V} 
\\[1ex]
&& (i=1,2).
\end{array}
\hspace{1em}
\right\}
\label{isospin_ff}
\eeq
For the nucleon this corresponds to the standard definition of the isoscalar and isovector 
form factors as $\{ F_i^{N, S}, F_i^{N, V} \} = {\textstyle\frac{1}{2}}(F_i^p \pm F_i^n)$.
The $\Lambda$ form factors are pure isoscalar.
In the $\Sigma^0$ form factors the isovector component is absent because the transition
$|I=1, I_3=0\rangle_{\textrm{current}} 
\rightarrow |I=1, I_3=0\rangle_B |I=1, I_3=0\rangle_{\bar{B}}$ is forbidden by
isospin symmetry. The $\Lambda$-$\Sigma^0$ transition form factors $F_i^{\Lambda-\Sigma}$ are 
pure isovector. Equation~(\ref{isospin_ff}) embodies the constraints imposed
on the octet baryon form factors by isospin symmetry. The same decomposition applies
to the spectral functions and the transverse densities.
\section{Spectral functions}
\label{Sec:Spectral_Functions}
\subsection{Chiral effective field theory}
\label{SubSec:ChiralEFT}
For the calculation of the baryon isovector spectral functions we use a manifestly Lorentz-covariant 
version of $\chi$EFT with $SU(3)$ flavor group including spin-1/2 flavor-octet and 
spin-3/2 flavor-decuplet baryons. The pseudoscalar mesons and spin-1/2 baryons
are described by fields in the $SU(3)$ octet representation (rank-2 tensors)
\begin{align}
\phi&=
\left(\begin{array}{ccc}
\frac{1}{\sqrt{2}}\pi^0 + \frac{1}{\sqrt{6}}\eta &\pi^+ &K^+\\
\pi^- & -\frac{1}{\sqrt{2}}\pi^0 + \frac{1}{\sqrt{6}}\eta & K^0\\
K^- & \bar{K}^0 & - \frac{2}{\sqrt{6}}\eta
\end{array}
\right),
\label{phi_octet}
\\
B&=
\left(\begin{array}{ccc}
\frac{1}{\sqrt{2}}\Sigma^0 + \frac{1}{\sqrt{6}}\Lambda &\Sigma^+ &p\\
\Sigma^- & -\frac{1}{\sqrt{2}}\Sigma^0 + \frac{1}{\sqrt{6}}\Lambda & n\\
\Xi^- & \Xi^0 & - \frac{2}{\sqrt{6}}\Lambda
\end{array}
\right) .
\label{B_octet}
\end{align}

The spin-3/2 baryons are described by fields in the $SU(3)$ decuplet 
representation (rank-3 totally symmetric tensor)
\beq
\left.
\begin{array}{lll}
T^{111}=\Delta^{++}, &T^{112}=\frac{1}{\sqrt{3}}\Delta^{+}, &T^{122}=\frac{1}{\sqrt{3}}\Delta^{0},\\[1ex]
T^{222}=\Delta^{-},  &T^{113}=\frac{1}{\sqrt{3}}\Sigma^{*+},&T^{123}=\frac{1}{\sqrt{6}}\Sigma^{*0},\\[1ex]
T^{223}=\frac{1}{\sqrt{3}}\Sigma^{*-}, &T^{133}=\frac{1}{\sqrt{3}}\Xi^{*0}, & \\[1ex]
T^{233}=\frac{1}{\sqrt{3}}\Xi^{*-}, & T^{333}=\Omega^{-}. &
\end{array}
\right\}
\eeq
The spin-1/2 fields $B$ are introduced as relativistic bispinor fields (Dirac fields). The spin-3/2 
fields are introduced as 4-vector-bispinor fields $T \equiv T_\mu$, which have to be subjected to 
relativistically covariant constraints to eliminate spurious spin-1/2 degrees of freedom. 
In the present formulation of the EFT the projection on spin-3/2 is implemented through the
use of consistent interaction vertices (see below); i.e., the spin-1/2 degrees of freedom are allowed 
to propagate but are filtered out in the interaction \cite{Pascalutsa:1998pw,Pascalutsa:1999zz, Pascalutsa:2000kd,Krebs:2008zb}. Electromagnetic interactions 
are introduced by the coupling to a 4-vector background field in the $SU(3)$ octet representation,
\begin{align}
v_\mu&=e\epsilon_\mu Q=
e\epsilon_\mu\left(\begin{array}{ccc}
\frac{2}{3} &0 &0\\
0 & -\frac{1}{3} & 0\\
0 & 0 & - \frac{1}{3}
\end{array}
\right) .
\end{align}

The construction of the chiral Lagrangian with spin-3/2 fields and the formulation of a
small-scale expansion have been described in Refs.~\cite{Hemmert:1996xg,Hemmert:1997ye}. 
In the $\epsilon$-expansion the octet-decuplet baryon mass splitting is counted at the same 
order as the pseudoscalar octet meson mass and momentum,
\beq
p, M_\phi \; \sim \; m_T - m_B \; \sim \; \epsilon ,
\eeq 
where $\epsilon$ denotes the generic expansion parameter.
The chiral order $N$ of a diagram is given by
\begin{align}
N=4L-2P_\phi-P_B-P_T+\sum_k k V_k,
\end{align}
where $L$ is the number of loops, $P_H$ the number of propagators of hadrons of the type $H$ 
and $V_k$ the number of vertices of a Lagrangian of order $k$. The standard power counting of 
$\chi$EFT is recovered with the extended-on-mass-shell (EOMS) scheme \cite{Fuchs:2003qc}, 
whereby the divergent 
parts of loop diagrams, along with the power-counting-breaking terms, are absorbed into the 
low-energy constants following the $\overline{MS}$ prescription.

\begin{figure}
\includegraphics[width=0.48\textwidth]{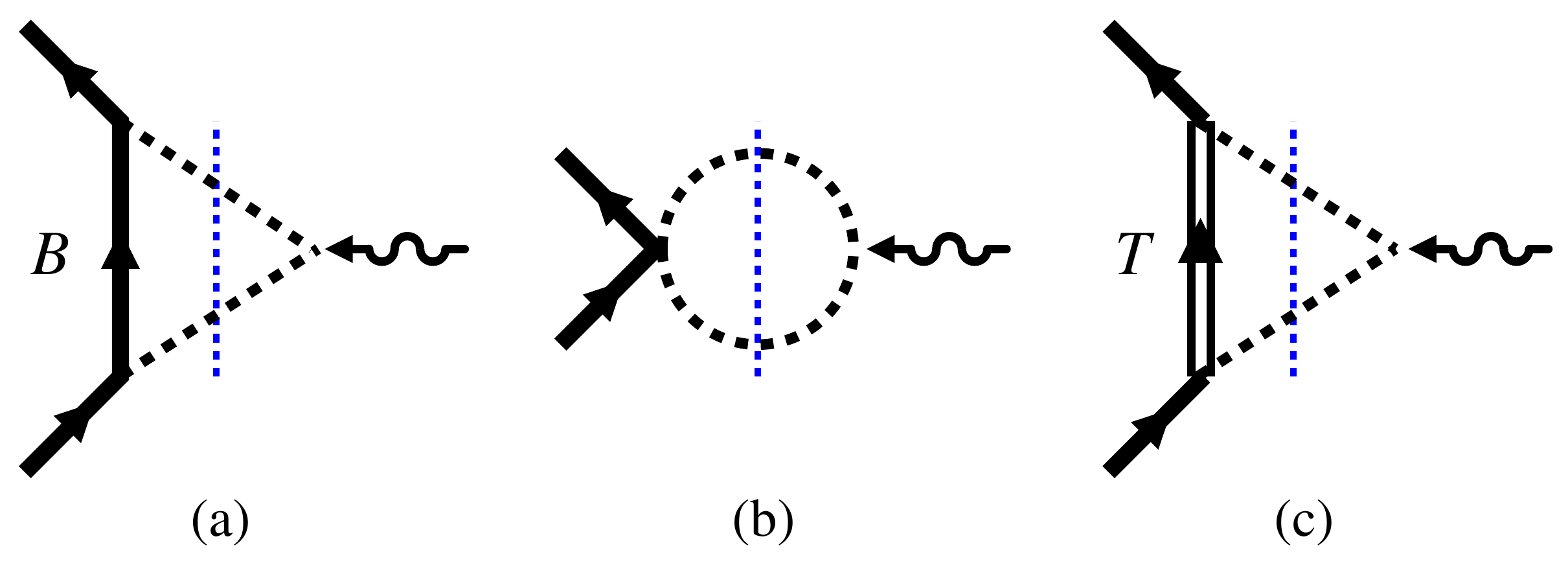}
\caption[]{
$\mathcal{O}(\epsilon^3)$ $\chi$EFT diagrams contributing to the spectral functions of 
the octet baryon isovector form factors on the two-pion (and two-kaon) cut. 
Dashed lines: pions (kaons); solid lines: octet baryons;
double lines: decuplet baryons; wavy lines: electromagnetic field. 
The two-pion (or two-kaon) cut is indicated by the dotted blue line.
Diagrams in which the electromagnetic field couples to 
the baryon or the meson-baryon vertices do not contribute to the two-pion (or two-kaon) 
cut and are not shown here.}
\label{Fig:LoopsOctet}
\end{figure}

The set of $\mathcal{O}(\epsilon^3)$ $\chi$EFT diagrams contributing to the 
spectral functions of the form factors on the 
two-pion cut at $t > 4 M_\pi^2$ is shown in 
Fig.~\ref{Fig:LoopsOctet}.\footnote{To clarify the parametric counting, we note that
$\mathcal{O}(\epsilon^3)$ refers to the order of the baryon 
electromagnetic vertex function, in which the background field 
is counted as $\mathcal{O}(\epsilon)$. The actual order of the spectral function 
extracted in this way is therefore $\mathcal{O}(\epsilon^2)$. Following
standard usage we refer to our calculation as $\mathcal{O}(\epsilon^3)$,
keeping in mind this distinction. \label{footnote:order}} 
These are the loop diagrams in which the electromagnetic field 
couples to the baryon through two-pion exchange 
in the $t$-channel: the ``triangle'' diagrams (a) and (c) with octet and decuplet 
intermediate states, and the ``tadpole'' diagram (b). Within the dispersive representation 
Eq.~(\ref{Eq:ff-spectral-rep}) these diagrams account for the densities at peripheral distances 
$b = \mathcal{O}(M_\pi^{-1})$. The two-kaon cut of the form factor at 
$t > 4 M_K^2$, which results from the corresponding two-kaon exchange diagrams 
in Fig.~\ref{Fig:LoopsOctet}, contributes only to the densities at much shorter distances
$b = \mathcal{O}(M_K^{-1})$ and can be neglected in the periphery (see below). 
The $\chi$EFT diagrams in which the electromagnetic field 
couples to the baryon or the meson-baryon vertices 
(not shown in Fig.~\ref{Fig:LoopsOctet}) have cuts only at $t > 4 \, m_{B, T}^2$, or no cuts
at all (polynomials); their contributions modify the densities only at distances 
$b = \mathcal{O}(M^{-1}_{B,T})$ or through delta functions at $\bm{b} = 0$ and can
be neglected in the periphery.
%
%

For reference we list here the Lagrangians generating the propagators and vertices in the $\mathcal{O}(\epsilon^3)$
diagrams of Fig.~\ref{Fig:LoopsOctet}. The lowest-order Lagrangian for the interaction between 
octet mesons and the electromagnetic field is
\begin{align}
\mathcal{L}_{\phi\phi}^{(2)}=\frac{f_0^2}{4}\mathrm{Tr}(u_\mu u^\mu+\chi_+),
\end{align}
where the vielbein is given by
$u_\mu=i\left\{u^\dagger,\nabla_\mu u\right\}$,
$u^2=U=\exp (i\sqrt{2}\phi / f_0)$,
$\nabla_\mu u=\partial_\mu
u-i(v_\mu+a_\mu)u+iu(v_\mu-a_\mu)u$; 
and the mass term is given by
$\chi_+=\chi U^\dagger + U \chi^\dagger$ with
$\chi=2B_0\, \text{diag}(m_u,m_d,m_s)$, where $m_f \; (f = u, d, s)$ are the quark masses;
and $f_0$ is the meson decay constant in the chiral limit. For the
couplings with the octet baryons we need the leading-order $SU(3)$
meson-baryon Lagrangian
\begin{align}
\mathcal{L}^{(1)}_{B\phi}&=\mathrm{Tr}\left(\bar{B}(i\slashed{\mathcal{D}}-m_{B0})B\right)
+\frac{D}{2}\mathrm{Tr}\left(\bar{B}\gamma^\mu \gamma_5 \left\{u_\mu,B\right\}\right) \nonumber\\
&+\frac{F}{2}\mathrm{Tr}\left(\bar{B}\gamma^\mu \gamma_5 \left[u_\mu,B\right]\right),\label{Lag_Bphi}
\end{align}
where the covariant derivative acts as $\mathcal{D}_\mu B=\partial_\mu
B + [\Gamma_\mu,B]$, with the chiral connection given by
$\Gamma_\mu=\frac{1}{2}[u^\dagger,\partial_\mu
  u]-\frac{i}{2}u^\dagger (v_\mu+a_\mu)
u-\frac{i}{2}u (v_\mu-a_\mu) u^\dagger$. The axial-vector
fields $a_\mu$ are set to zero in our case, and the two axial couplings
are fixed as $D = 0.80$ and $F = 0.46$ \cite{Alarcon:2012nr}. Finally, when introducing the 
decuplet baryons, the Lagrangians that are needed are
\begin{align}
\mathcal{L}_{T\phi}^{(1)}=&\bar{T}_\mu^{abc}(i\gamma^{\mu\nu\alpha}\mathcal{D}_\alpha-m_{T0}\gamma^{\mu\nu})T_\nu^{abc} \label{lag_T_phi}, \\
\mathcal{L}_{TB\phi}^{(1)}&=\frac{i\mathcal{C}}{m_{T0}}\epsilon^{ilm}\left[(\partial_\mu\bar{T}_\nu^{ijk})\gamma^{\mu\nu\rho}u_\rho^{jl}B^{km}+\mathrm{H.c.}\right],
\label{lag_T_B_phi}
\end{align}
where $\mathcal{C}= - h_A/(2\sqrt{2})$ is the octet-decuplet axial coupling,
$\gamma^{\mu\nu}=\frac{1}{2}\left[\gamma^{\mu},\gamma^{\nu}\right]$,
$\gamma^{\mu\nu\rho}=\frac{1}{2}(\gamma^\mu\gamma^\nu\gamma^\rho-\gamma^\rho\gamma^\nu\gamma^\mu)$,
and
$\gamma^{\mu\nu\rho\sigma}=\frac{1}{2}\left[\gamma^{\mu\nu\rho},\gamma^{\sigma}\right]$. The
covariant derivative acts on the decuplet as $\mathcal{D}_\alpha
T_\nu^{abc}=\partial_\alpha T_\nu^{abc} +
(\Gamma_\alpha,T_\nu)^{abc}$, where
$(X,Y)^{abc}=X^{ad}Y^{dbc}+X^{bd}Y^{adc}+X^{cd}Y^{abd}$. The vertices in Eqs.~(\ref{lag_T_phi})
and (\ref{lag_T_B_phi}) are consistent vertices (they transform covariantly under point 
transformations), which effectively implement the projection on spin-3/2 degrees of freedom.

The spectral functions (imaginary parts) resulting from the $t$-channel cut of the diagrams 
of Fig.~\ref{Fig:LoopsOctet} can be obtained by applying cutting 
rules \cite{Granados:2013moa} and are given by finite phase-space integrals,
which do not require renormalization. We have also calculated the 
form factors themselves from the full set of $\mathcal{O}(\epsilon^3)$ diagrams for
the baryon electromagnetic vertex function (not shown in Fig.~\ref{Fig:LoopsOctet}), 
including loop diagrams and their renormalization, 
and verified electromagnetic gauge invariance of the result.
We have confirmed that our expressions reproduce the 
ones of Ref.~\cite{Ledwig:2011cx} when reduced to the $SU(2)$ flavor group. 
Further tests validating our calculation are described in \ref{App:Validation}.

For the numerical evaluation we consider a range of values for the
couplings and baryon masses that goes from the $SU(2)$ values ($m_{B}
= 939$~MeV, $m_{T} = 1232$~MeV and $f_\pi = 92.2$~MeV, $h_A = 2.85$; see Ref.~\cite{Ledwig:2011cx})
to the $SU(3)$ average values ($\bar{m}_B = 1151$~MeV, $\bar{m}_T =
1382$~MeV, $\bar{f}_\phi = 1.17 f_\pi$, $h_A = 2.40$; see Refs.~\cite{Ledwig:2014rfa,Alarcon:2012nr}).  
This provides an uncertainty band associated with the systematic errors 
and gives an estimate of the expected higher-order corrections.
The meson masses are taken at their physical values, $M_\pi = 139$~MeV, 
$M_K = 494$~MeV. According to the chiral expansion the meson mass differences 
are of the same order as the meson masses themselves, whereas the baryons 
have a common mass in the chiral limit and the mass splittings are suppressed.
This circumstance has important consequences for the description of peripheral 
baryon structure, as it ensures that the pion and kaon exchange contributions to
the densities are evaluated with the physical masses and possess the correct
ranges $1/(2 M_\pi)$ and $1/(2 M_K)$ already in the lowest order.

%
%
\begin{figure}[t]
\epsfig{file=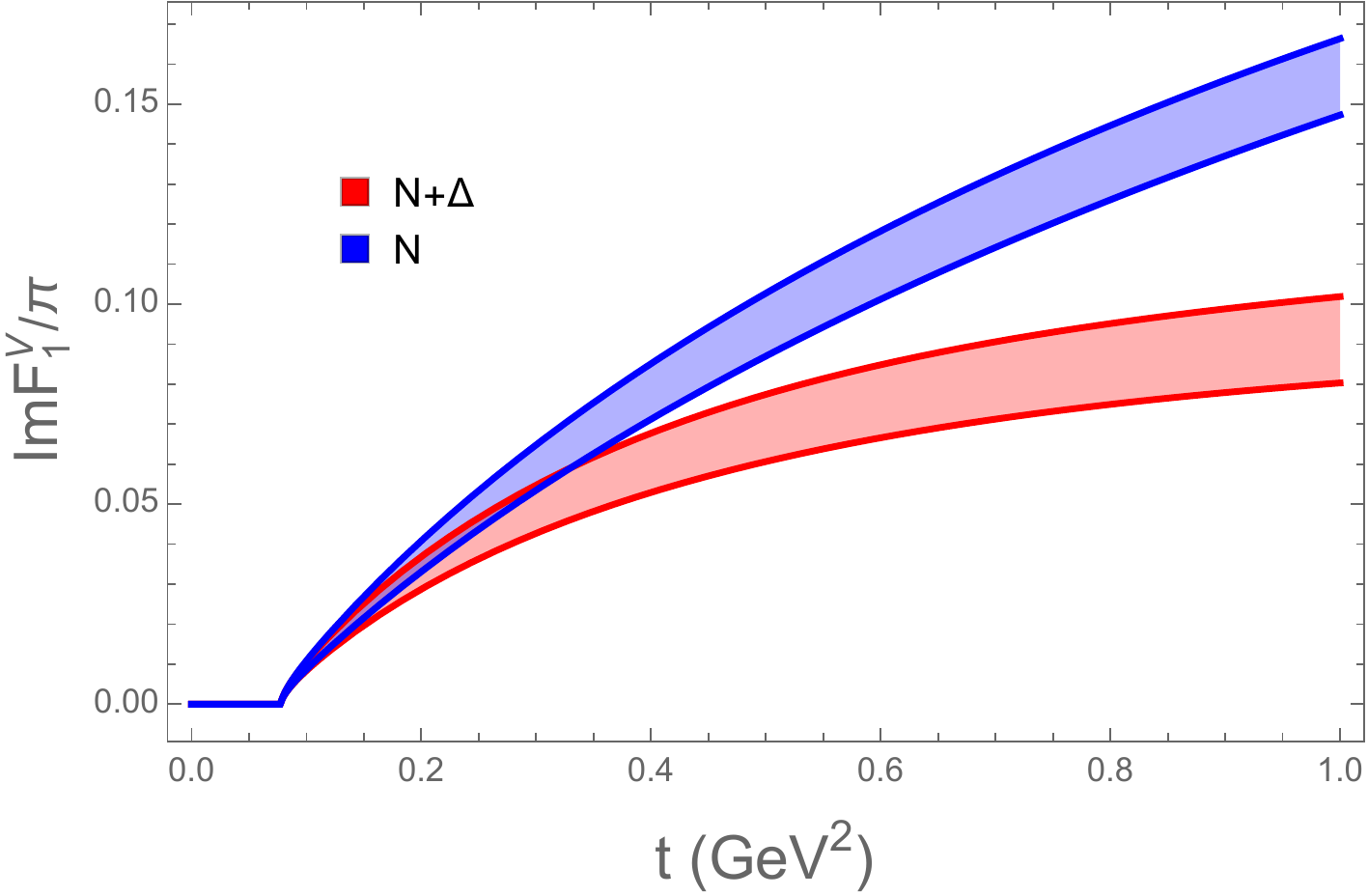,width=.45\textwidth,angle=0}
\\[-2ex]
{\footnotesize (a)}
\\[3ex]
\epsfig{file=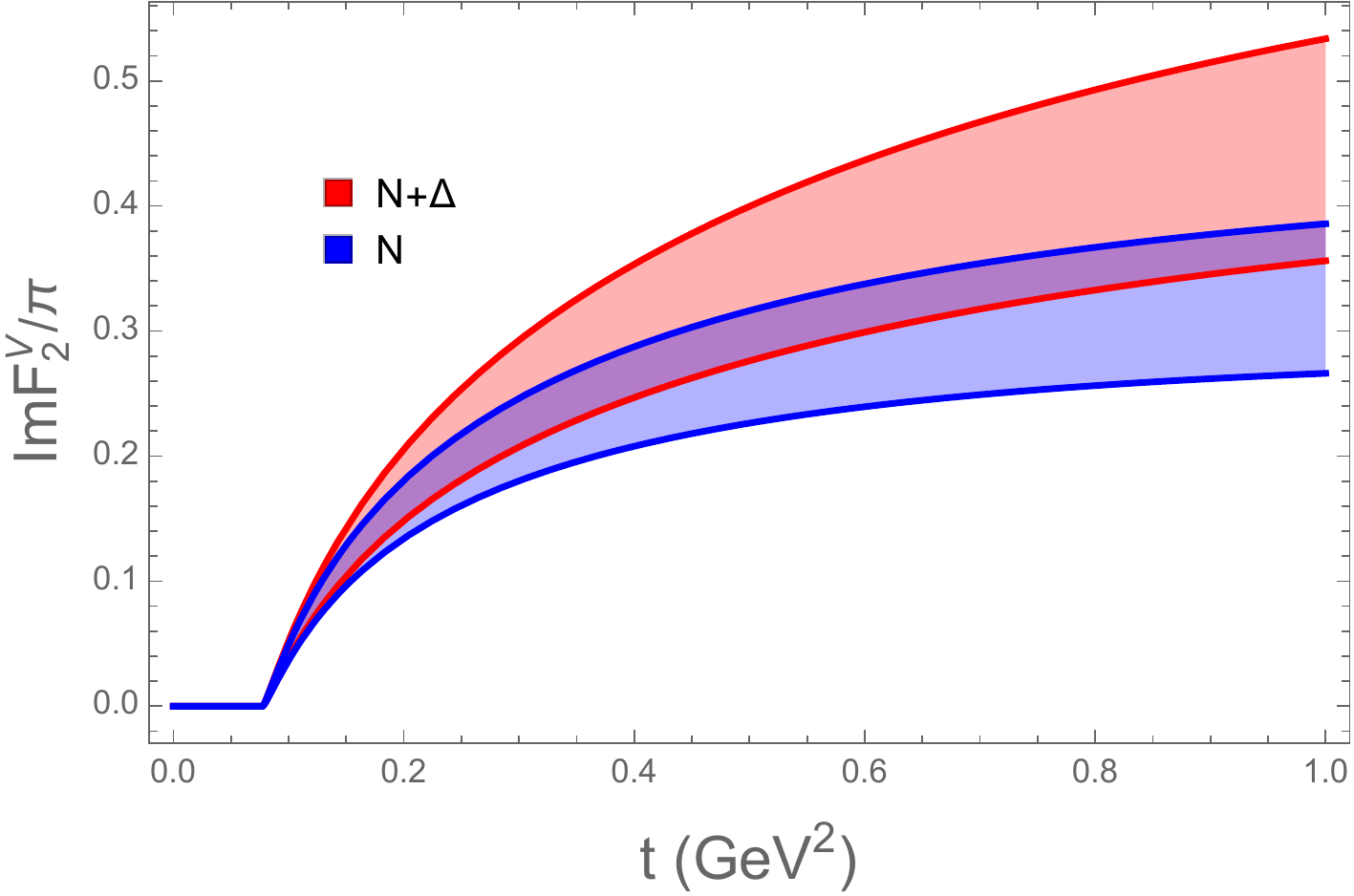,width=.45\textwidth,angle=0}
\\[-2ex]
{\footnotesize (b)}
\caption[]{\small $\mathcal{O}(\epsilon^3)$ $\chi$EFT results for the nucleon isovector 
spectral functions, with and without the intermediate $\Delta(1232)$ contribution, 
in the $SU(2)$ flavor limit.}
\label{Fig:ImFV}
\end{figure} 
The $\mathcal{O}(\epsilon^3)$ $\chi$EFT 
results for the nucleon isovector spectral functions are shown in Fig.~\ref{Fig:ImFV}.
The steep rise above the threshold at $t = 4 \, M_\pi^2$ is caused by the subthreshold singularity at 
$t = 4 \, M_\pi^2 - M_\pi^4/m_N^2$ (on the unphysical sheet of the nucleon form factor), which results 
from the triangle diagram Fig.~\ref{Fig:LoopsOctet} (a) with an intermediate nucleon and is a general 
feature of the analytic structure. In the diagram (c) with intermediate $\Delta$ isobar the subthreshold
singularity is further removed from threshold, resulting in a smaller contribution. The intermediate
$\Delta$ contribution reduces the intermediate $N$ result in the case of $\textrm{Im}\, F_1^V$,
and enhances it in the case of $\textrm{Im}\, F_2^V$. We note that this pattern ensures the proper 
scaling behavior of the form factors in the large-$N_c$ limit of QCD, where the $N$ and $\Delta$ 
become degenerate \cite{Granados:2013moa,Granados:2016jjl}.

In \ref{App:Delta} we compare the results of the present $\mathcal{O}(\epsilon^3)$ 
calculation with those of Ref.~\cite{Granados:2013moa,Granados:2016jjl}, which used
an ``inconsistent'' form of the $\pi N \Delta$ vertex. The results for the $\Delta$
contribution agree at $\mathcal{O}(\epsilon^3)$ but differ by higher-order terms, 
which cause numerical differences of 20\% in 
$\textrm{Im}\, F_1$ (50\% in $\textrm{Im}\, F_2$) at $t \sim 0.6\, \textrm{GeV}^2$
(see Fig.~\ref{Fig:Off-shell-effects}).
\subsection{Improvement through unitarity}
\label{Subsec:Improvement}
The $\chi$EFT expressions by themselves describe the baryon isovector spectral functions only in the 
near-threshold region $t = 4 \, M_\pi^2 + \textrm{few} \, M_\pi^2$. This can be seen clearly in the 
case of the nucleon, where the $\chi$EFT expressions can be compared with the spectral functions
obtained from dispersion theory (see below) and was noticed in earlier $\chi$EFT 
calculations \cite{Gasser:1987rb,Bernard:1996cc,Becher:1999he,Kubis:2000zd,Kaiser:2003qp}.
The reason for the discrepancy is the strong rescattering within the $I = J = 1$ $\pi\pi$ system 
in the $t$-channel, which manifests itself in the $\rho$ meson resonance at $t \sim 30 \, M_\pi^2 = 0.6 \, \textrm{GeV}^2$. 
In order to construct the baryon densities down to distances $b \sim 1$ fm we need to extend
the calculation of isovector spectral functions into the $\rho$ meson mass region 
(see Sec.~\ref{Subsec:dispersive_representation} and Fig.~\ref{Fig:rho_integrand}). 
This can be accomplished in an approach that combines the $\chi$EFT calculations with 
dispersion theory. 
We describe this approach first for the baryon form factors in which the 
two-pion cut starts at the normal threshold $t_{\rm thr} = 4 M_\pi^2$ (such as the nucleon
$B = N$); the role of anomalous thresholds with $t_{\rm thr} < 4 M_\pi^2$ in the strange 
baryon form factors will be considered subsequently.

%
%
\begin{figure}[t]
\begin{center}
\epsfig{file=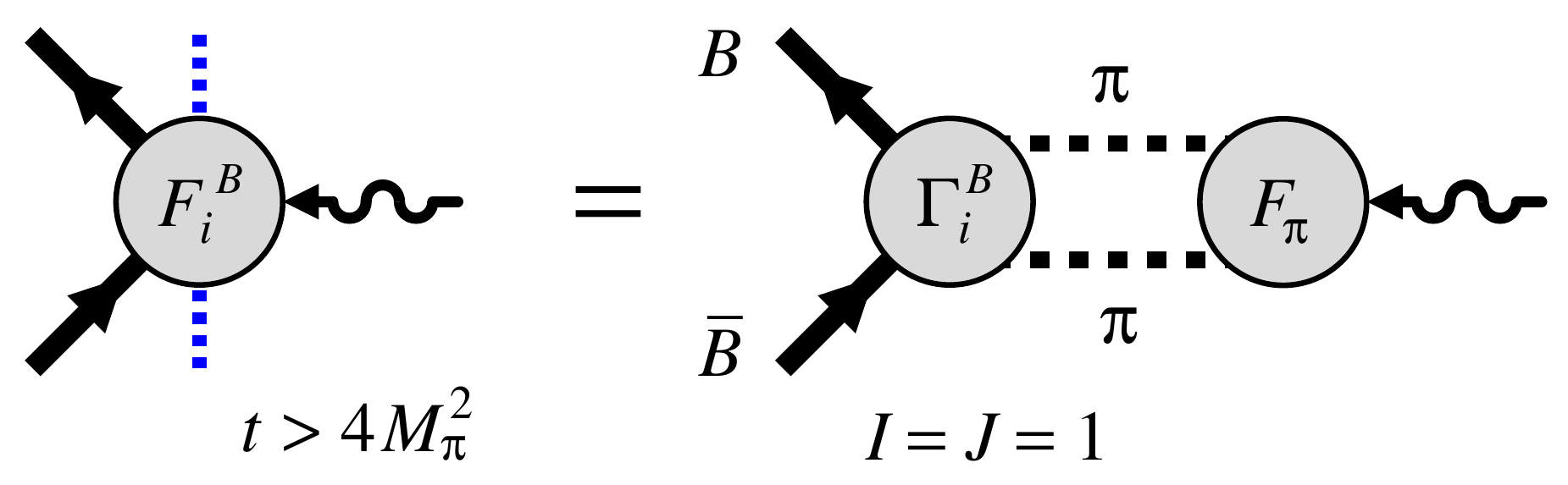,width=.45\textwidth,angle=0} 
\end{center}
\caption[]{Unitarity relation for the isovector spectral function on the two-pion cut,
Eq.~(\ref{Eq:Disp_rep_ImFV}).}
\label{fig:unitarity}
\end{figure}
On general grounds the baryon isovector spectral function on the two-pion cut can be expressed 
as (here $t$ is real and $t > 4 \, M_\pi^2$) \cite{Frazer:1960zza,Frazer:1960zzb,Hohler:1974ht}
\begin{equation}
\label{Eq:Disp_rep_ImFV}
\textrm{Im} F_i^B(t) = \frac{k_{\rm cm}^3}{\sqrt{t}} \; \Gamma_i^B(t) \; F^*_{\pi}(t) 
\hspace{2em} (i=1,2),
\end{equation}
where $k_{\rm cm}=\sqrt{t/4 - M_\pi^2}$ is the center-of-mass momentum of the $\pi\pi$ system
in the $t$-channel, $\Gamma_i^B(t)$ is the complex $I=J=1$  $\pi \pi \rightarrow B\bar{B}$ partial wave 
amplitude, and $F_{\pi}(t)$ is the complex pion form factor in the timelike region 
(see Fig.~\ref{fig:unitarity}). 
Equation~(\ref{Eq:Disp_rep_ImFV}) follows from the unitarity condition in the $t$-channel and is 
valid strictly in the region up to the four-pion threshold, $4 \, M_\pi^2 < t < 16 \, M_\pi^2$; if contributions 
from four-pion states are neglected it can effectively be applied up to 
$t \sim 50 \, M_\pi^2 = 1 \textrm{GeV}^2$.
The expression on the right-hand side of Eq.~(\ref{Eq:Disp_rep_ImFV}) is real 
because the complex functions $\Gamma_i^B(t)$ and $F_\pi(t)$ have the same phase on the two-pion 
cut (Watson theorem \cite{Watson:1954uc}). 
It is convenient to rewrite Eq.~(\ref{Eq:Disp_rep_ImFV}) in the form
\begin{equation}
\label{Eq:Disp_rep_ImFV_ratio}
\textrm{Im} F_i^B(t) = \frac{k_{\rm cm}^3}{\sqrt{t}} \; \frac{\Gamma_i^B(t)}{F_{\pi}(t) } \; 
|F_\pi(t)|^2 \hspace{2em} (i=1,2).
\end{equation}
This representation has two advantages: (a) The function $\Gamma_i^B(t)/F_{\pi}(t)$ is real 
at $t > 4 \, M_\pi^2$ and therefore has no two-pion cut; (b) the squared modulus $|F_\pi(t)|^2$ can 
be extracted directly from the $e^+e^- \rightarrow \pi^+\pi^-$ exclusive annihilation cross section, 
without determining the phase of the complex form factor.
Equation~(\ref{Eq:Disp_rep_ImFV_ratio}) is a variant of the $N/D$ method of amplitude analysis \cite{Chew:1960iv}. 
The $\pi\pi \rightarrow B\bar B$
$t$-channel partial-wave amplitude is represented in the form $\Gamma_i^B (t) = N(t)/D(t)$, 
such that the right-hand cut (related to the $t$-channel exchanges) appears only in the factor 
$1/D(t)$ and the left-hand cut (related to the $s$-channel intermediate states) in the factor $N(t)$.
In the case at hand the $D$ function is naturally chosen as the inverse pion form factor,
$D(t) = 1/F_\pi(t)$ \cite{Hohler:1974ht}.

%
%
\begin{figure*}[t]
\begin{tabular}{ll}
\epsfig{file=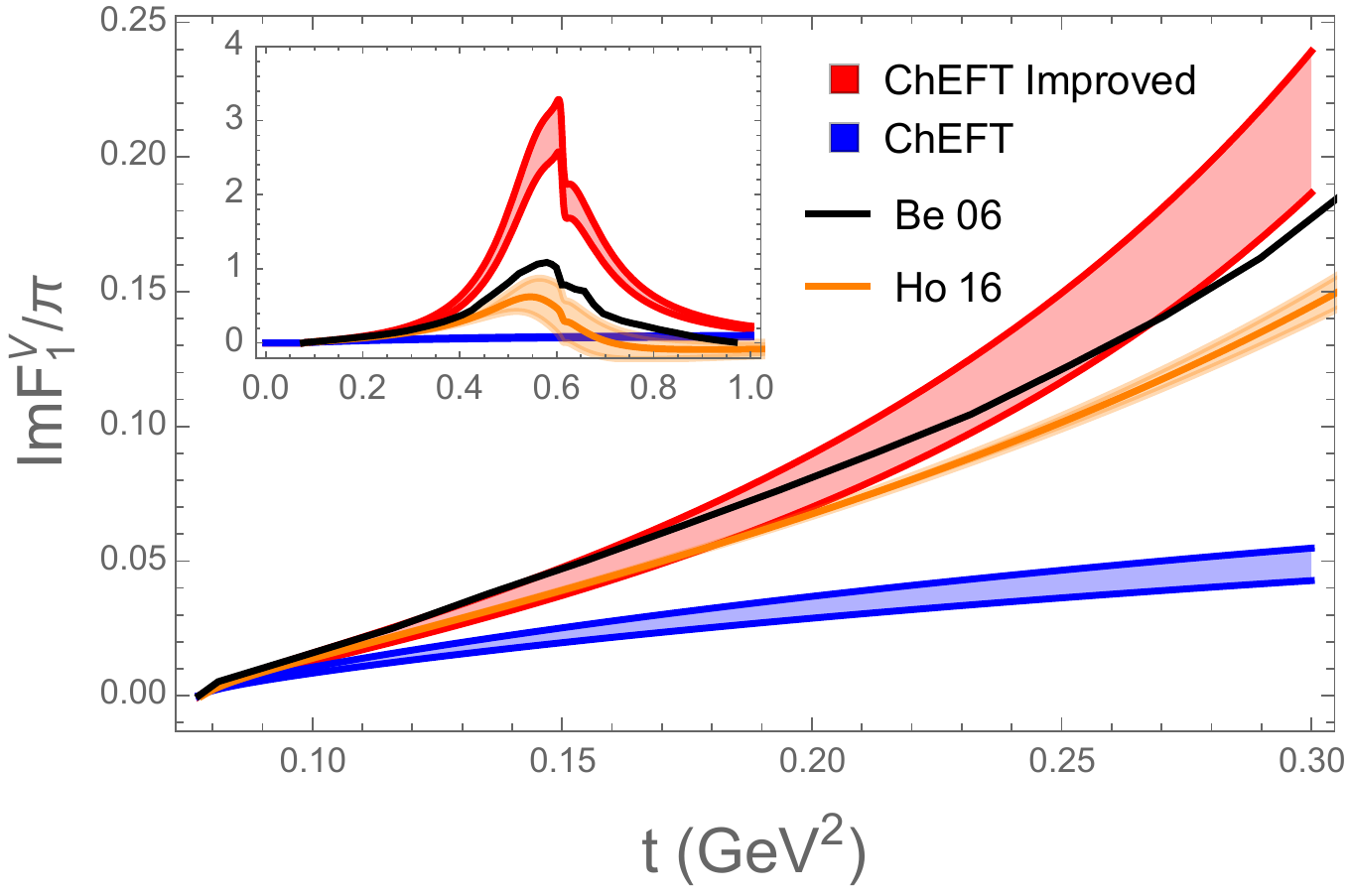,width=.45\textwidth,angle=0} &
\epsfig{file=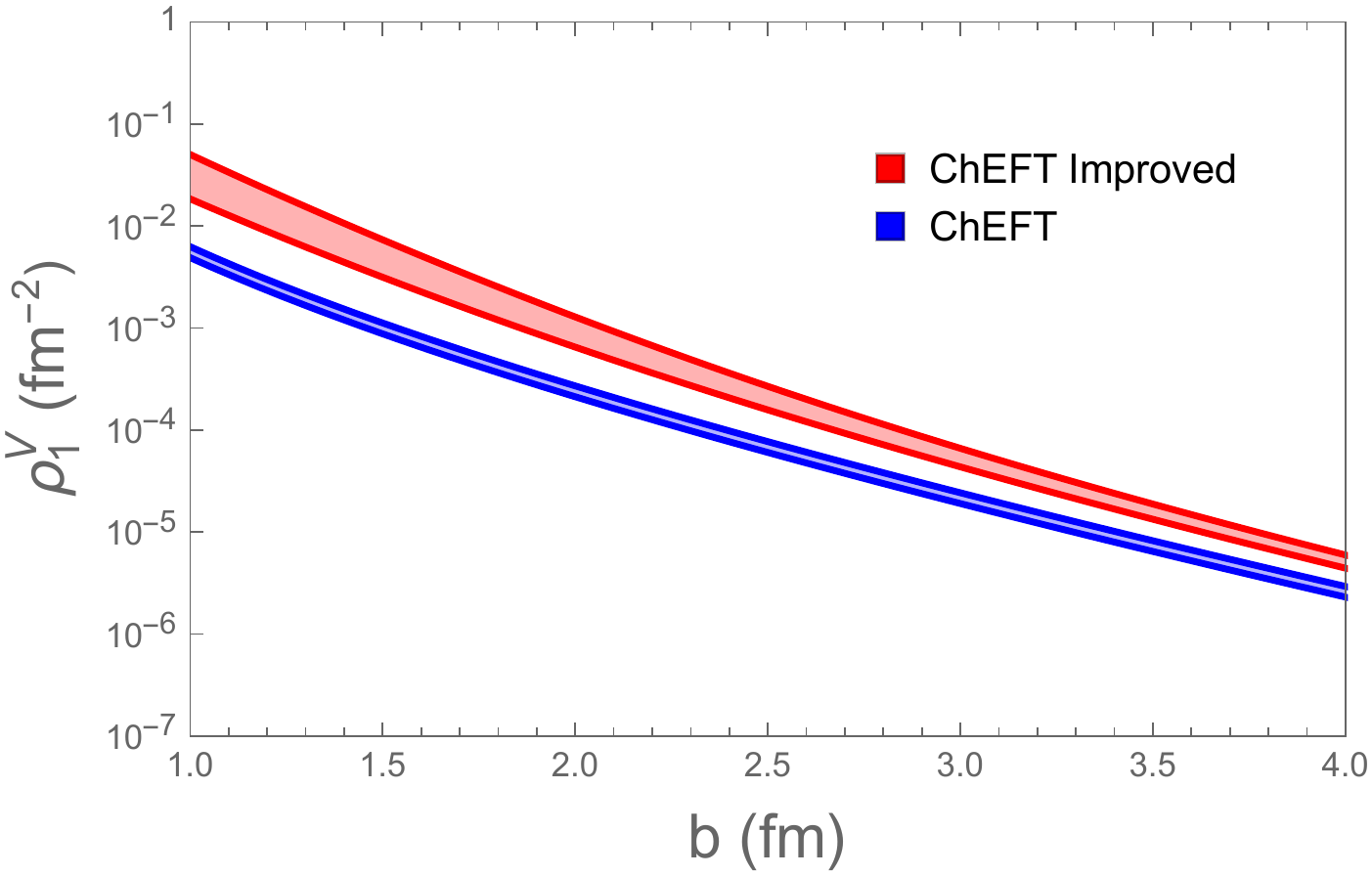,width=.45\textwidth,angle=0}
\\[-2ex]
{\footnotesize (a)} & {\footnotesize (b)}
\\[1ex]
\epsfig{file=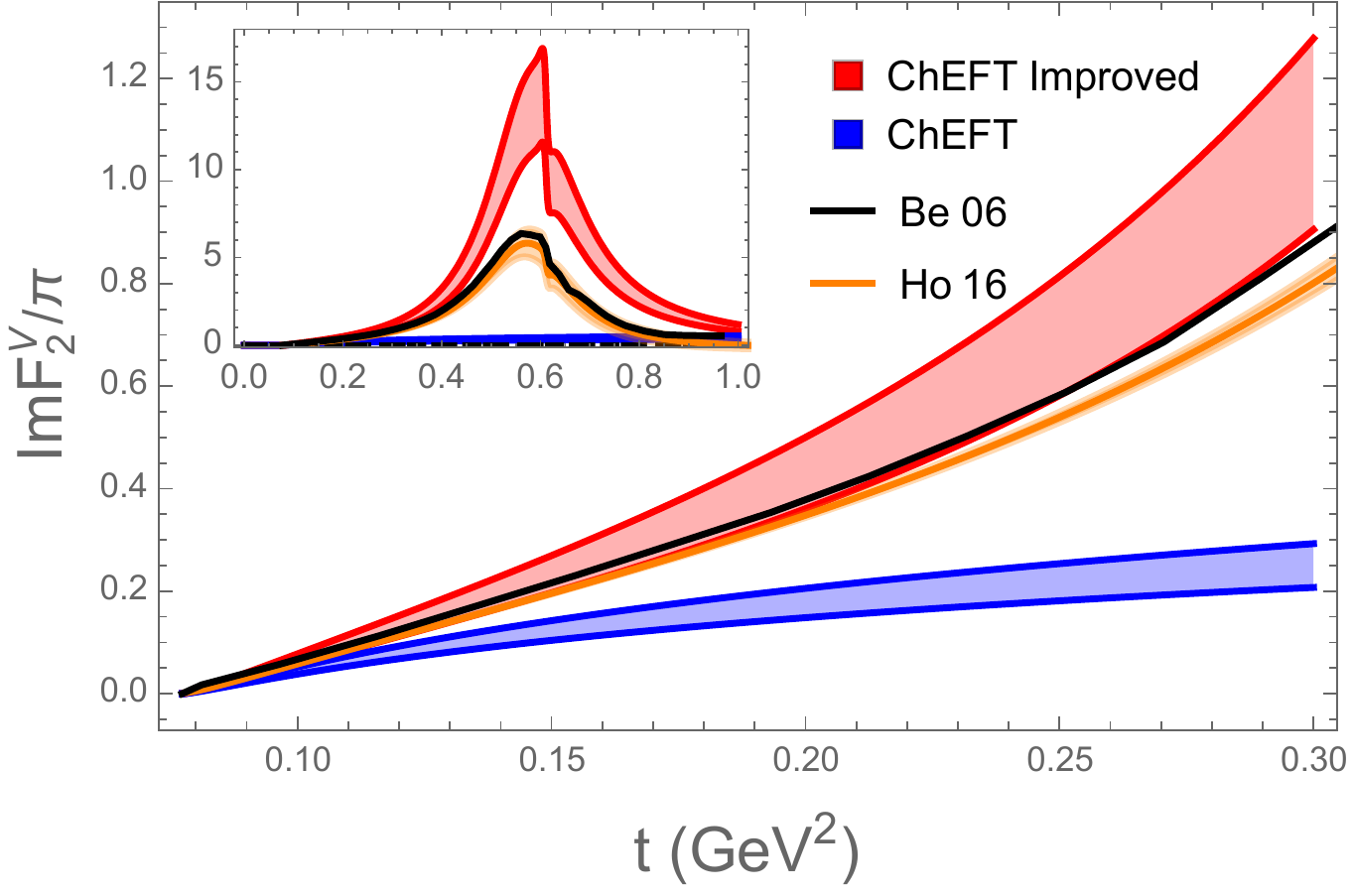,width=.45\textwidth,angle=0} &
\epsfig{file=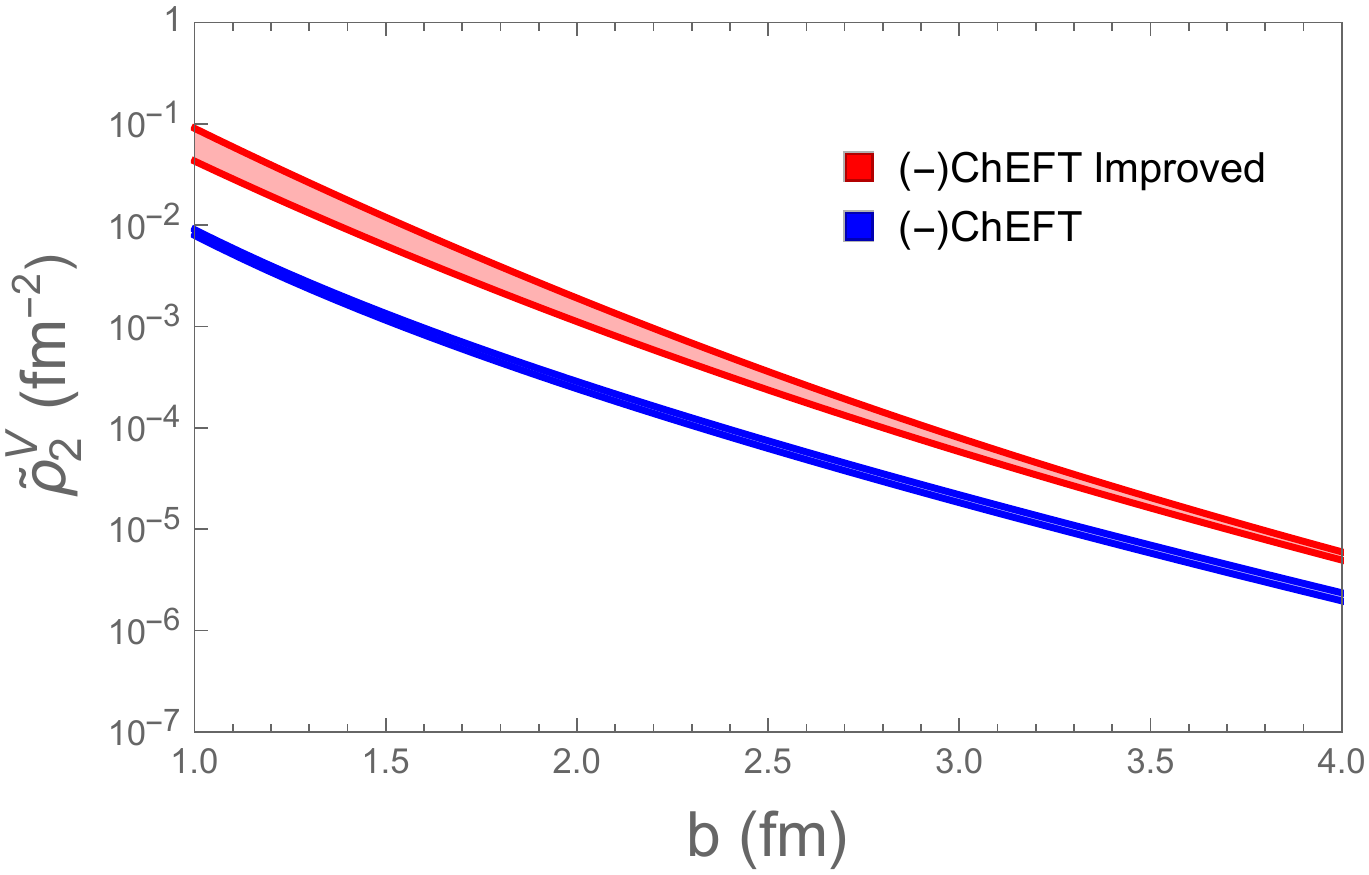,width=.45\textwidth,angle=0}
\\[-2ex]
{\footnotesize (c)} & {\footnotesize (d)}
\\[1ex]
\end{tabular}
\caption[]{\small (a), (c) Nucleon isovector spectral functions obtained from 
$\mathcal{O}(\epsilon^3)$ $\chi$EFT and the improvement through unitarity, Eq.~(\ref{Eq:improvement}).
Blue bands: $\mathcal{O}(\epsilon^3)$ $\chi$EFT results, including contributions from 
both $N$ and $\Delta$ intermediate states, cf.~Fig.~\ref{Fig:ImFV}. 
Red bands: Results after improvement through unitarity,
Eq.~(\ref{Eq:improvement}). Brown bands and black line: Spectral functions obtained 
from Roy-Steiner equations \cite{Hoferichter:2016duk} and analytic continuation of 
the $\pi\pi \rightarrow N\bar N$ 
amplitudes \cite{Belushkin:2005ds,Hohler:1976ax}. The main plot shows the 
functions up to $t = 0.3\, \textrm{GeV}^2$; the inset shows them up to $t = 1\, \textrm{GeV}^2$.
(b), (d) Effect of improvement on the peripheral transverse densities. The isovector magnetic 
density is shown with opposite sign $(-)$ on the logarithmic scale.}
\label{Fig:Improvement_ImF_V}
\end{figure*} 
The representation Eq.~(\ref{Eq:Disp_rep_ImFV_ratio}) suggests a new approach to calculating the 
spectral function on the two-pion cut. We use $\chi$EFT to calculate the real function 
$\Gamma_i^B(t)/F_{\pi}(t)$ at $t > 4 \, M_\pi^2$ to a fixed order. We then multiply the result
with the empirical $|F_\pi(t)|^2$, which contains the effects of $\pi\pi$ rescattering and
the $\rho$ meson resonance. The major advantage of this approach is that the $\chi$EFT
calculation is not affected by the strong $\pi\pi$ rescattering, which would require higher-order 
unitarity corrections when treated within the $\chi$EFT. We therefore expect this approach to
show much better convergence than direct $\chi$EFT calculations of the spectral function.
From a general perspective, the organization according to Eq.~(\ref{Eq:Disp_rep_ImFV_ratio}) is 
consistent with the 
idea of separation of scales basic to $\chi$EFT. The function $\Gamma_i^B(t)/F_{\pi}(t)$
is dominated by the singularities of the baryon Born diagrams (or diagrams with $\pi B$ inelastic
intermediate states in higher orders), which are governed by the scales $M_\pi$ and $m_T - m_B$
(the octet-decuplet mass difference). The $t$-dependence of the pion form factor, in contrast, 
is governed by the chiral-symmetry-breaking scale $\Lambda_\chi$, which is of the order of the
vector meson mass.

In our $\chi$EFT calculation of the spectral function at $\mathcal{O}(\epsilon^3)$
in Sec.~\ref{SubSec:ChiralEFT} the function $\Gamma_i^B(t)/F_{\pi}(t)$ is effectively 
evaluated at $\mathcal{O}(\epsilon^0)$ (cf.\ Footnote~\ref{footnote:order}).
At this level the pion form factor in the denominator enters at leading order,
$F_\pi(t) \equiv 1$, so that the $\chi$EFT result for $\Gamma_i^B(t)/F_{\pi}(t)$ 
is the same as that for $\Gamma_i^B(t)$ itself. The prescription of Eq.~(\ref{Eq:Disp_rep_ImFV_ratio})
therefore simply amounts to multiplying our $\mathcal{O}(\epsilon^3)$ results 
for the spectral functions by $|F_\pi(t)|^2$,
\newline
\begin{eqnarray}
\label{Eq:improvement}
\textrm{Im} F_i^B(t) \, [\textrm{improved}] &=&
\textrm{Im} F_i^B(t) \, [\textrm{$\chi$EFT}] \, \times \, |F_\pi(t)|^2 
\nonumber \\[1ex] && (i=1,2).
\end{eqnarray}
This formula permits an extremely simple implementation of unitarity at $\mathcal{O}(\epsilon^3)$.
For the pion form factor we use the Gounaris-Sakurai parametrization including 
effects of $\rho$-$\omega$ mixing \cite {Gounaris:1968mw,Barkov:1985ac}, with the 
parameters determined in Ref.~\cite{Lorenz:2012tm}.
The prescription Eq.~(\ref{Eq:improvement}) results in a remarkable improvement of the $\chi$EFT 
predictions for the spectral functions. The improved $\chi$EFT results for the nucleon
($B = N$) reproduce the spectral functions obtained from amplitude analysis
(analytic continuation of the $\pi\pi \rightarrow N\bar N$ partial-wave amplitudes
\cite{Belushkin:2005ds,Hohler:1976ax}, Roy-Steiner equations \cite{Hoferichter:2016duk})
up to $t \sim 16 \, M_\pi^2 = 0.3 \, \textrm{GeV}^2$ within errors 
(see Fig.~\ref{Fig:Improvement_ImF_V}a and c). Note that this is achieved without adding
any free parameters and represents a genuine prediction of $\chi$EFT. The improved
$\mathcal{O}(\epsilon^3)$
results have qualitatively correct behavior even in the $\rho$ meson mass region. 
We expect that higher-order $\chi$EFT
corrections to the ratio $\Gamma_i^B(t)/F_{\pi}(t)$ would further improve the agreement
with the dispersion-theoretical results at $t > 0.3$ GeV$^2$ \cite{InPreparation}.

The improvement according to Eq.~(\ref{Eq:improvement}) has a dramatic effect on the
peripheral transverse densities of the nucleon 
(see Fig.~\ref{Fig:Improvement_ImF_V}b and d). At distances 
$b \sim 1$ fm the improved predictions are up to an order of magnitude larger than what would
be obtained with the original $\chi$EFT results. At distances $b \sim 3$ fm the change
is still by a factor $\sim 3$. At asymptotically large distances (much larger than those
shown in the figure) the improvement would change the densities by the constant factor 
$|F_\pi(t = 4 \, M_\pi^2)|^2 \approx 1.3$.

In order to extend the dispersive improvement to the other octet baryons it is necessary to 
discuss the role of anomalous thresholds in the strange baryon form 
factors \cite{Karplus:1958zz,Landau:1959fi}.
In the normal situation considered so far, the Born terms in the $\pi\pi \rightarrow B\bar B$ 
amplitudes give rise to singularities on the unphysical sheet of the baryon form factor,
and the principal cut starts at the normal threshold 
$t_{\rm thr} = 4 M_\pi^2$. Such is the case with the 
$N$ and $\Delta$ Born terms in the $\pi\pi \rightarrow N\bar N$ amplitudes. 
In certain special situations the Born term singularities in the $\pi\pi \rightarrow B\bar B$ 
amplitudes move onto the physical sheet of the form factor and give rise to cuts with 
anomalous thresholds at $t_{\rm thr} < 4 M_\pi^2$. This occurs if \cite{Karplus:1958zz}
\beq
m_B^2 \; > \; m_{B'}^2 + M_\pi^2 ,
\label{anomalous_condition}
\eeq
where $m_{B'}$ is the mass of the baryon pole in the Born term, and the anomalous 
threshold is located at
\beq
t_{\rm thr} \; = \; 4 M_\pi^2 - (m_B^2 - m_{B'}^2 - M_\pi^2)^2/m_{B'}^2 \; < \; 4 M_\pi^2.
\label{anomalous_t}
\eeq
Such is the case for the $\Lambda$ pole in the $\pi\pi \rightarrow \Sigma \bar\Sigma$ 
amplitudes, which produces an anomalous threshold at $t_{\rm thr} = 3.08 \, M_\pi^2$. 
In a general dispersion-theoretical treatment with exact masses the contribution from 
the anomalous cut would need to be considered explicitly in the spectral representation 
of the form factors and the densities. In the present treatment based on leading-order $\chi$EFT
anomalous thresholds do not occur, as the octet baryon masses are taken at a common value 
in the chiral limit (this is required by the small-scale expansion)
and therefore $m_{B'} = m_B$ in Eq.~(\ref{anomalous_condition}). 
We can thus use Eq.~(\ref{Eq:improvement}) at leading order for the entire baryon octet. 
This feature is a consequence of our particular combination of $\chi$EFT and dispersion 
theory and permits a major simplification of the description of the
strange baryon form factors. The anomalous threshold in the $\Sigma$ form 
factors appears only in higher-order $\chi$EFT calculations and affects the behavior
of the densities at very large distances; its contribution at realistic distances 
$b \sim 1-4$ fm is expected to be small (cf.\ the discussion in 
Sec.~\ref{Subsec:dispersive_representation} and Fig.~\ref{Fig:rho_integrand}).

An anomalous threshold would also appear in the $K\bar K \rightarrow \Sigma\bar\Sigma$ 
amplitudes through the $N$ Born term, at 
$t_{\rm thr} = 4 M_K^2 - (m_\Sigma^2 - m_N^2 - M_K^2)^2/m_N^2 = 3.6 \, M_K^2 < 4 M_K^2$, 
cf.~Eq.~(\ref{anomalous_t}), if the exact baryon masses were used. Again this feature is absent 
in our leading-order $\chi$EFT calculation, where the $N$ and $\Sigma$ are taken at a common mass.
In any case this anomalous threshold is only marginally lower than the normal $K\bar K$ 
threshold, so that its cut corresponds to a high-mass contribution that can be neglected
in the peripheral densities on the same grounds as the normal $K\bar K$ cut.

%
%
\begin{figure*}[t]
\begin{tabular}{ll}
\epsfig{file=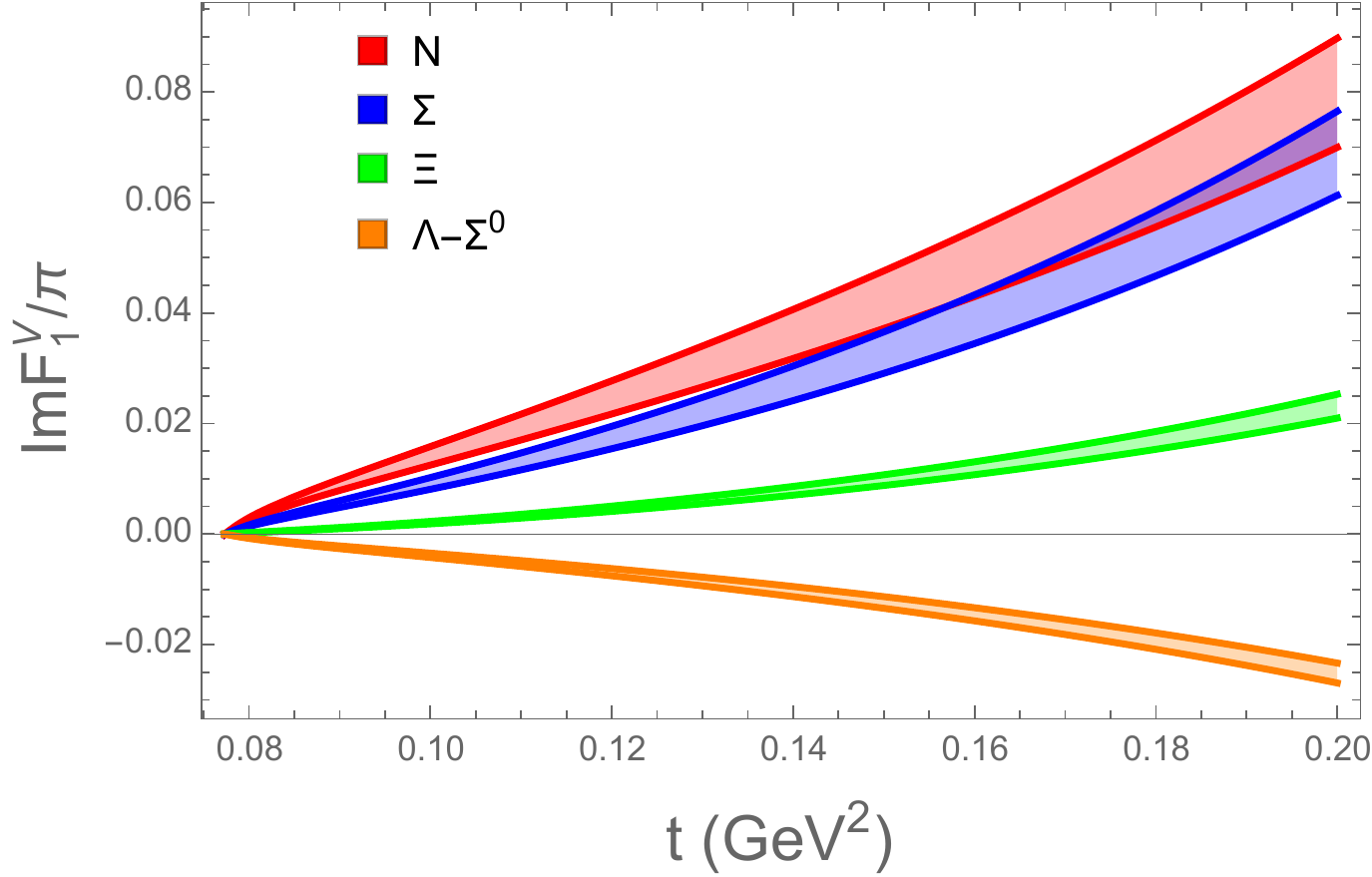,width=.45\textwidth,angle=0} &
\epsfig{file=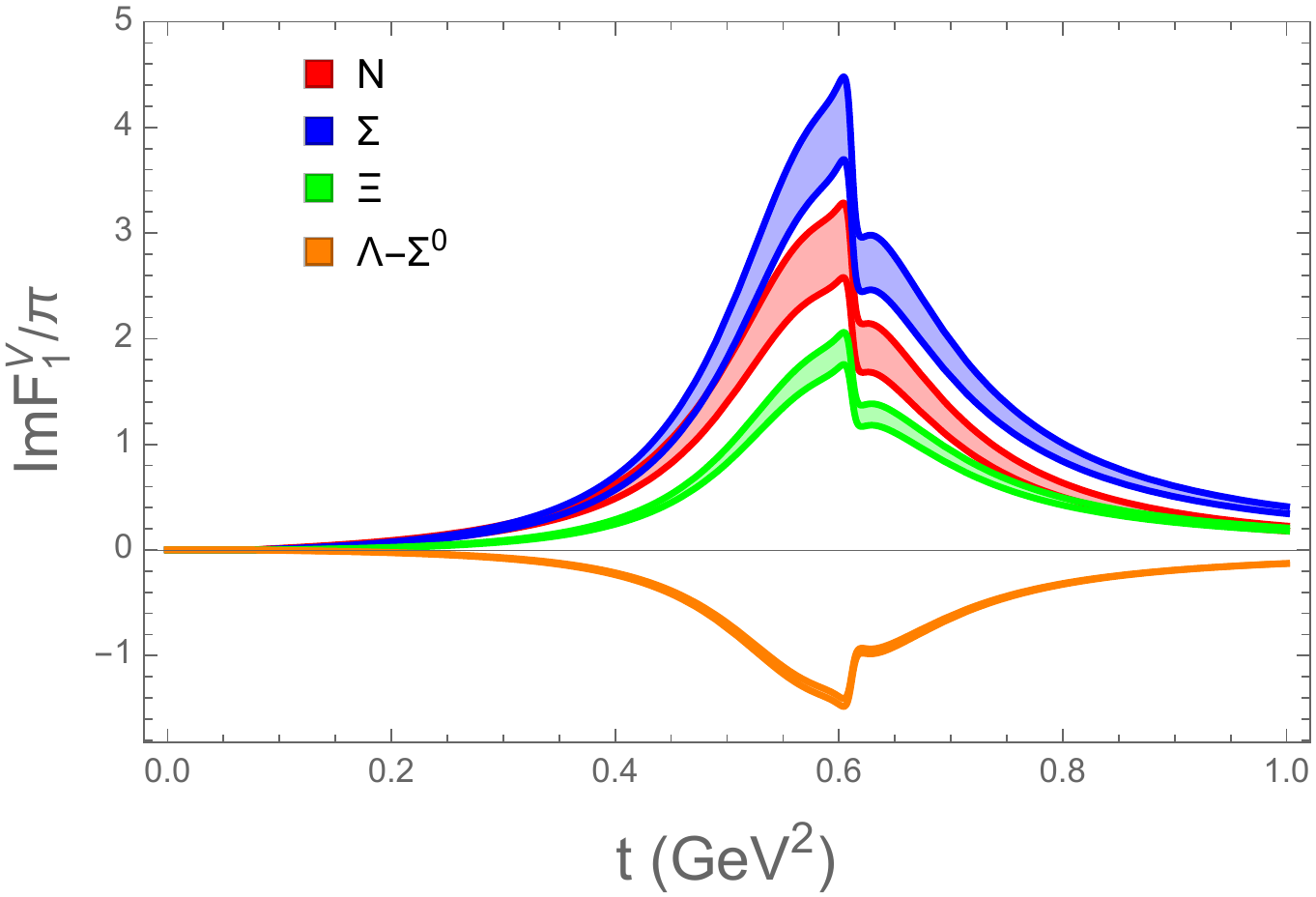,width=.45\textwidth,angle=0}
\\[-2ex]
{\footnotesize (a)} & {\footnotesize (b)}
\\[1ex]
\epsfig{file=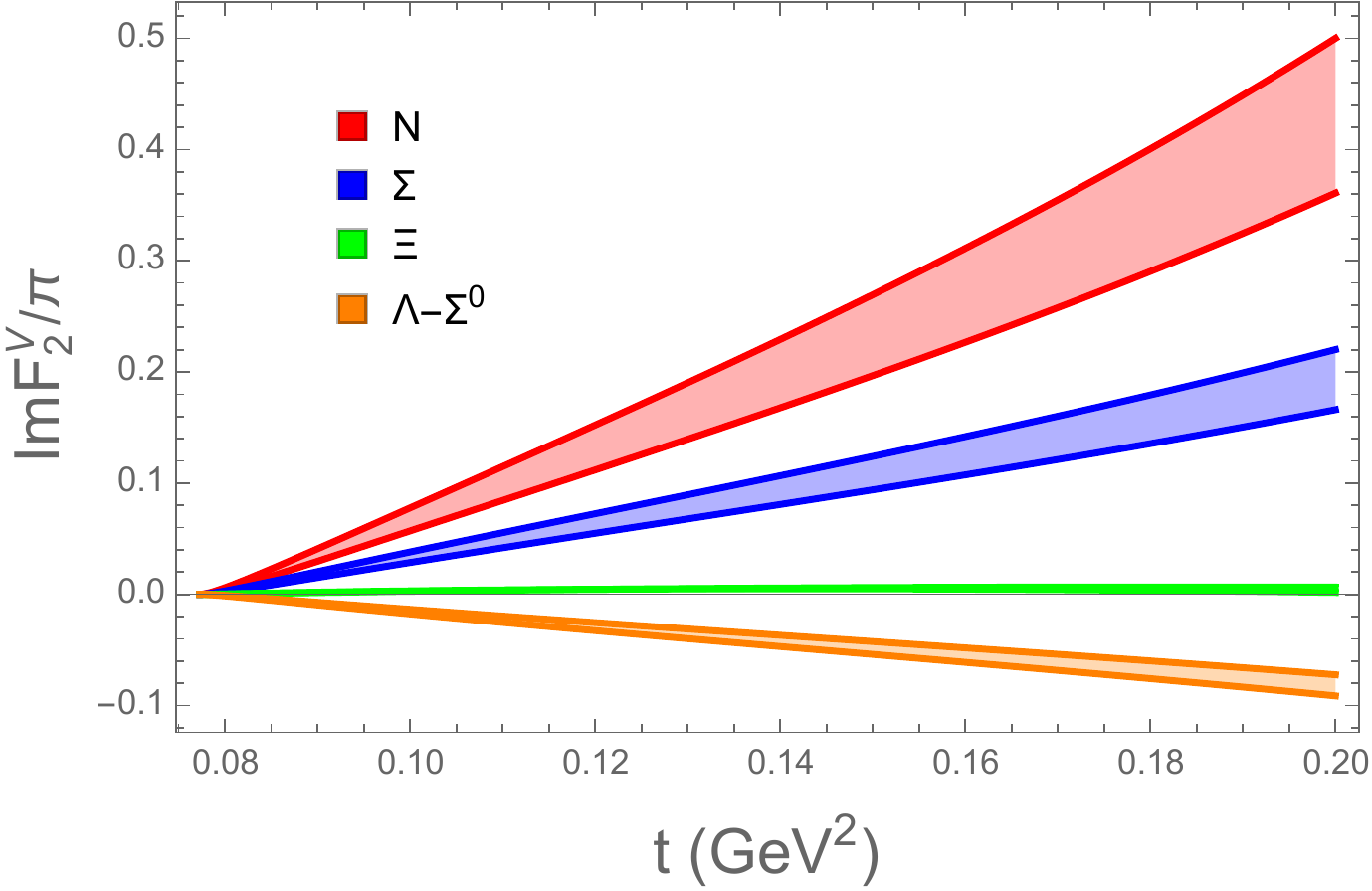,width=.45\textwidth,angle=0} &
\epsfig{file=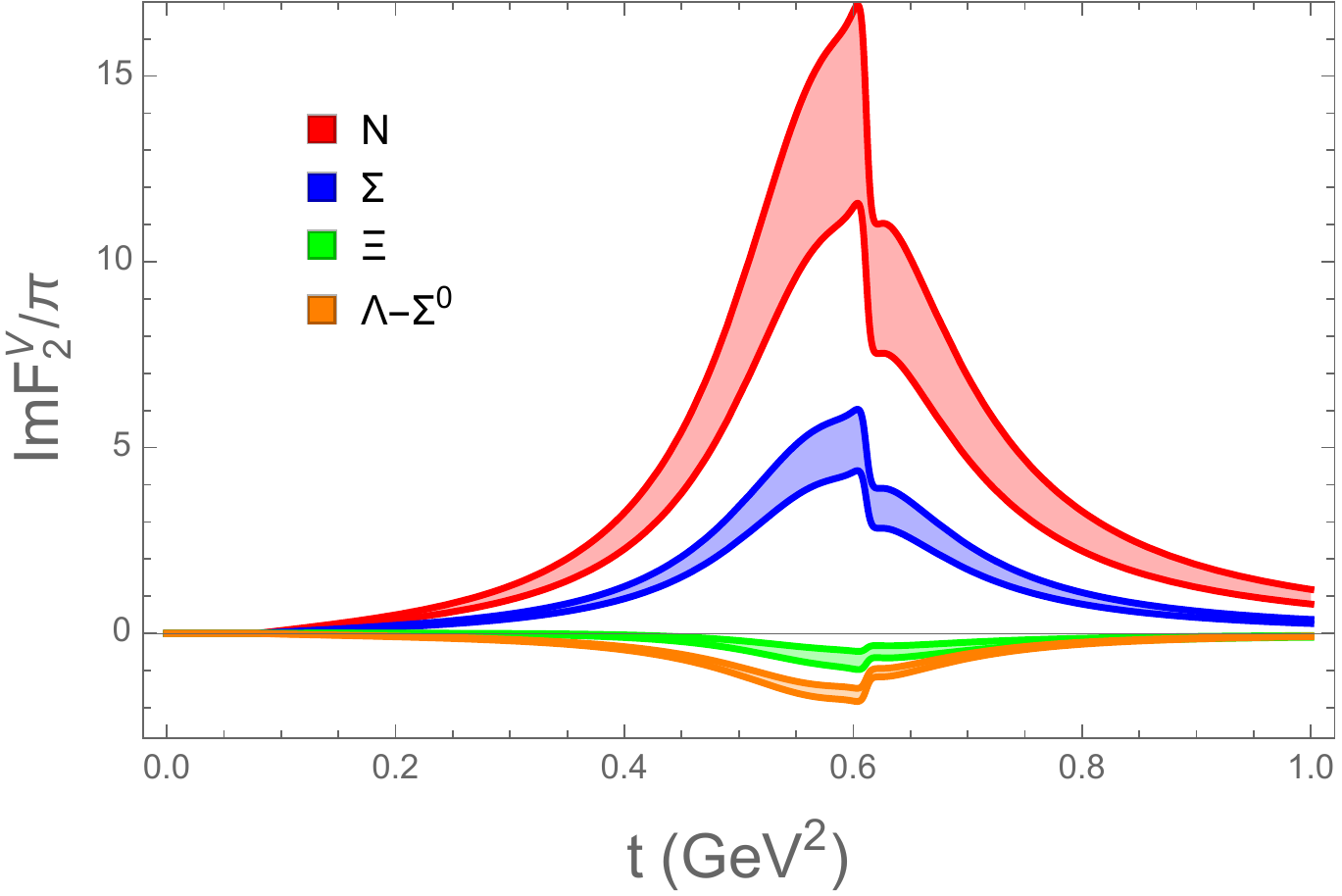,width=.45\textwidth,angle=0}
\\[-2ex]
{\footnotesize (c)} & {\footnotesize (d)}
\end{tabular}
\caption[]{\small Isovector component of the spectral functions of the octet baryons,
cf.\ Eq.~(\ref{isospin_ff}), as obtained from $\mathcal{O}(\epsilon^3)$ $\chi$EFT 
with dispersive improvement, Eq.~(\ref{Eq:improvement}).
(a), (b) Dirac form factors $F_1^V$. (c), (d) Pauli form factors $F_2^V$. Plots (a), (c) show the
near-threshold region; plots (b), (d) show the region up to $t = 1$ GeV$^2$.}
\label{Fig:ImFV-Octet}
\end{figure*} 
Based on the preceding considerations we now use the prescription Eq.~(\ref{Eq:improvement}) to 
calculate the isovector 
spectral functions of the strange octet baryons on the two-pion cut. While we cannot compare with
a dispersion-theoretical result in this case, we expect the approximation to be of similar
quality as in the case of the nucleon. The results for $\textrm{Im}\, F_1^V$ and $\textrm{Im}\, F_2^V$
are shown in Fig.~\ref{Fig:ImFV-Octet} [see Eq.~(\ref{isospin_ff})
for the definition of the isovector component in the octet baryon states]. They exhibit several 
interesting features. (a) In the near-threshold region $t = 4 \, M_\pi^2 + \textrm{few} \, M_\pi^2$
(see Fig.~\ref{Fig:ImFV-Octet}a, c)
the dominant contribution to the spectral functions comes from the triangle diagram with intermediate 
octet baryons, Fig.~\ref{Fig:LoopsOctet}a. The ratios of the spectral functions for the different baryons 
are therefore determined approximately by the ratios of the products of the $B \rightarrow B' \pi$ 
couplings for the relevant intermediate states, which follow directly from the $SU(3)$ chiral
Lagrangian Eq.~(\ref{Lag_Bphi}):
\beq
\begin{array}{cccccccc}
& \textrm{Im} F_i^{N, V} & \!\! : \!\! & \textrm{Im} F_i^{\Sigma, V} & \!\! : \!\! & \textrm{Im} F_i^{\Xi, V} 
& \!\! : \!\! &
\textrm{Im} F_i^{\Lambda -\Sigma, V} \\[1ex]
= \!\! &(F+D)^2 & \!\! : \!\! & (2F^2 + \frac{2}{3}D^2) & \!\! : \!\! & (F-D)^2 & \!\! : \!\! 
& (-\frac{2}{\sqrt{3}}FD) \\[1ex]
= \!\! & 1.59 & \!\! : \!\! & 0.85 & \!\! : \!\! & 0.12 & \!\! : \!\! & (-0.42) \\[1ex]
\end{array}
\label{Eq:ImF_ratios}
\eeq
(b) At larger values of $t$ (see Fig.~\ref{Fig:ImFV-Octet}b, d) the spectral functions 
receive sizable contributions from the triangle diagram with intermediate decuplet baryons
and the contact terms, Fig.~\ref{Fig:LoopsOctet}c and b. These contributions change the
relative order of the different baryons compared to the near-threshold region dominated
by intermediate octet baryons. The signs of the intermediate octet and decuplet contributions
are summarized in Table~\ref{table:octet_decuplet}.
In the Dirac form factor the intermediate decuplet contributes to Im $F_1^{N, V}$ with {\em opposite} 
sign to the intermediate octet, and to Im $F_1^{\Sigma, V}$ and Im $F_1^{\Xi, V}$ with the
{\em same} sign. As a result, Im $F_1^{\Sigma, V}$ is now larger than Im $F_1^{N, V}$, and Im $F_1^{\Xi, V}$ 
is comparable to Im $F_1^{N, V}$. In the Pauli form factor the intermediate decuplet contributes
to Im $F_2^{N, V}$ with the {\em same} sign as the intermediate octet, and to Im $F_2^{\Sigma, V}$ 
and Im $F_2^{\Xi, V}$ with {\em different} sign. As a result the relative order of Im $F_2^{N, V}$ 
and Im $F_2^{\Sigma, V}$ remains the same, while Im $F_2^{\Xi, V}$ becomes negative at
larger values of $t$. These changes between the near-threshold region and larger values of $t$ 
are a non-trivial consequence of the inclusion of decuplet degrees of freedom in the $\chi$EFT.
\begin{table}
\[
\begin{array}{l|cc|cc|cc|cc|}
     & \multicolumn{2}{|c|}{N} & \multicolumn{2}{|c|}{\Sigma} & \multicolumn{2}{|c|}{\Xi} & \multicolumn{2}{|c|}{\Lambda-\Sigma}  \\[.5ex]
\hline
& \bf{8} & \bf{10} & \bf{8} & \bf{10} & {\bf 8} & \bf{10} & {\bf 8} & \bf{10} \\[.5ex]
\hline
\textrm{Im}\, F_1^V  & + & - & + & + & + & +  & - & -\\[.5ex]
\hline
\textrm{Im}\, F_2^V  & + & + & + & - & + & - & - & + \\[.5ex]
\hline
\end{array}
\]
\caption[]{Signs of intermediate octet and decuplet contributions to the isovector spectral functions 
of the octet baryon form factors, $\textrm{Im}\, F_i^V \; (i = 1, 2)$ on the two-pion cut.}
\label{table:octet_decuplet}
\end{table}

To complete our assessment we want to quantify also the contribution of the two-kaon cut to 
the peripheral isovector densities in $\chi$EFT. Numerical evaluation shows that in
the nucleon at $b = 1\, \textrm{fm}$ the isovector density from the two-kaon cut is smaller
than that from the two-pion cut by an order of magnitude; at $b = 2\, \textrm{fm}$ it is 
already smaller by two orders of magnitude. Similar ratios are obtained for the strange octet baryons.
We can therefore neglect the contributions of the two-kaon cut in the peripheral isovector densities. 
The quoted kaon/pion ratios refer to the $\chi$EFT densities before the dispersive improvement.
An interesting theoretical question is how the dispersive improvement could be extended to 
the kaon sector through a coupled-channel formalism.

\subsection{Isoscalar spectral functions}
\label{Sec:Modeling_isoscalar_VM}
The isoscalar component of the baryon spectral functions behaves very differently from the 
isovector one and requires separate treatment. In the limit of exact isospin symmetry the 
isoscalar spectral functions arise from the exchange of an odd number of pions in the $t$-channel (G-parity).
The lowest-mass exchange with vector quantum numbers is the three-pion exchange ($t > 9 \, M_\pi^2$).
In $\chi$EFT this contribution appears only at $\mathcal{O}(\epsilon^7)$ and is very small \cite{Bernard:1996cc}. 
Inclusion of the $\omega$ resonance through unitarity in the three-pion channel, in analogy to our 
treatment of the $\rho$ in the two-pion channel, would be possible in principle but requires 
three-body unitarity techniques that are not readily available.

The present $SU(3)$-flavor $\chi$EFT generates an isoscalar term in the spectral function
through two-kaon exchange. Its contribution to the peripheral densities is extremely small
compared to the isovector densities, or the isoscalar densities resulting from $\omega$ and
$\phi$ exchange (see below). Treatment of the $\phi$ as a resonance in the two-kaon channel,
along the lines of the $\rho$ in the two-pion channel, is impractical because of the mixing 
with the three-pion and other hadronic channels;
see Refs.~\cite{Hammer:1998rz,Hammer:1999uf} for a discussion.

In the present study we therefore model the isoscalar spectral functions of the octet baryons
at $t < 1 \, \textrm{GeV}^2$ through phenomenological vector meson exchange ($\omega, \phi$),
\begin{align}
\text{Im}\, F_i^{B, S} (t) = \pi \sum_{V = \omega, \phi} a_i^{VBB} \, \delta(t-M^2_{V})
\hspace{1em} (i = 1, 2).
\label{Eq:isoscalar_poles}
\end{align}
The vector meson couplings of the octet baryons are obtained from $SU(3)$ symmetry, certain 
assumptions about the $F/D$ ratio, and the empirical vector meson couplings to the nucleon;
see \ref{App:Vector}. The contribution from states with $t > 1 \, \textrm{GeV}^2$ is
strongly suppressed in the peripheral densities, so that the two-pole parametrization
Eq.~(\ref{Eq:isoscalar_poles}) is sufficient for our purposes. We do not aim for a precise 
description of the isoscalar sector here, as the peripheral densities are dominated by the 
isovector component.
\section{Octet baryon densities}
\label{Sec:Transverse_densities_octet}
\subsection{Charge and magnetization densities}
Using the spectral functions calculated in Sec.~\ref{Sec:Spectral_Functions} we now calculate
the peripheral transverse densities of the octet baryons and study their properties.
The studies of Sec.~\ref{Subsec:dispersive_representation} have shown that at distances 
$b > 1\, \textrm{fm}$ the dispersion integrals Eqs.~(\ref{Eq:rho1-spectral-rep}) and 
(\ref{Eq:rho2tilde-spectral-rep}) are dominated by masses $t < 1 \, \textrm{GeV}^2$, 
where our approximations in the spectral functions are justified.

In order to quantify the uncertainties of the peripheral densities we propagate the estimated
theoretical uncertainties of the spectral functions through the dispersion integrals 
Eqs.~(\ref{Eq:rho1-spectral-rep}) and (\ref{Eq:rho2tilde-spectral-rep}).
In the isovector spectral functions we use the estimated $\chi$EFT errors in the region
$t < 16 \, M_\pi^2$ (as shown by the error bands in Fig.~\ref{Fig:Improvement_ImF_V}a and c) 
and assign an additional error of $\pm 50\%$ in the $\rho$ meson mass region (these error bands are
not shown in the insets of Fig.~\ref{Fig:Improvement_ImF_V}a and c but inferred from
the comparison of the improved $\chi$EFT results with the amplitude analysis results
\cite{Belushkin:2005ds,Hoferichter:2016duk} in this region). 
In the isoscalar spectral functions we generate an error 
band from the uncertainties of the empirical $\omega NN$ couplings extracted from fits to 
the isoscalar nucleon form factor data \cite{Belushkin:2006qa}
(see \ref{App:Vector} and the parameters in Table~\ref{Table:VBB-coefficients-2-var}).
Note that the error estimates performed here are specific to the peripheral densities
and do not take into account constraints resulting from the total isovector/isoscalar 
charges of the baryons, which would involve the densities at central as well as peripheral
distances; these constraints would significantly reduce the error of the overall densities
at $b \sim 1 \, \textrm{fm}$.

\begin{figure*}
\begin{center}
\epsfig{file=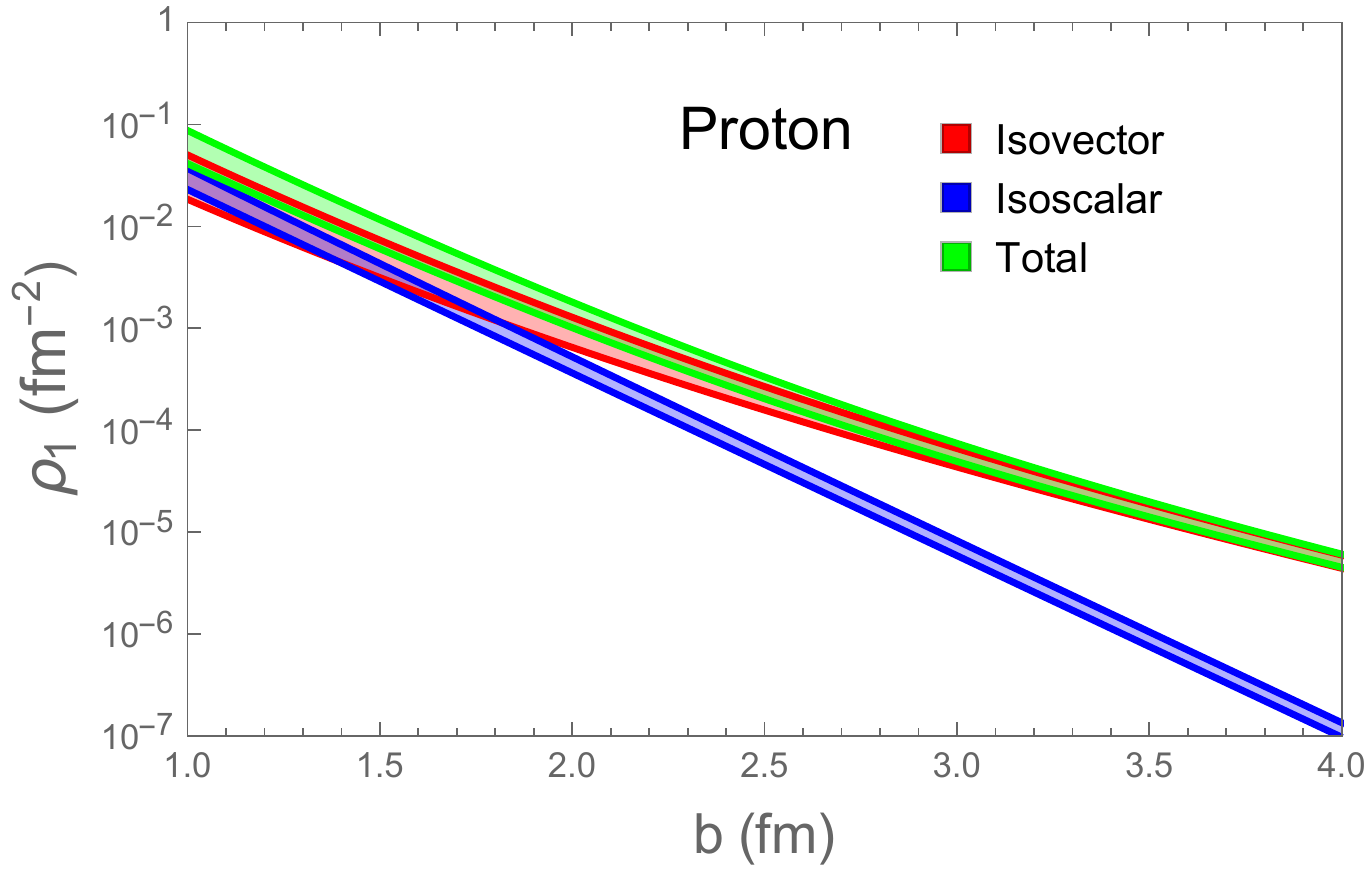,width=.45\textwidth,angle=0}
\epsfig{file=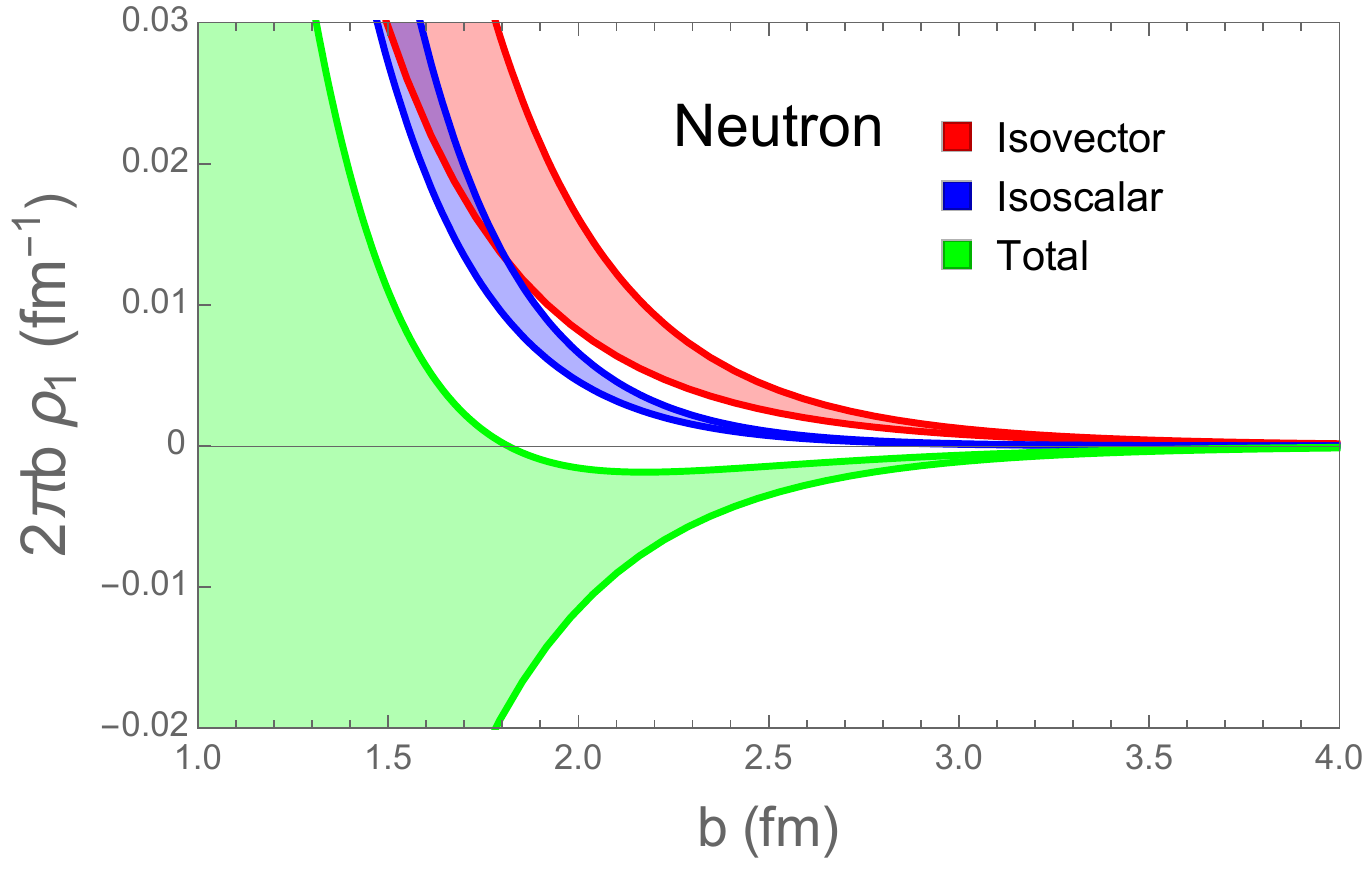,width=.45\textwidth,angle=0}\\
\epsfig{file=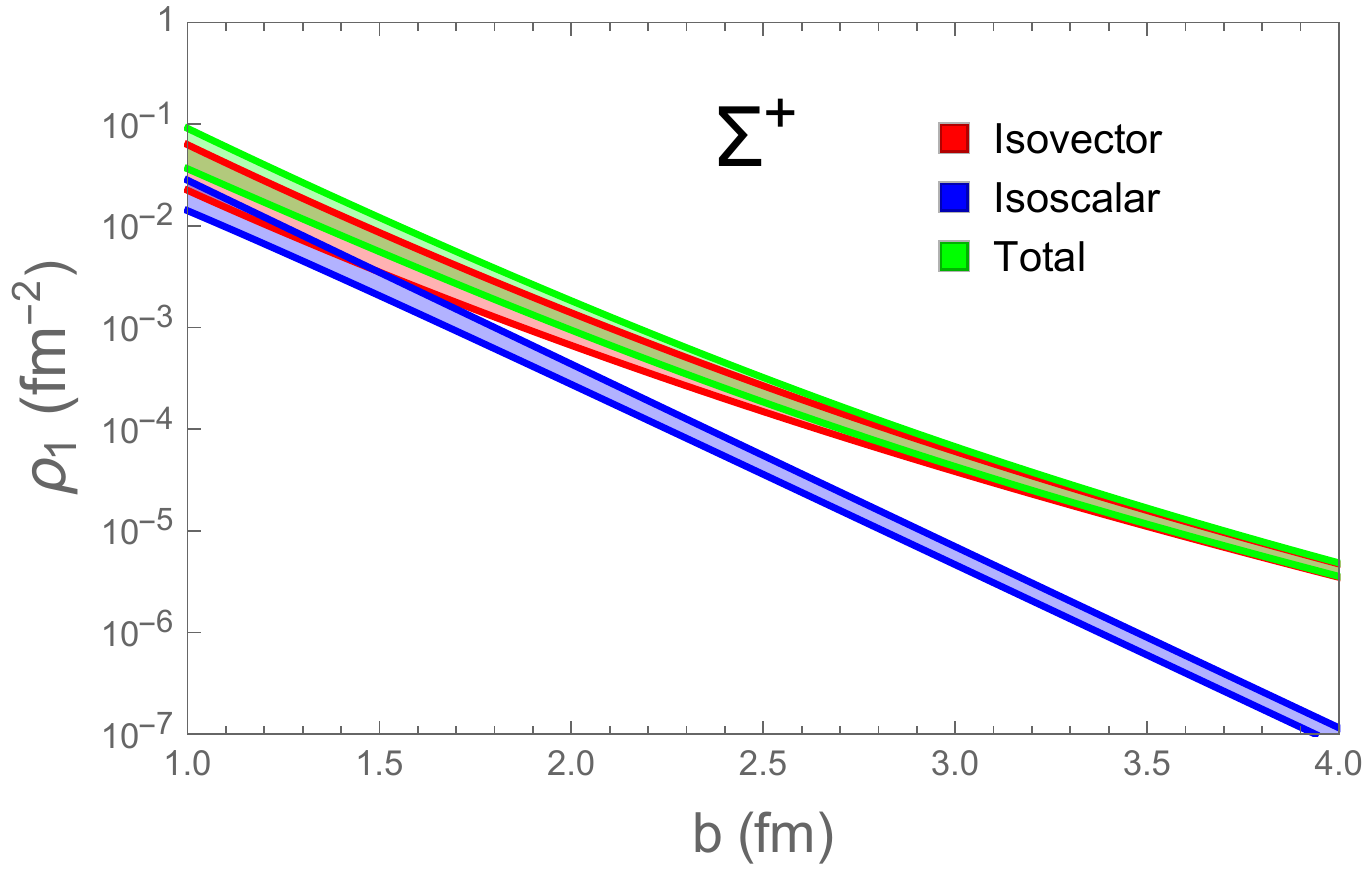,width=.45\textwidth,angle=0}
\epsfig{file=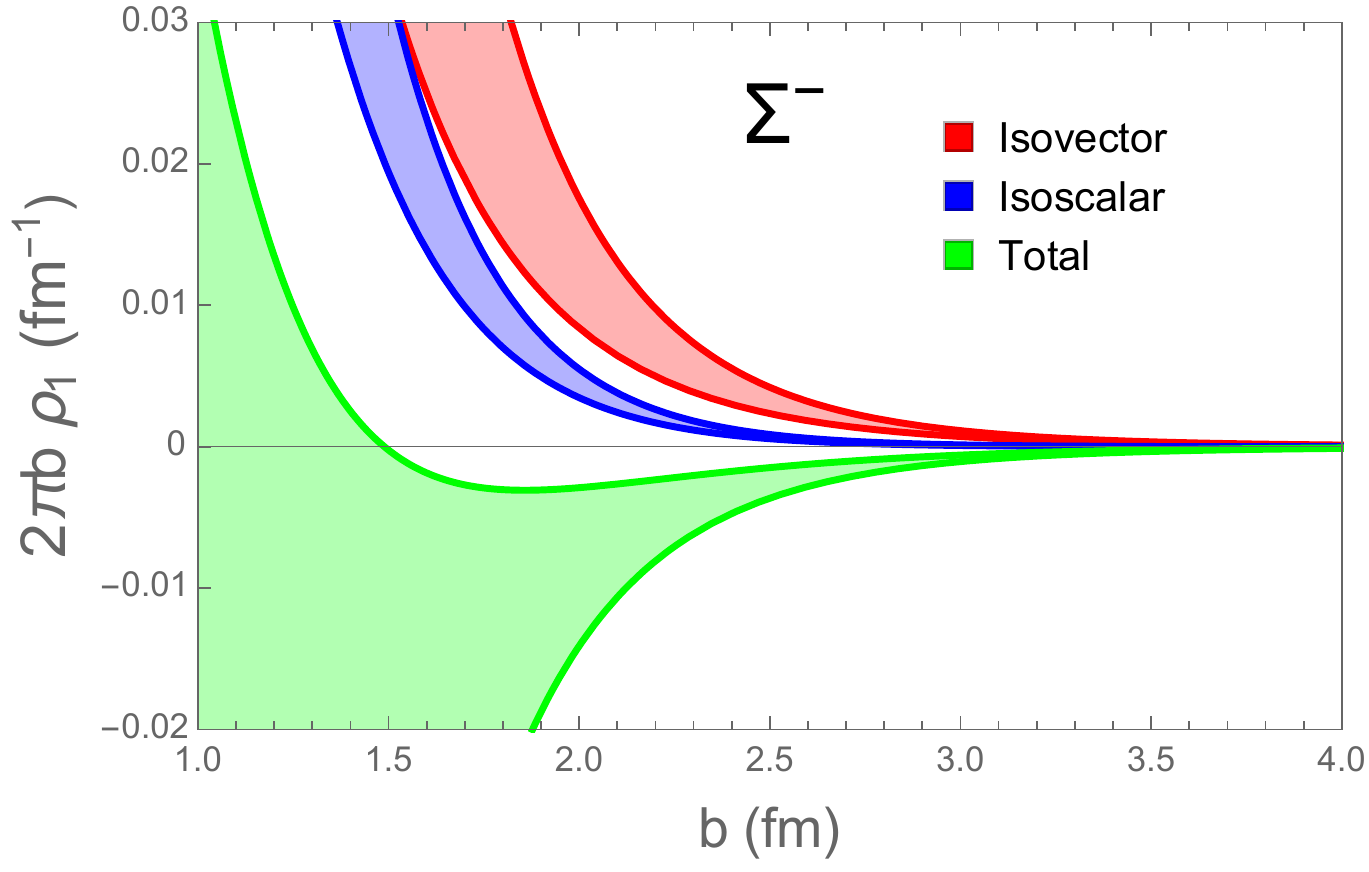,width=.45\textwidth,angle=0}\\
\epsfig{file=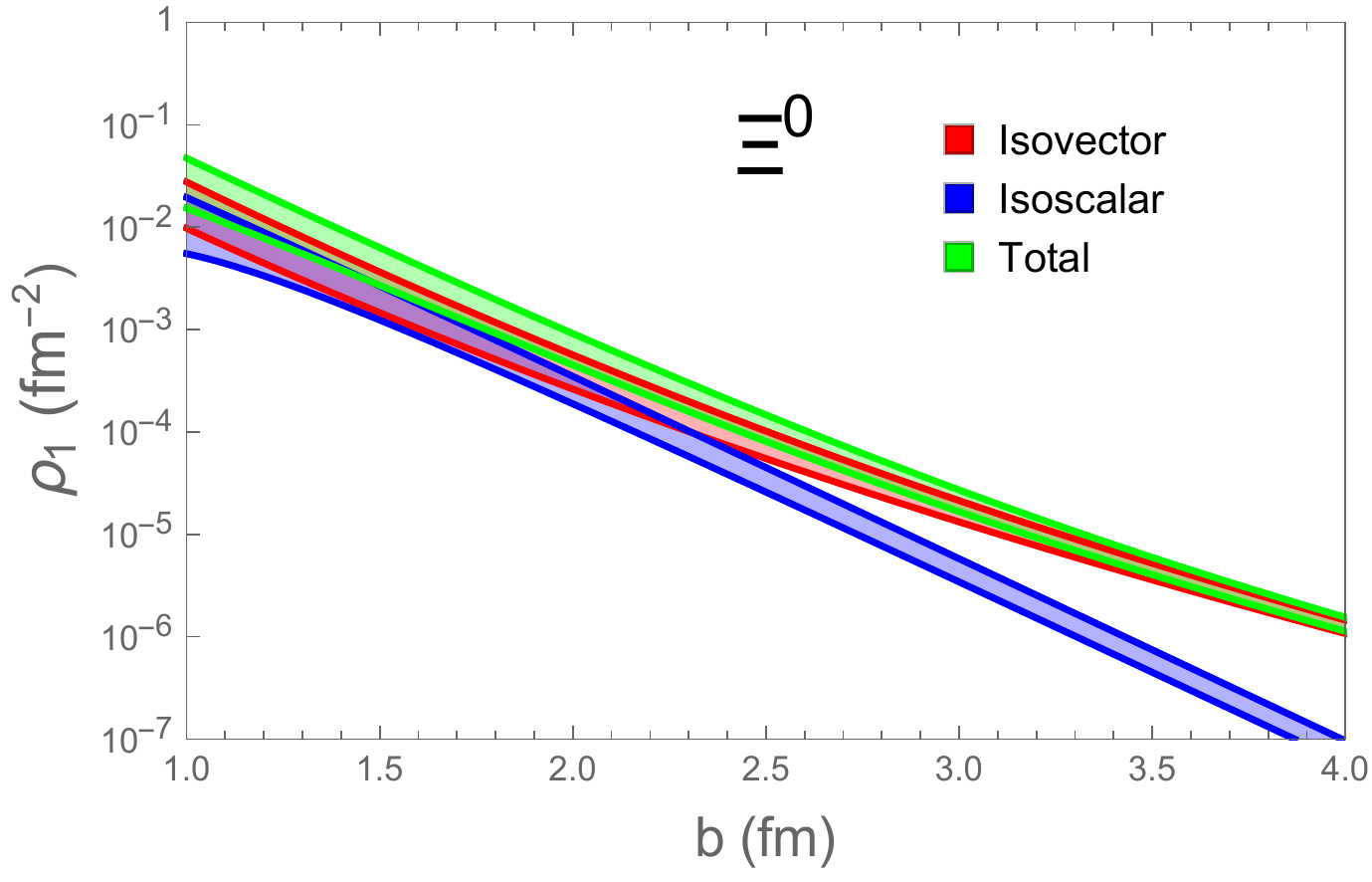,width=.45\textwidth,angle=0}
\epsfig{file=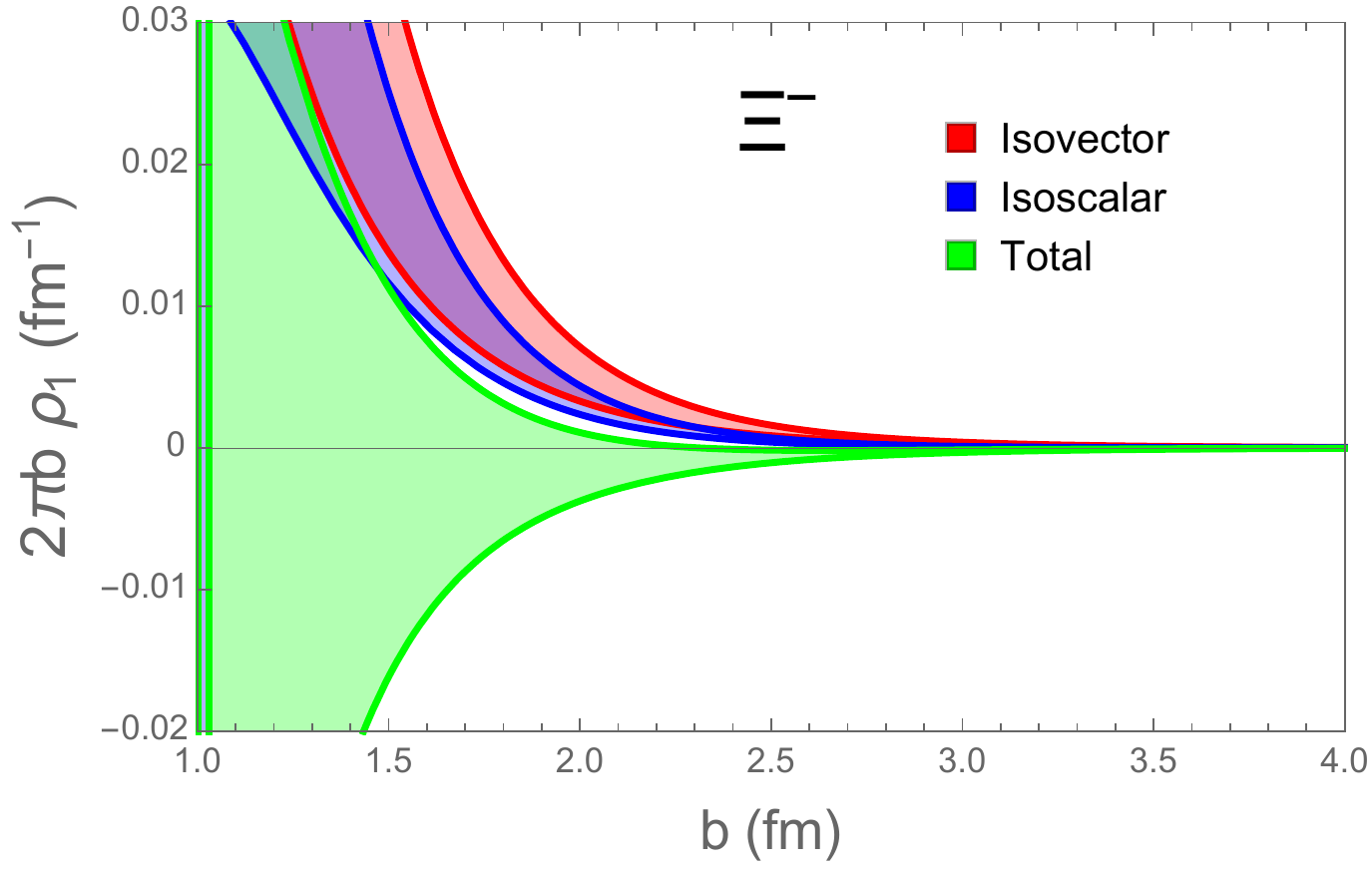,width=.45\textwidth,angle=0} \\
\epsfig{file=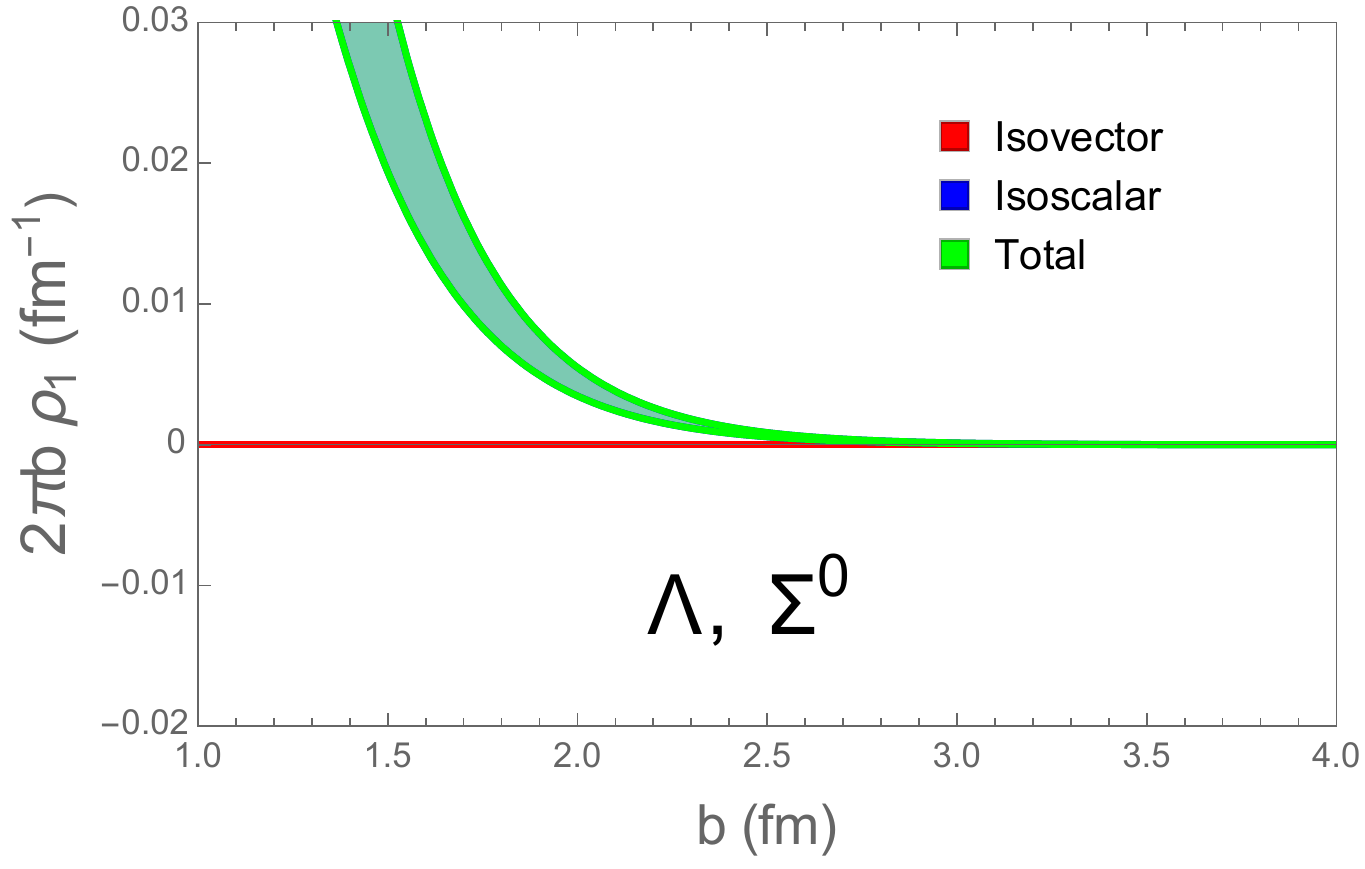,width=.45\textwidth,angle=0}
\epsfig{file=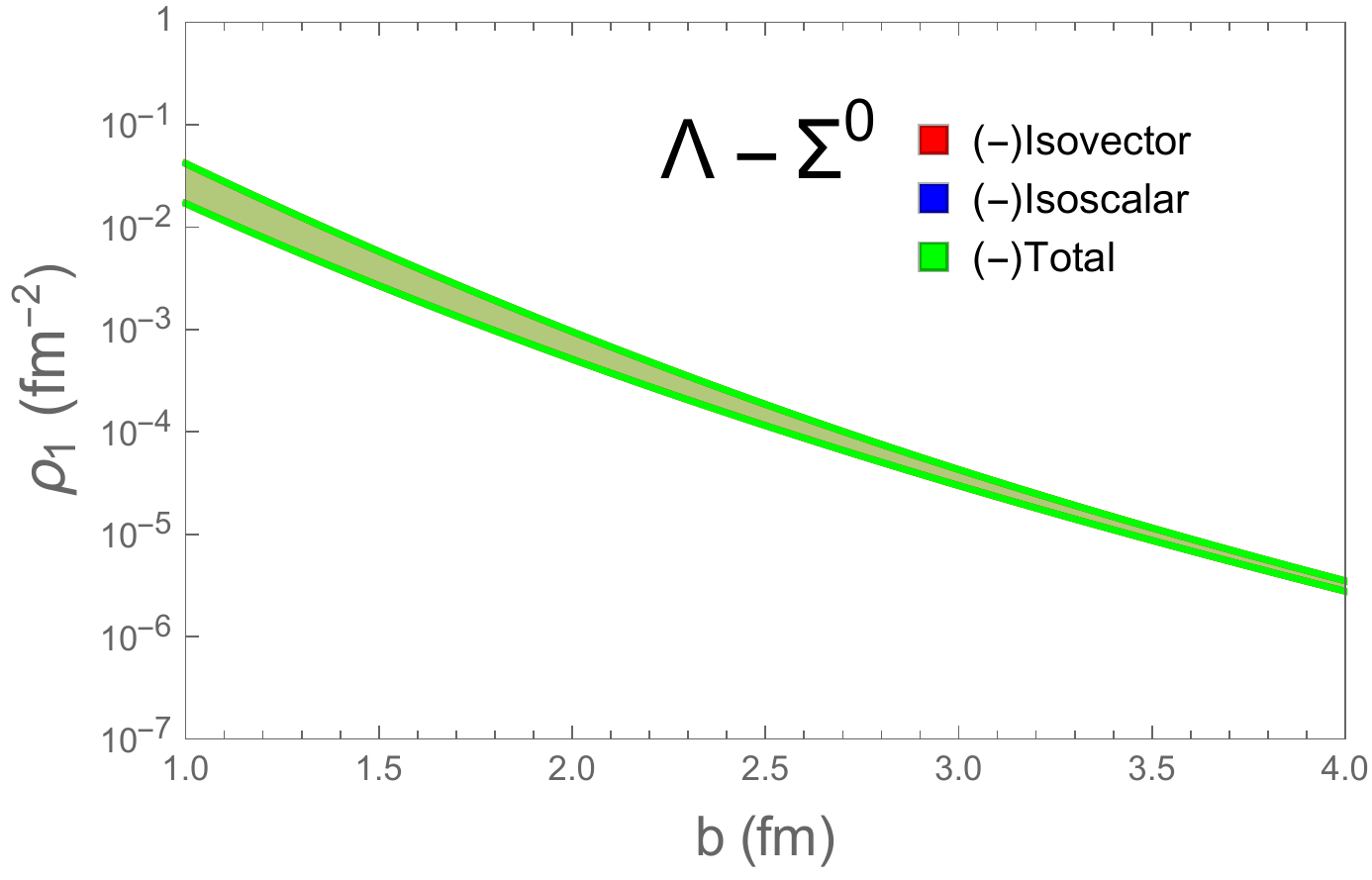,width=.45\textwidth,angle=0}
\caption[]{Transverse charge densities of the octet baryons. Red: Isovector component
calculated using $\chi$EFT and dispersive improvement. Blue: Isoscalar component 
estimated from vector meson poles. Green: Total density (sum or difference of isoscalar
and isovector components). For the densities with fixed sign we
plot $\rho_1(b)$ on a logarithmic scale (the signs are indicated in the legends of the
plots); for those with changing sign we plot the
radial densities $2\pi b \rho_1(b)$ on a linear scale.}
\label{Fig:rho1-Octet}
\end{center}
\end{figure*} 
%
%
\begin{figure*}
\begin{center}
\epsfig{file=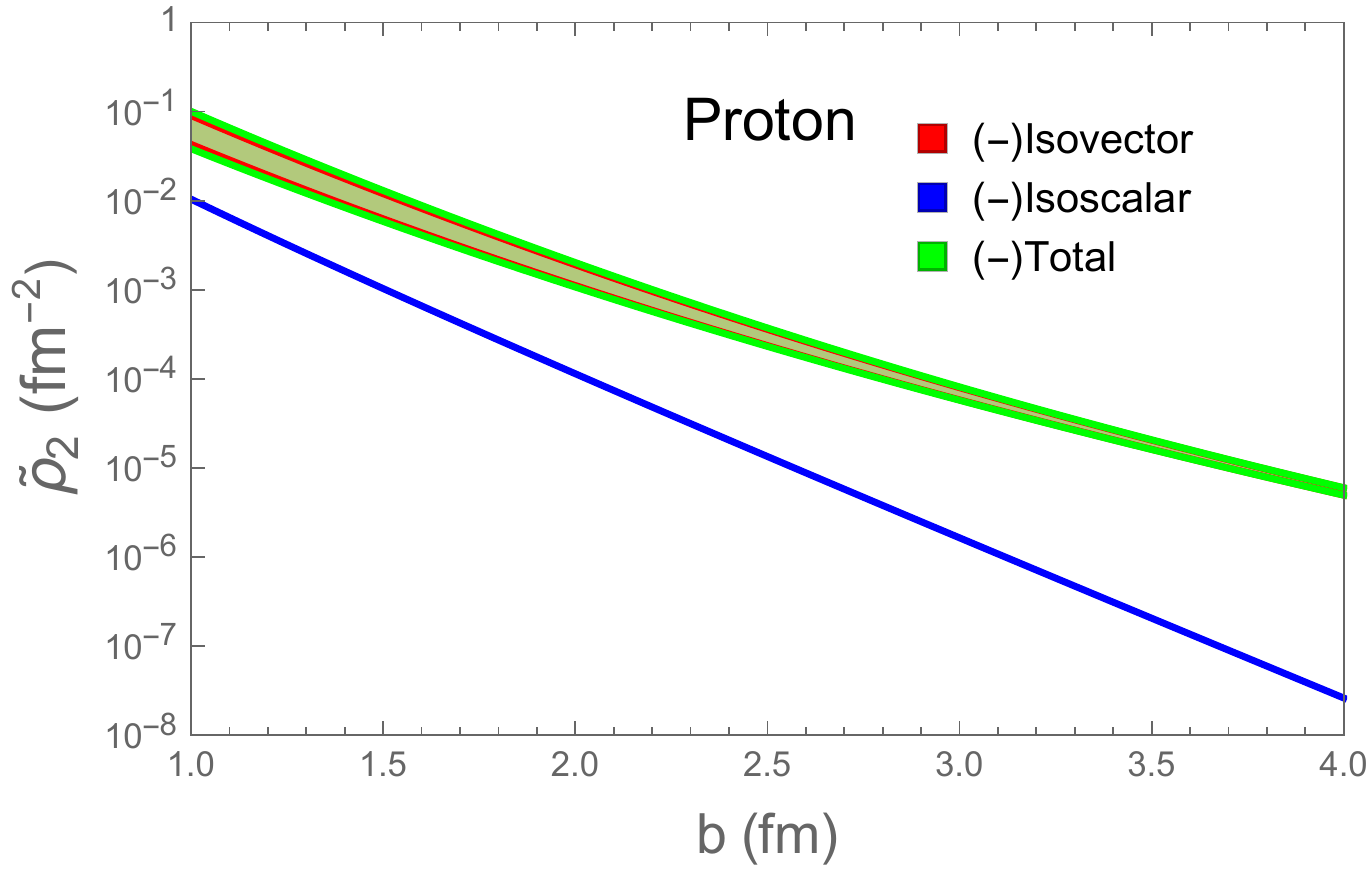,width=.45\textwidth,angle=0}
\epsfig{file=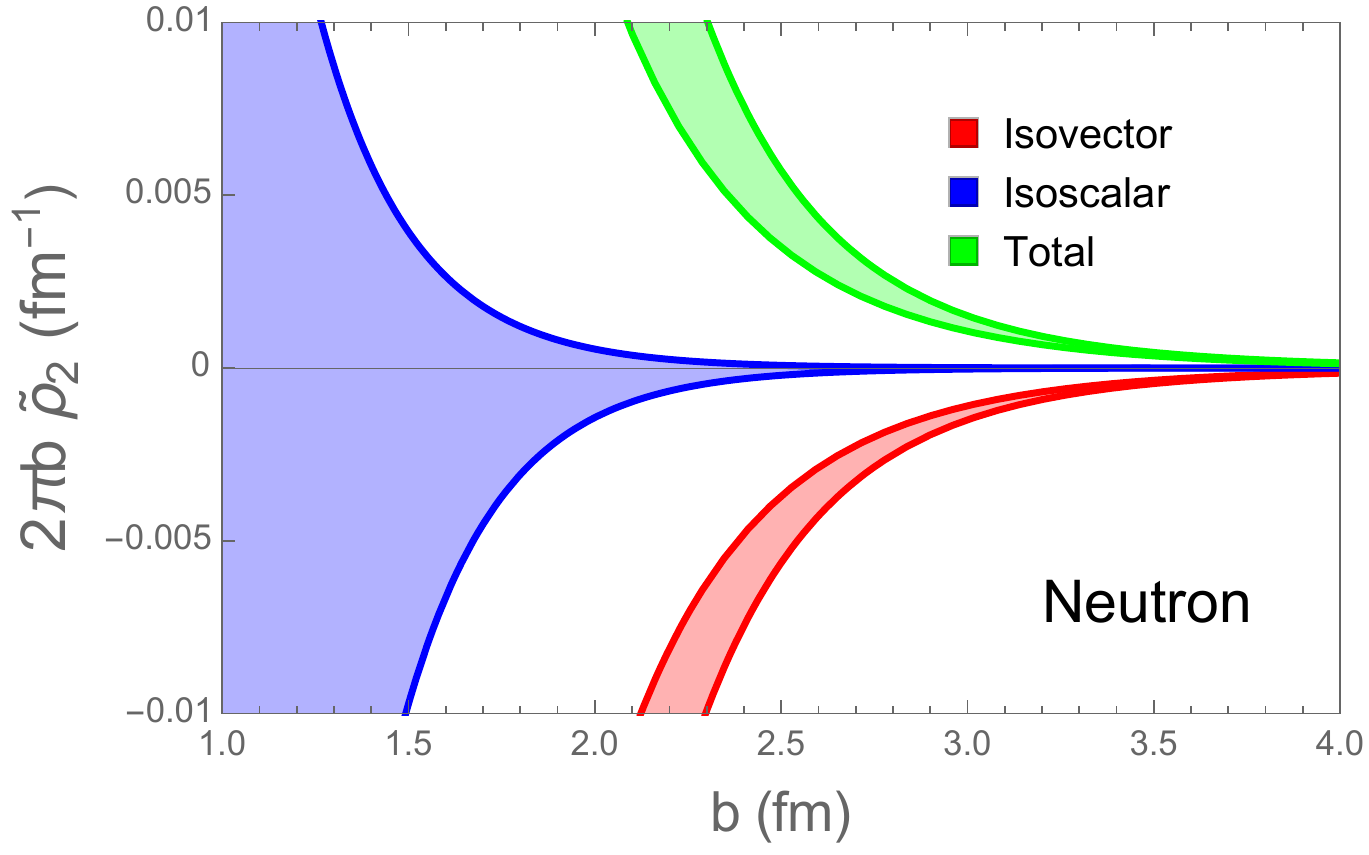,width=.45\textwidth,angle=0}\\ 
\epsfig{file=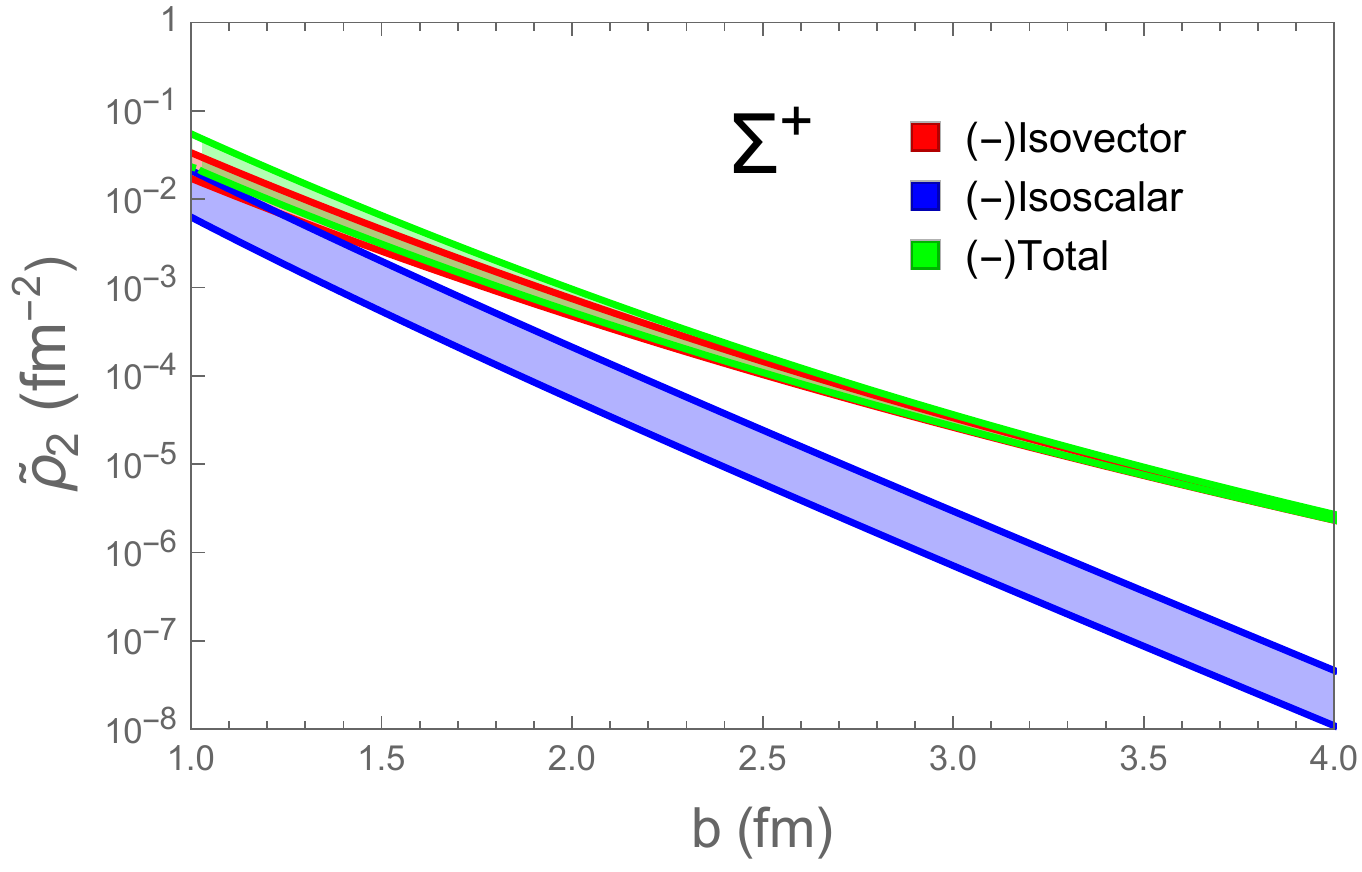,width=.45\textwidth,angle=0}
\epsfig{file=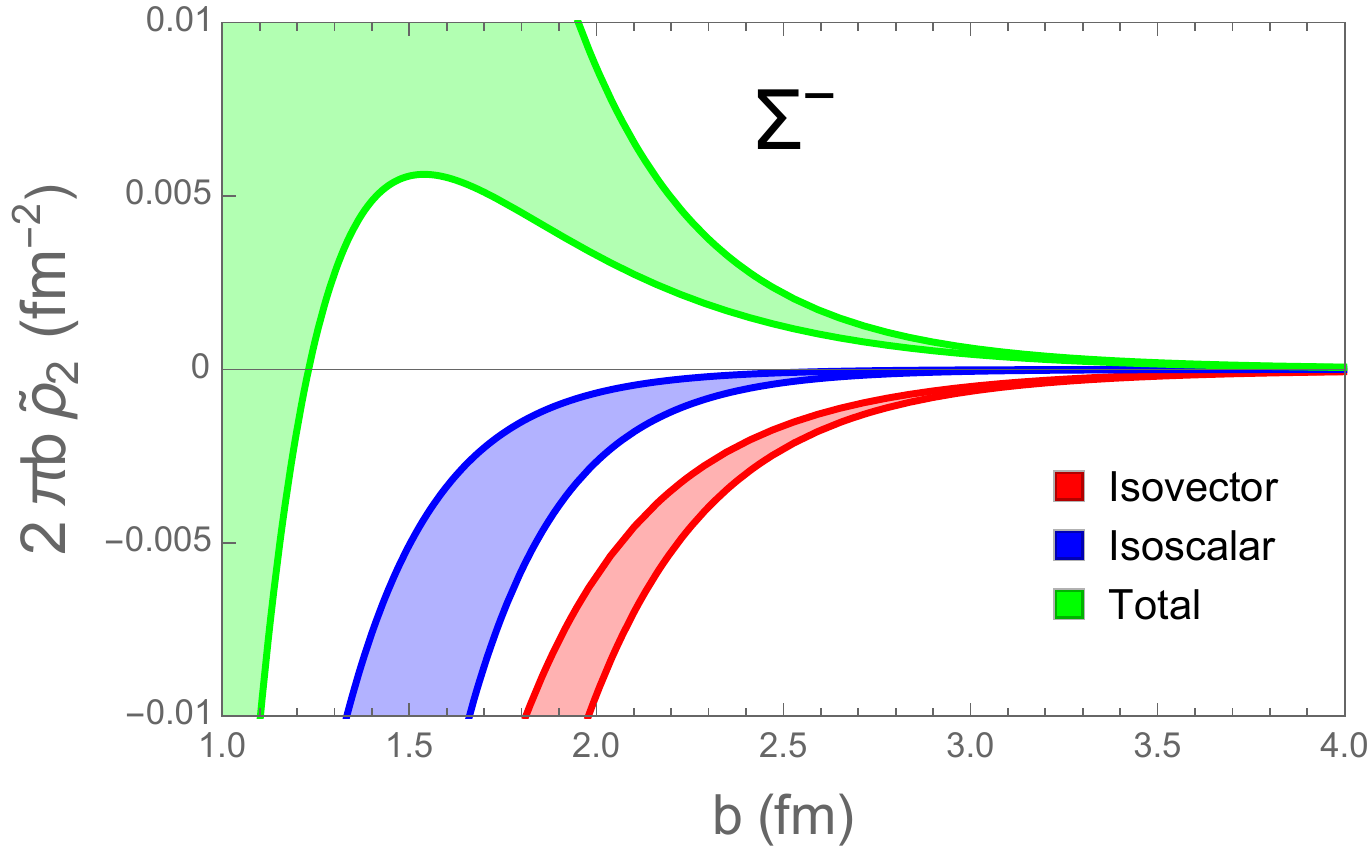,width=.45\textwidth,angle=0}\\ 
\epsfig{file=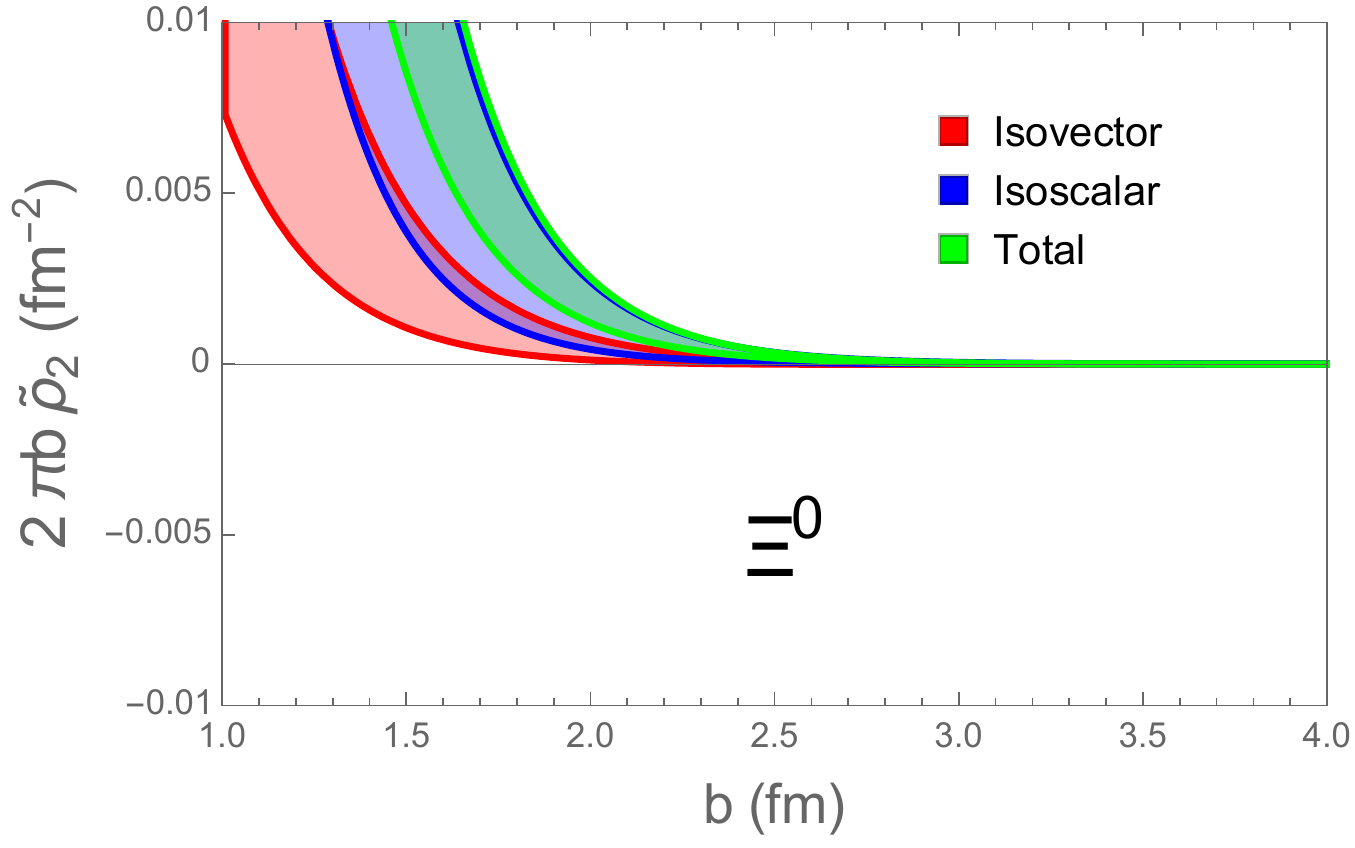,width=.45\textwidth,angle=0}
\epsfig{file=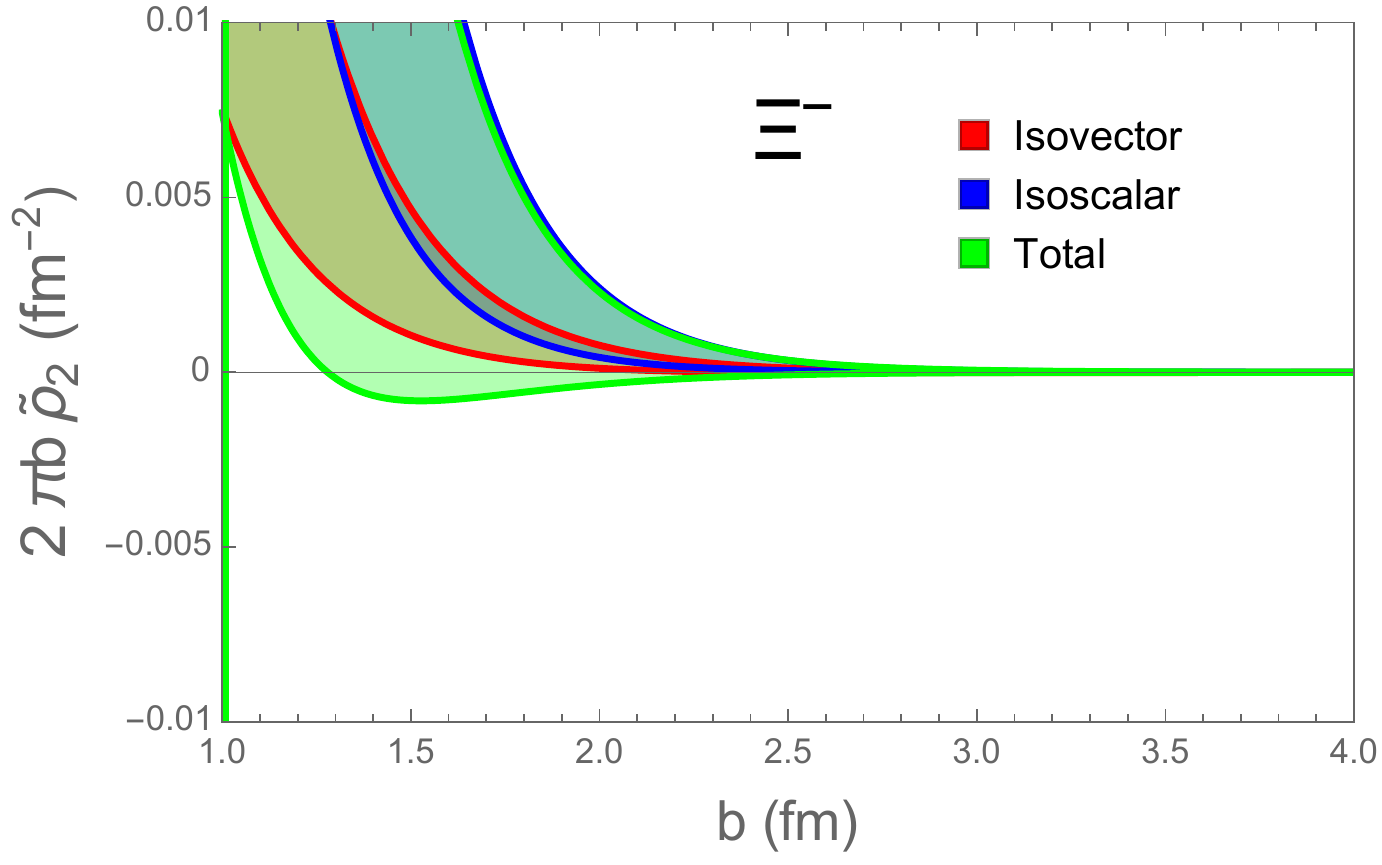,width=.45\textwidth,angle=0}\\
\epsfig{file=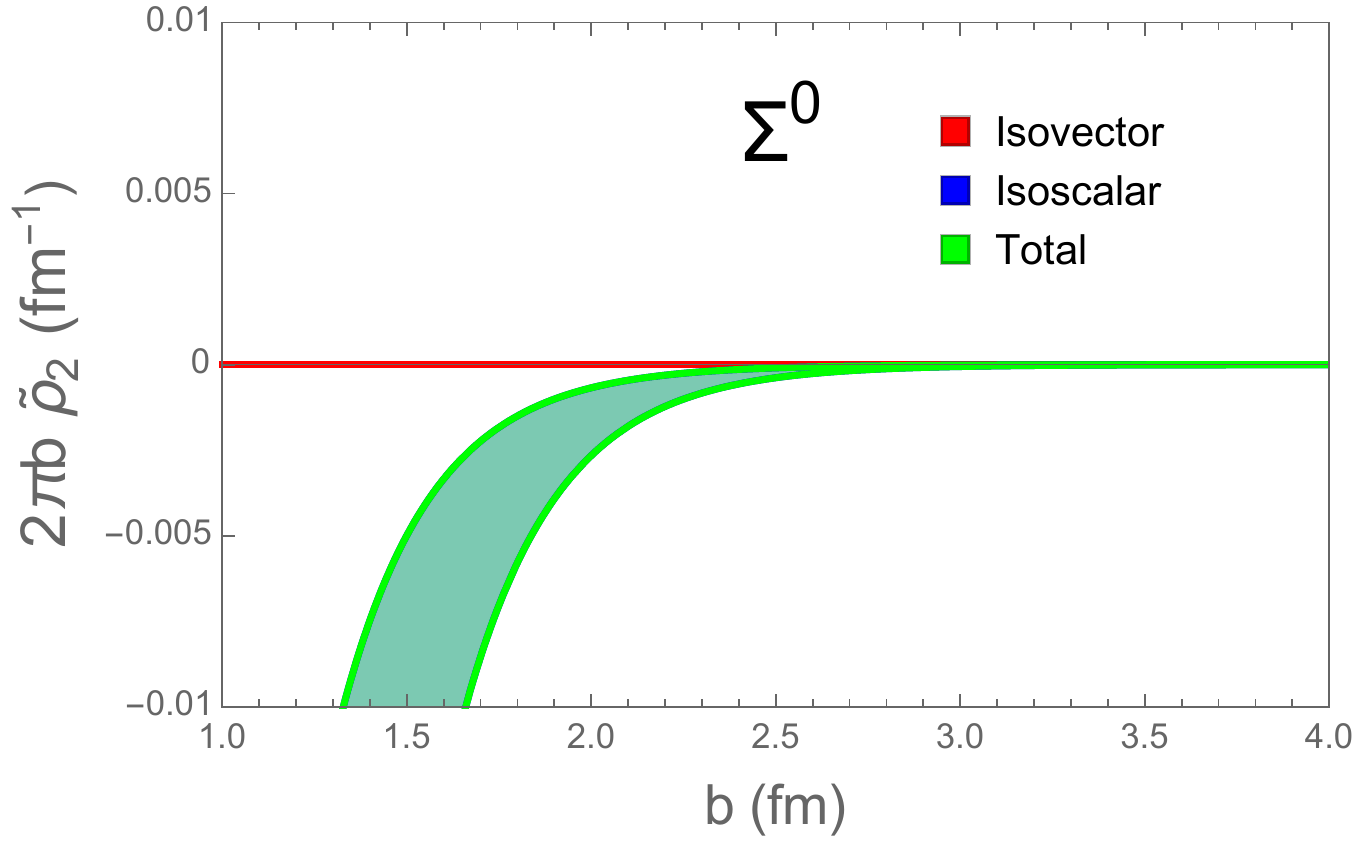,width=.45\textwidth,angle=0}
\epsfig{file=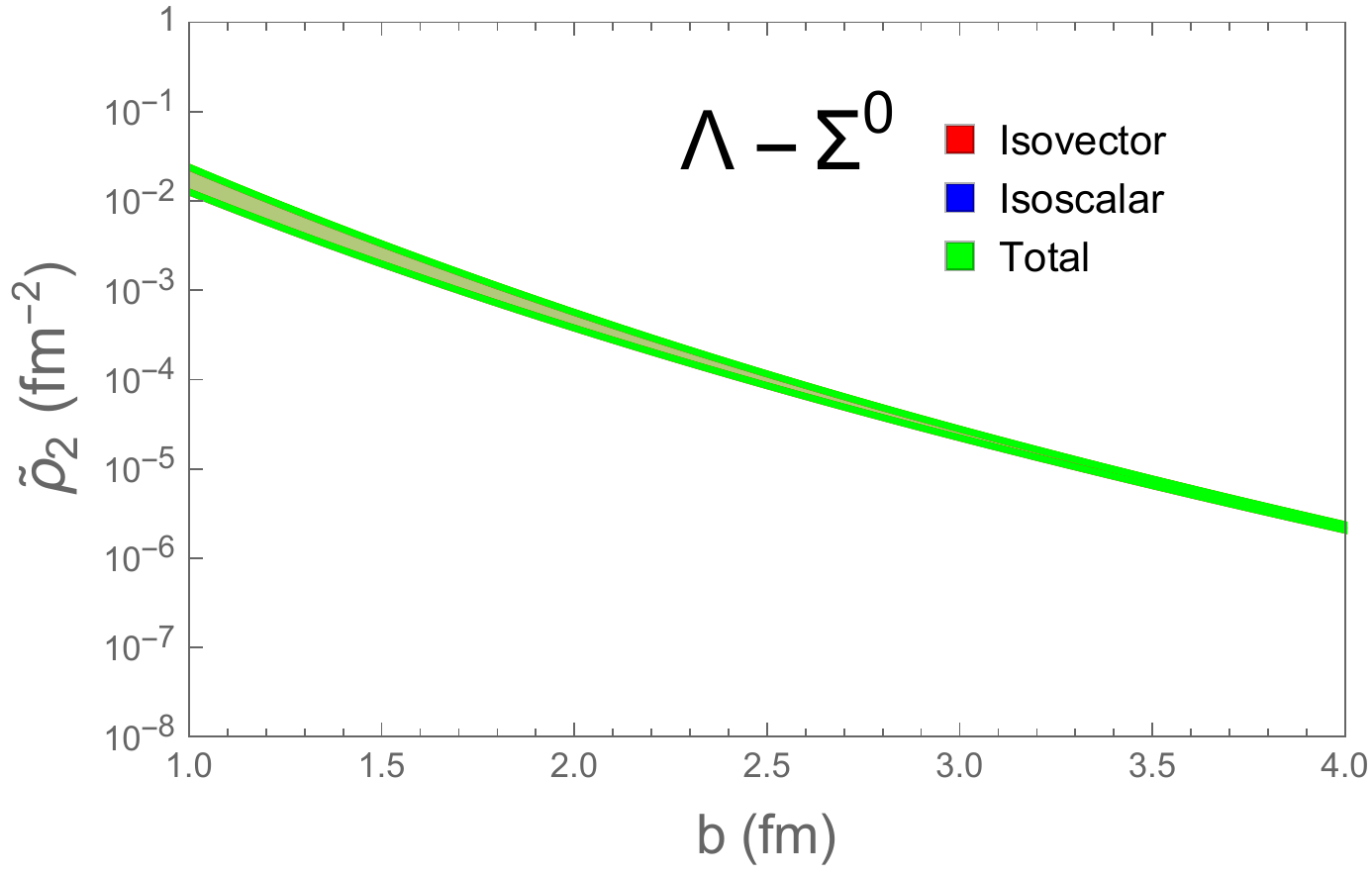,width=.45\textwidth,angle=0}
\caption[]{\small Transverse magnetization densities of the octet baryons. Red: Isovector component
calculated using $\chi$EFT and dispersive improvement. Blue: Isoscalar component 
estimated from vector meson poles. Green: Total density (sum or difference of isoscalar
and isovector components). For the densities with fixed sign we plot 
$\widetilde\rho_2(b)$ on a logarithmic scale (the signs are indicated in the legends of the
plots); for those with changing sign we plot the
radial densities $2\pi b \widetilde\rho_2(b)$ on a linear scale. \label{Fig:rho2-Octet}}
\end{center}
\end{figure*} 
The resulting charge densities $\rho_1^B(b)$ and magnetization densities $\widetilde\rho_2^B(b)$
are summarized in Figs.~\ref{Fig:rho1-Octet} and \ref{Fig:rho2-Octet}. For each octet baryon $B$
we show the total densities as well as their isovector and isoscalar components, defined according to
Eq.~(\ref{isospin_ff}),
\beq
\left.
\begin{array}{rcl}
\{ \rho^p, \rho^n \} &=& \rho^{N, S} \pm \rho^{N, V} ,
\\
\rho^{\Lambda} &=& \rho^{\Lambda, S} ,
\\
\{ \rho^{\Sigma^+}, \rho^{\Sigma^-} \} &=& \rho^{\Sigma, S} \pm \rho^{\Sigma, V} ,
\\
\rho^{\Sigma^0} &=& \rho^{\Sigma, S} ,
\\
\rho^{\Lambda - \Sigma} &=& \rho^{\Lambda-\Sigma, V} ,
\\
\{ \rho^{\Xi^0}, \rho^{\Xi^-} \} &=& \rho^{\Xi, S} \pm \rho^{\Xi, V} .
\end{array}
\hspace{1em}
\right\}
\label{isospin_rho}
\eeq
The densities with definite sign (positive or negative) are plotted on a logarithmic scale; 
for the densities with changing sign the plots show the radial densities 
$2\pi b \rho_1(b)$ and $2\pi b \widetilde\rho_2(b)$ on a linear scale.
Several general features of the results are worth noting:
(a)~The densities decay exponentially at large $b$, as dictated by the analytic properties 
of Eqs.~(\ref{Eq:rho1-spectral-rep}) and (\ref{Eq:rho2tilde-spectral-rep}).
The decay rate depends on the effective masses in the spectral integral. 
The isovector densities decay approximately as $\sim \exp(- M_\rho b)$ at 
$b \sim 1$ fm, and with a smaller effective mass at $b > 2 \, \textrm{fm}$, reflecting
the changing distribution of strength in the spectral integrals 
(see Fig.~\ref{Fig:rho_integrand}).\footnote{A decay of the isovector 
densities as $\sim \exp(- 2 M_\pi b)$ is observed only at 
extremely large distances $b \sim m_N^2/M_\pi^3$, where the dispersive integrals are
dominated by the extreme near-threshold region $t - 4 M_\pi^2 \sim M_\pi^4/m_N^2$;
see Ref.~\cite{Granados:2013moa} for discussion.} The isoscalar densities decay as $\exp(- M_\omega b)$
at all $b > 1 \textrm{fm}$.
(b)~At distances $b > 3$ fm the overall densities are dominated by the isovector component (if present).
Isoscalar and isovector densities become comparable only at distances $b < 2$ fm.
(c)~Our approximations allow us to reconstruct the isovector density with an uncertainty
of $\sim \pm 30\%$ at $b = 2$ fm, and with significantly lower uncertainty at larger distances.
We now want to inspect the densities of the individual baryon multiplets.

\underline{{\it Proton and Neutron.}} 
The nucleon isovector densities were studied in $\chi$EFT in Ref.~\cite{Granados:2013moa}. 
The dispersive improvement of the $\chi$EFT spectral functions carried out in the present work
(see Sec.~\ref{Subsec:Improvement}) significantly increases the peripheral densities 
(see Fig.~\ref{Fig:ImFV-Octet}b and d) and allows us to construct them down to much smaller 
distances $b \gtrsim 1 \, \textrm{fm}$.
The nucleon isovector and isoscalar charge densities are both positive (see Fig.~\ref{Fig:rho1-Octet}). 
In the proton charge density the isovector and isoscalar contribute with the same sign, 
leading to a positive total density with smooth behavior. In the neutron charge 
density the isovector and isoscalar contribute with opposite sign, causing cancellations and
varying behavior of the total density. At distances $b > 2$ fm the isovector dominates and
the total neutron charge density is negative, as expected from the intuitive picture of the
``pion cloud'' resulting from $n \rightarrow \pi^- + p$ transitions (this interpretation
can be demonstrated rigorously in the light-front representation of chiral 
dynamics \cite{Granados:2015rra,Granados:2015lxa,Granados:2016jjl}). At shorter distances
$b \sim 1$ fm our results are consistent with a positive charge density in the neutron, 
as is observed in the empirical densities and has been discussed in the 
literature \cite{Miller:2007uy,Miller:2011du}; however, the uncertainties of our present
calculation are large and do not allow us to predict the sign at $b < 2$ fm.

In the nucleon magnetization density the isoscalar component is significantly smaller 
than in the charge density, due to the comparatively small value of the $\omega NN$ 
tensor coupling (see Fig.~\ref{Fig:rho2-Octet}). As a result, the isovector magnetization 
density dominates down to much smaller distances $b \sim 1\, \textrm{fm}$, and the proton
and neutron have approximately opposite magnetization densities in the periphery.

In Refs.~\cite{Granados:2013moa,Granados:2015rra} it was observed that the peripheral
isovector nucleon charge and magnetization densities satisfy an approximate inequality 
$|\widetilde\rho_2^N (b)| < \rho_1^N (b)$. It applies to the densities arising from 
the triangle diagrams with nucleon intermediate states and can be derived rigorously
in the light-front representation of chiral dynamics 
\cite{Granados:2015rra,Granados:2015lxa,Granados:2016jjl}. The contributions from
$\Delta$ intermediate states violate the inequality (in the large-$N_c$ limit of
QCD they restore the hierarchy $|\widetilde\rho_2^N (b)| \gg \rho_1^N (b)$ required
by general scaling arguments). The results of the present calculation confirm 
these observations.

\underline{{\it $\Lambda$ and $\Sigma^0$.}} In the $\Lambda$ and $\Sigma^0$ charge
and magnetization densities the isovector component is absent, cf.~Eq.~(\ref{isospin_rho}). 
In the $\chi$EFT description this happens by way of cancellation of the $\pi^+$ and $\pi^-$ 
contributions from the triangle diagrams with $\pi^\pm \Sigma^\mp$ intermediate states,
and similarly for the diagrams with $\pi^\pm \Sigma^{*\mp}$ intermediate states. 
The $\Lambda$ and $\Sigma^0$ densities
are thus pure isoscalar. They are dominated by $\omega$ and $\phi$ exchange down to distances 
$b \sim 1$ fm and are an order of magnitude smaller than the densities of the other baryons 
at $b > 2$ fm. In our parametrization based on $SU(3)$ symmetry the vector couplings are the 
same for the $\Lambda$ and $\Sigma^0$ , $a_1^{\omega \Lambda\Lambda} = a_1^{\omega \Sigma^0\Sigma^0}$
(and similarly for $\phi$) (see \ref{App:Vector} and 
Table~\ref{Table:VBB-coefficients-2}), resulting in identical peripheral charge densities
(see Fig.~\ref{Fig:rho1-Octet}). In contrast, the tensor couplings have opposite signs,
$a_2^{\omega \Lambda\Lambda} \approx -a_2^{\omega \Sigma^0\Sigma^0}$, giving rise to
approximately opposite magnetization densities (see Fig.~\ref{Fig:rho2-Octet}).

\underline{{\it $\Lambda$-$\Sigma^0$ transition.}}
The $\Lambda$-$\Sigma^0$ transition densities are pure isovector, cf.~Eq.~(\ref{isospin_rho}).
They receive sizable peripheral contributions from the chiral processes with $\pi^\pm \Sigma^\mp$ 
(octet) and $\pi^\pm \Sigma^{*\mp}$ (decuplet) intermediate states. The signs of the transition
charge and magnetization densities are opposite to those of the nucleon densities
(see Figs.~\ref{Fig:rho1-Octet} and \ref{Fig:rho2-Octet}), as can be inferred already
from the spectral functions (see Fig.~\ref{Fig:ImFV-Octet}). The relative magnitude
of the isovector densities at distances $b \gtrsim 3$ fm is determined by the ratio of the
spectral functions near threshold, cf.~Eq.~(\ref{Eq:ImF_ratios}) and Fig.~\ref{Fig:ImFV-Octet}a and c; 
at distances $b \sim 1$ fm it is given by the ratio of the spectral functions 
in the vector meson mass region, cf.~Fig.~\ref{Fig:ImFV-Octet}b and d.

\underline{{\it $\Sigma^+$ and $\Sigma^-$.}} The charge densities of the $\Sigma^+$ and $\Sigma^-$ 
baryons are similar to the ones in the proton and neutron (see Fig.~\ref{Fig:rho1-Octet}).
In difference to the $\Sigma^0$, in the
$\Sigma^\pm$ densities both isovector and isoscalar components are present, cf.~Eq.~(\ref{isospin_rho}).
The isovector charge densities $\rho^{\Sigma, V}$ and $\rho^{N, V}$ are similar at all distances 
$b > 1$ fm, because the isovector $\Sigma$ spectral function is close to the nucleon one both near 
threshold and in the vector meson region, cf.~Fig.~\ref{Fig:ImFV-Octet}a and b. The isoscalar charge 
densities $\rho^{\Sigma, S}$ and $\rho^{N, S}$ are likewise similar, because the $\omega$ couplings
of the two baryons are close (see Table~\ref{Table:VBB-coefficients-2}). Altogether therefore the
$\Sigma^+$ charge density is very close to that of the proton, and the $\Sigma^-$ charge density shows
a sign change similar to the neutron; the details depend on the accuracy of our calculation of 
the isovector spectral functions and the modeling of the isoscalar ones.

The $\Sigma^+$ and $\Sigma^-$ magnetization densities are again similar to the ones in the proton
and neutron, with some notable differences compared to the charge densities
(see Fig.~\ref{Fig:rho2-Octet}). In the magnetization
densities the isovector $\widetilde\rho_2^{\Sigma, V}$ is smaller than $\widetilde\rho_2^{N, V}$
by about a factor $\sim 1/2$, while the isoscalar $\widetilde\rho_2^{\Sigma, S}$ is larger in
magnitude than $\widetilde\rho_2^{N, S}$.

\underline{{\it $\Xi^0$ and $\Xi^-$.}} In the $\Xi$ baryons the peripheral isovector charge density is 
substantially smaller than in the nucleon or charged $\Sigma$ states
(see Fig.~\ref{Fig:rho1-Octet}). This reflects the fact that
the isovector spectral function of the $\Xi$ is relatively suppressed near threshold, 
cf.~Eq.~(\ref{Eq:ImF_ratios}) and Fig.~\ref{Fig:ImFV-Octet} a and c, because the intermediate 
octet contribution to the $\Xi$ is small and comparable to the decuplet one. (In the other baryons
the octet intermediate states clearly dominates in the periphery.) At the same time, the isoscalar
density of the $\Xi$ has normal size. As a result, in the $\Xi$ charge density the isoscalar and 
isovector become comparable at slightly larger distances than in the nucleon and $\Sigma$. This has interesting 
consequences for the charge density of the $\Xi^-$ baryon, which is the difference of the isoscalar 
and isovector components. It suggests a sign change from a negative charge density at large $b$
to a positive one at intermediate $b$, similar to the neutron, but with the transition happening
at larger $b$ than in the neutron (we cannot confirm this behavior as the estimated uncertainty
of the total $\Xi^-$ charge density is larger than that of the neutron in the present
calculation).

The results for the $\Xi$ magnetization density resemble those for the $\Sigma^0$ and $\Lambda$ 
baryons (see Fig.~\ref{Fig:rho2-Octet}). 
Although an isovector component is present in the $\Xi$ spectral function, it is extremely small
(see Fig.~\ref{Fig:ImFV-Octet}), resulting in a tiny contribution to the magnetic density,
(see Fig.~\ref{Fig:rho2-Octet}). On the other hand, the contribution of the isoscalar vector mesons 
is larger for the $\Xi$ baryons than for the rest of the baryons in the octet.  
This results in a magnetic density clearly dominated by the isoscalar component, even in the
peripheral region. Given that the isoscalar component contributes in the same way for both, 
the $\Xi^0$ and $\Xi^-$, and that the isovector component is negligible, the magnetic density 
profile comes out almost exactly the same for both baryons.
\subsection{Flavor decomposition}
\label{subsec:flavor}    
It is interesting to study the quark flavor decomposition of the transverse charge 
and magnetization densities in the octet baryons \cite{Miller:2011du}. 
For each baryon $B$ the densities can be decomposed as
\beq
\rho_i^B \; = \; e_u \rho^{B, u} + e_d \rho^{B, d} + e_s \rho^{B, s} \hspace{2em} (i = 1,2),
\label{eq:flavor_decomposition_B}
\eeq
where $\rho_i^{B, f}(b) \; (f = u, d, s)$ represent the densities associated with the vector currents 
of the quark fields with flavor $f$, $J^\mu_f = \bar\psi_f \gamma^\mu \psi_f$; the quark charge
factors are included explicitly in Eq.~(\ref{eq:flavor_decomposition_B}). The flavor densities 
$\rho_1^{B, f}$ satisfy the sum rules
\beq
\int d^2 b \; \rho_1^{B, f}(b) \; = \; N_f,
\eeq
where $N_f$ represents the total quark flavor content of the baryon as determined by the isospin and
hypercharge; i.e., the number of ``valence quarks'' in the baryon. In the context of GPDs the 
densities $\rho_1^{B, f}$ represent the
integral of the quark and antiquark distributions at transverse distance $b$ over the light-front
momentum fraction $x$,
\beq
\rho_1^{B, f}(b) \; = \; \int_0^1 dx \; [f^{B, f}(x, b) - f^{B, \bar f}(x, b)] .
\eeq
The flavor densities $\rho_2^{B, f}$ are related in a similar manner to the quark distributions
associated with the baryon helicity-flip GPDs. Note that the densities $\rho_1^{B, f}$ can
be positive or negative, depending on whether there are more quarks or antiquarks of flavor $f$ 
at a given transverse position $b$. The flavor densities within the isospin multiplets
are related by isospin symmetry,
\beq
\left.
\begin{array}{rclrcl}
\rho_i^{p, u} & \! = \! & \rho_i^{n, d}, & \rho_i^{p, s} & \! = \! & \rho_i^{n, s}, 
\\
\rho_i^{\Sigma^+, u} & \! = \! & \rho_i^{\Sigma^-, d}, & \rho_i^{\Sigma^+, s} & \! = \! & \rho_i^{\Sigma^-, s} 
= \rho_i^{\Sigma^0, s}, 
\\
\rho_i^{\Sigma^0, u} & \! = \! & \rho_i^{\Sigma^0, d}, &
\\
\rho_i^{\Xi^0, u} & \! = \! & \rho_i^{\Xi^-, d}, & \rho_i^{\Xi^0, s} & \! = \! & \rho_i^{\Xi^-, s} .
\end{array}
\right\}
\label{rho_flavor_isospin}
\eeq

For the flavor decomposition of the form factors and densities we need to separate the 
contributions of non-strange ($u, d$) and strange ($s$) quarks in the isoscalar 
electromagnetic current [cf.~Eq.~(\ref{isospin_rho})]
\beq
\rho_i^{B, S} = \rho_i^{B, u+d} + \rho_i^{B, s} \hspace{2em} (i = 1, 2).
\label{isoscalar_ud_s}
\eeq
In our model of the isoscalar spectral functions, Eq.~(\ref{Eq:isoscalar_poles}),
the $u + d$ densities are identified with the contribution of the $\omega$ pole, 
while the $s$ densities are identified with the $\phi$ pole. We emphasize that this 
identification is based on the assumptions of vector dominance and ideal mixing (see \ref{App:Vector}) 
and can only provide a rough estimate of the strange-nonstrange separation of the isoscalar current
(for a discussion of the uncertainties of the flavor separation of the isoscalar
nucleon form factors, see Refs.~\cite{Hammer:1998rz,Hammer:1999uf,Meissner:1997qt}).
The results presented in the following are intended only to illustrate the basic 
features of the spatial dependence of the flavor densities in the peripheral region.

Using Eqs.~(\ref{rho_flavor_isospin}) and (\ref{isoscalar_ud_s}) it is straightforward to
calculate the quark flavor densities in terms of the baryon isovector, isoscalar up-down,
and isoscalar strange densities. The quark flavor densities drop exponentially at large $b$ 
with a rate determined by the relevant $t$-channel exchanges. In the $u$ and $d$ quark 
densities the isovector two-pion exchange dominates at large $b$ (if present) and causes 
the densities to be of equal magnitude and opposite sign,
\beq
\rho_1^{B, u}(b) \, = \, -\rho_1^{B, d}(b) \hspace{1em} 
(b \rightarrow \infty, \; B = p, n, \Sigma^\pm, \Xi^0, \Xi^-).
\label{rho1_u_d_infinity}
\eeq
In the $p, \Sigma^+$ and $\Xi^0$ the peripheral $u$ density is positive, and the $d$ density is negative,
as they result from processes with emission of a $\pi^+$ contributing to the $u$ quark and
$d$ antiquark densities; in the $n, \Sigma^-$ and $\Xi^-$ the signs are opposite.
At shorter distances $b < 2\, \textrm{fm}$ the $u$ and $d$ densities can show more 
complex behavior, as the isovector and isoscalar non-strange exchanges become comparable
($\rho$ and $\omega$). The strange quark density drops with the range of $\phi$ exchange
and has no long-range component.

%
%
\begin{figure}
\begin{tabular}{l}
\epsfig{file=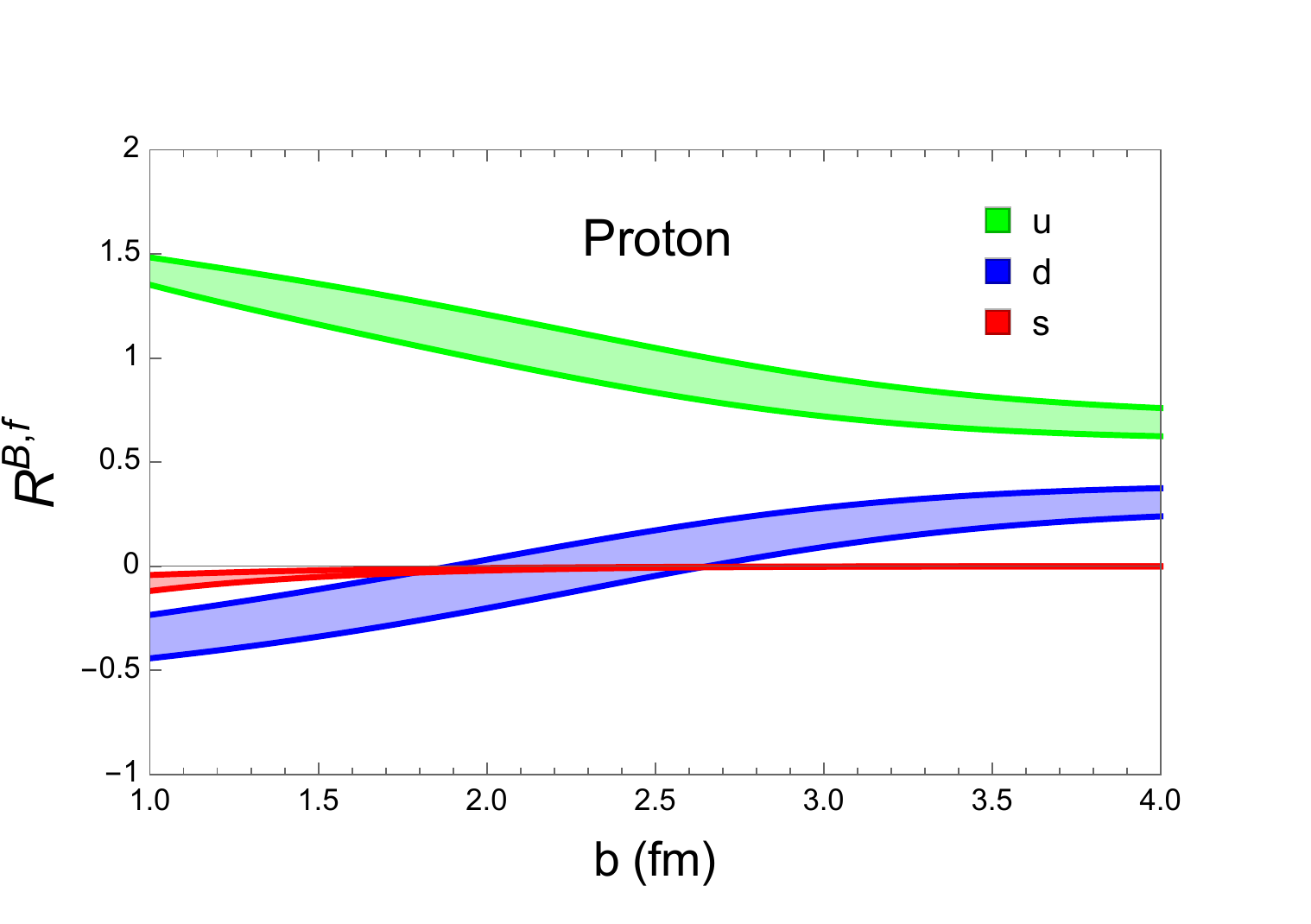,width=.45\textwidth,angle=0} 
\\[-3ex]
{\footnotesize (a)}
\\[-3ex]
\epsfig{file=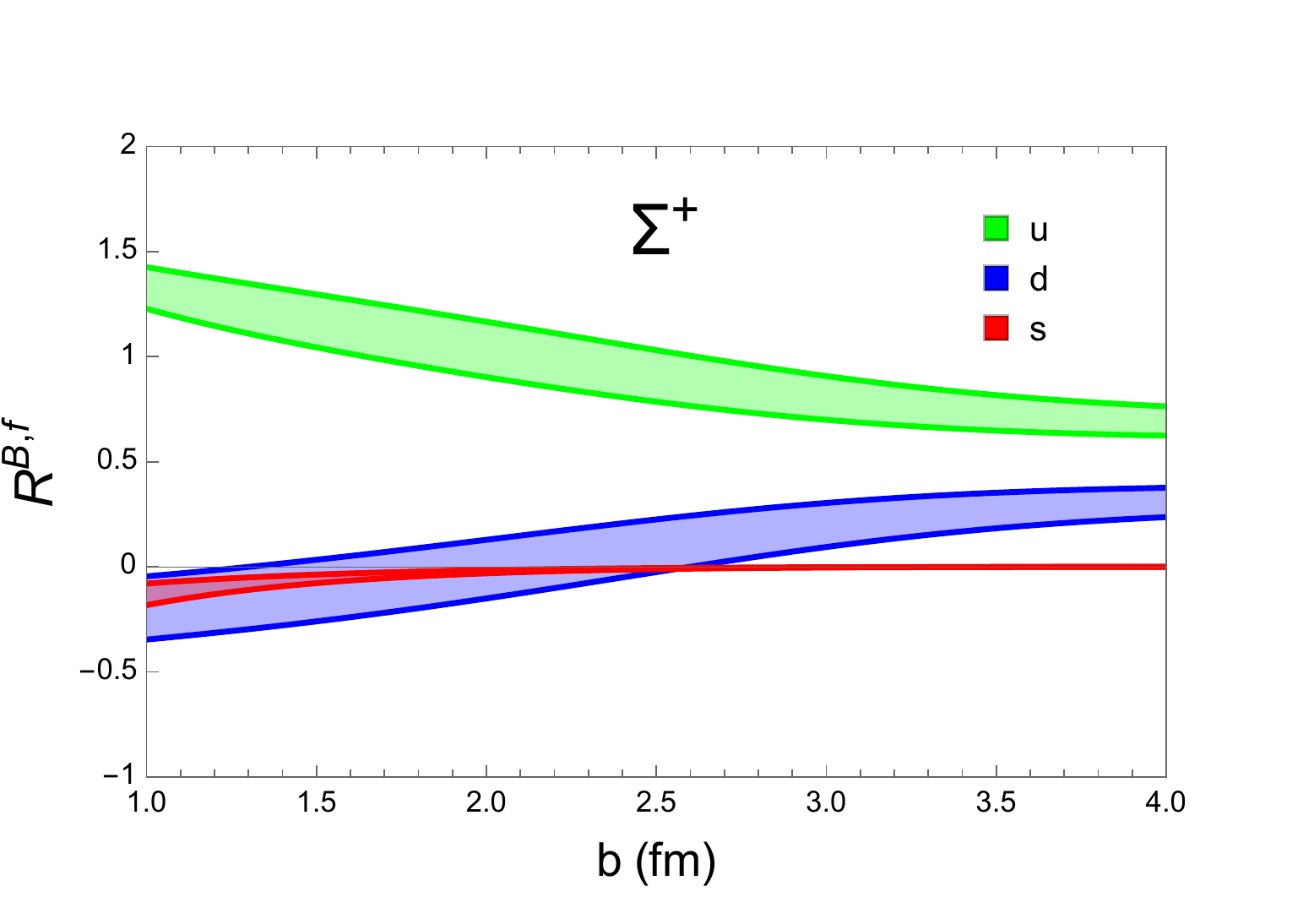,width=.45\textwidth,angle=0}
\\[-3ex]
{\footnotesize (b)}
\\[-3ex]
\epsfig{file=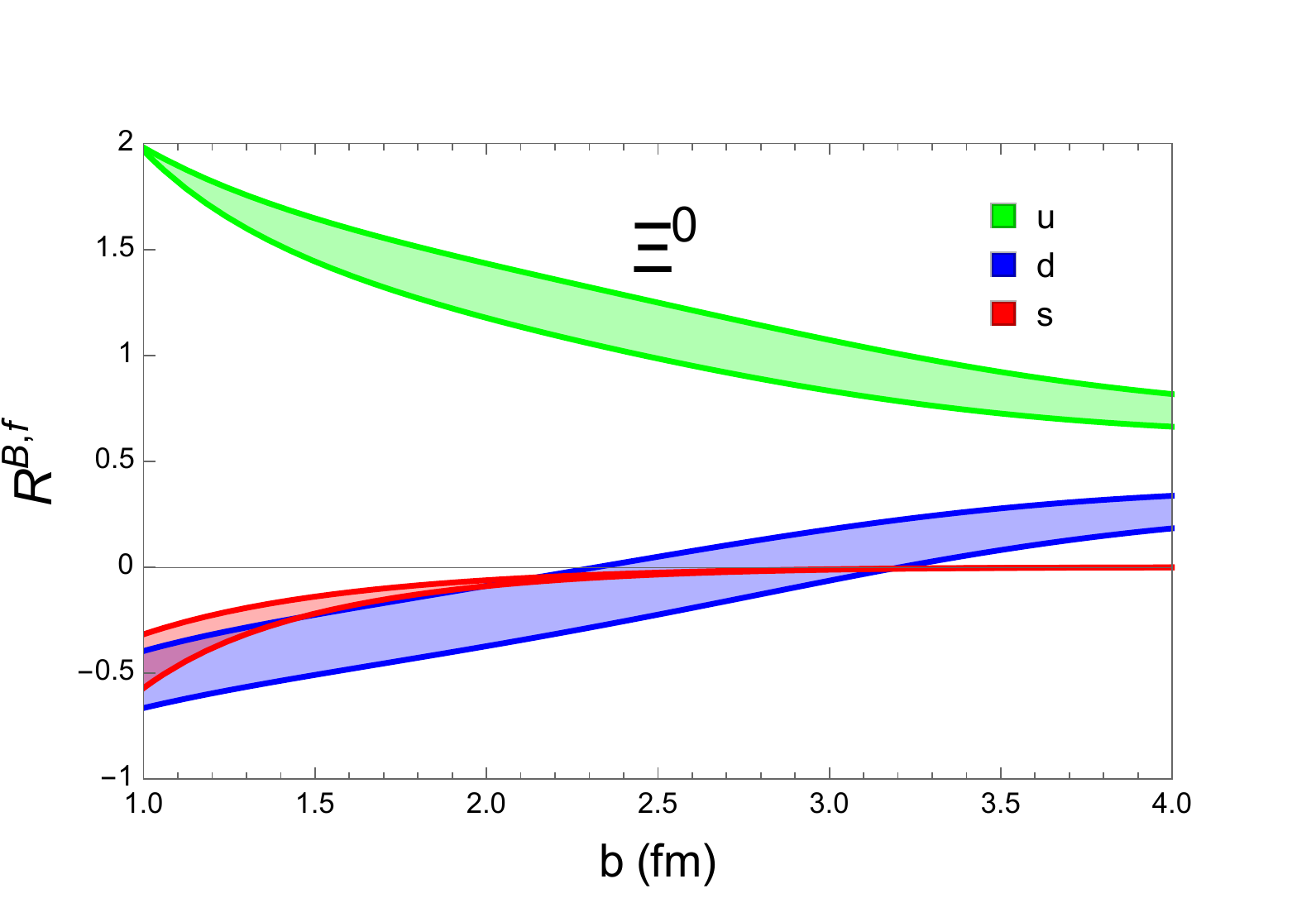,width=.45\textwidth,angle=0} 
\\[-3ex]
{\footnotesize (c)}
\end{tabular}
\caption[]{\small The ratios $R^{B, f}(b) \; (f = u,d,s)$, Eq.~(\ref{R_def}), describing the 
relative contribution of each flavor to the total baryon charge density at a given distance $b$.} 
\label{Fig:quark_fraction}
\end{figure} 
Of particular interest for baryon structure is the relative contribution of the various quark flavors 
to the charge density in the baryon at a given distance. It is convenient to consider the ratios
\beq
R^{B, f}(b) \equiv \frac{e_f \rho_1^{B, f}(b)}{\sum_{f'} e_{f'} \rho_1^{B, f'}(b)}
= \frac{e_f \rho_1^{B, f}(b)}{\rho_1^{B}(b)} ,
\label{R_def}
\eeq
which at any $b$ satisfy
\beq
\sum_f R^{B, f}(b) = 1 ,
\eeq
and describe how the charge density at a given distance is decomposed over quark flavors. 
Because the exponential dependence of the individual densities on $b$ cancels between the numerator 
and denominator, the ratios are of order unity and vary only slowly with $b$. They can therefore
provide direct insight into the changing dynamics at different distances. Figure~\ref{Fig:quark_fraction} 
shows the calculated ratios for the $p, \Sigma^+$ and $\Xi^0$ baryons. The results exhibit several interesting
features: (a)~At large distances $b > 3\, \textrm{fm}$ the $u$ and $d$ ratios in all three baryons 
approach the values $R^{B, u} = 2/3$ and $R^{B, d} = 1/3$, corresponding to $u$ and $d$ quark densities of equal magnitude and opposite sign, Eq.~(\ref{rho1_u_d_infinity}), multiplied by the quark charges $2/3$ and $-1/3$.
(b)~The strange quark fraction drops rapidly with $b$ and is negligible above 2 fm.
(c)~At distances $b \sim 1$ fm the $u$ and $d$ ratios in the proton attain values that are
comparable with the number of $u$ and $d$ valence quarks multiplied by 
their charge, $R^{p, u} \sim 2 \times 2/3 = 4/3$ and $R^{p, d} \sim 1 \times (-1/3) = -1/3$.
Such ratios would be expected in a generic mean-field picture of the motion of valence quarks in
the nucleon, in which all quarks occupy the same orbital, and the ratio of the densities is
simply determined by their numbers and charges \cite{Miller:2011du}. We emphasize that the 
present calculation can predict only peripheral densities, and that the accuracy becomes marginal
at $b \sim 1\, \textrm{fm}$ (cf.\ the discussion in Sec.~\ref{Subsec:dispersive_representation}).
Still it is very reassuring that the ratios thus obtained match with the ``quark model''
expectations at shorter distances. (d)~In the $\Sigma^+$ the mean-field picture would predict
$R^{\Sigma^+, u} \sim 2 \times 2/3 = 4/3$ and $R^{\Sigma^+, s} \sim 1 \times (-1/3) = -1/3$
for the ratios at small $b$. Figure~\ref{Fig:quark_fraction}b shows that the calculated
$s$ ratio at $b = 1\, \textrm{fm}$ is smaller in magnitude than $-1/3$, indicating significant 
deviations from the mean-field picture at these distances. This conclusion is supported
by the presence of a substantial (within errors) non-valence $d$-quark density at 
$b = 1\, \textrm{fm}$. (e) Similarly, in the $\Xi^0$ the mean-field picture would imply
that $R^{\Xi^0, s} / R^{\Xi^0, u} \sim 2 \times (-1/3) / (2/3) = -1$.\footnote{The mean-field
picture cannot literally be applied to the ratios $R^{B, f}$ if the baryon 
charge is zero (such as for the $\Xi^0$), as the total charge density in the denominator of
Eq.~(\ref{R_def}) would be identically zero in this approximation.} Figure~\ref{Fig:quark_fraction}c
shows that the calculated $s/u$ ratio at $b = 1\, \textrm{fm}$ is substantially smaller in magnitude 
than this prediction, and that there is a large non-valence $d$-quark density. Both observations
point to the importance of $\pi^+\Xi^-$ and $\pi^+\Xi^{\ast -}$ components in the $\Xi^0$ at distances
$b = 1\, \textrm{fm}$.
\subsection{Isospin breaking effects}
Some comments are in order regarding the effect of isospin symmetry breaking on the peripheral 
baryon densities. In the present study we assume isospin symmetry, which allows us to strictly separate
the isovector and isoscalar cuts of the baryon form factors and model the spectral functions accordingly.
Baryon structure at distances $b > 2 \, \textrm{fm}$ is dominated by the isovector two-pion cut.
If isospin symmetry breaking were included, the isoscalar current would also couple to the two-pion 
cut ($\rho$-$\omega$ mixing), and the isoscalar density would develop a long-range component.
In the baryons where an isovector component is present already in the isospin-symmetric case
($p, n, \Sigma^+, \Sigma^-, \Xi^0, \Xi^-$),
the effect of this coupling on the peripheral densities would be negligible. However, in the
baryons where the isovector component is absent in the isospin-symmetric case ($\Sigma^0, \Lambda$), 
the isospin-breaking coupling would qualitatively change the asymptotic behavior of the peripheral 
densities, resulting in a small long-range component that falls off like the isovector densities
produced by the two-pion cut. A rough estimate suggests that this isospin-breaking component should 
be noticeable only at distances $b > 4\, \textrm{fm}$, which are irrelevant for most practical purposes.
From a theoretical perspective, though, the long-range structure of the $\Sigma^0$ and $\Lambda$ can
provide a sensitive test of isospin breaking effects in baryon structure.
\section{Summary and outlook}
\label{Sec:Summary}
In this work we have used methods of $\chi$EFT and dispersion analysis to calculate for the first time the peripheral 
transverse densities of all octet baryons and study their properties. We now want to summarize the main 
results and insights --- methodological and phenomenological --- and suggest directions for further study.

On the methodological side, we observed that the transverse densities offer a convenient framework
for the dispersive analysis of the baryon form factors, as the contributions from high-mass states at
$t \gtrsim 1\, \textrm{GeV}^2$ are exponentially suppressed and can be controlled by the parameter $b$.
The framework is ideally suited for a dispersive analysis with dynamical input from $\chi$EFT.

The combination of $\chi$EFT with $t$-channel unitarity significantly improves the $\chi$EFT 
predictions for the baryon spectral functions on the two-pion cut. The $N/D$ representation of the 
spectral function in Eq.~(\ref{Eq:Disp_rep_ImFV_ratio}) is natural from the $\chi$EFT perspective, 
as it separates dynamics governed by the scales $M_\pi$ and $m_T - m_B$ [in the modified 
$\pi\pi \rightarrow B \bar B$ amplitude $\Gamma_i^B(t)/F_\pi(t)$] from that at the scale 
$\Lambda_\chi$ [in the pion form factor $|F_\pi(t)|^2$]. The results can be improved systematically
by calculating the function $\Gamma_i^B(t)/F_\pi(t)$ to higher orders in the $\chi$EFT expansion.

It would be interesting to study the convergence of the results for the spectral functions when
including corrections of $\mathcal{O}(\epsilon^4)$ and $\mathcal{O}(\epsilon^5)$ in $\Gamma_i^B(t)/F_\pi(t)$. 
At $\mathcal{O}(\epsilon^4)$ the only modifications compared to $\mathcal{O}(\epsilon^3)$ are 
from new $\pi\pi B\bar B$ contact terms. At $\mathcal{O}(\epsilon^5)$ 
the $\pi\pi \rightarrow B \bar B$ amplitude involves processes with $\pi B$ intermediate states
in addition to the Born diagrams, resulting in a much richer structure. Also, at $\mathcal{O}(\epsilon^5)$ $\pi\pi$
rescattering in the $t$-channel becomes possible, and one should be able to explicitly observe 
the cancellation of the phase between the numerator and denominator. A calculation of the
$\Lambda$-$\Sigma$ transition spectral function in a similar approach was recently reported
in Ref.~\cite{Granados:2017cib}.

On the phenomenological side, we found that the octet baryons exhibit a rich peripheral structure 
as a result of two-pion exchange between the electromagnetic current and the baryon (the $\rho$ meson 
is included as a $\pi\pi$ resonance within our dispersive approach). Chiral processes with decuplet 
intermediate states contribute significantly to the isovector spectral functions at 
$t \sim 30 \, M_\pi^2 = 0.6\, \textrm{GeV}^2$
and the densities at $b < 2\, \textrm{fm}$. Their contribution qualitatively changes the results 
compared to octet intermediate states, including the relative strength of the spectral functions
or densities in the individual baryons. The inclusion of the decuplet baryons as explicit degrees of freedom
in the $\chi$EFT is therefore essential for a realistic description of peripheral baryon structure.

In the nucleon isovector spectral functions and densities, the interplay between $N$ and $\Delta$ 
intermediate states can be explained as a consequence of the large-$N_c$ limit of QCD.
$N$ and $\Delta$ contributions together are needed for the $\chi$EFT results to satisfy the
general $N_c$-scaling relations, and their relative sign can be inferred in this way
\cite{Granados:2013moa,Granados:2016jjl}. It would be interesting to extend these large-$N_c$ arguments 
to the spectral functions and densities of the $SU(3)$ octet baryons; see 
Refs.~\cite{Flores-Mendieta:2014vaa,Flores-Mendieta:2015wir} and references therein. In particular, this might explain 
why in the strange baryon spectral functions ($\Sigma, \Xi$) the contributions from decuplet 
intermediate states have opposite sign compared to the nucleon spectral functions ($N$), as seen 
in Table~\ref{table:octet_decuplet}. We plan to address this question in a 
forthcoming study.

Some comments are in order regarding the comparison of our results with baryon form factor data and 
empirical densities. For the nucleon (proton and neutron), empirical charge and magnetization densities
have been determined by Fourier-transforming parametrizations of the space-like form factor data;
see Ref.~\cite{Venkat:2010by} for a discussion of the associated uncertainties. In order to reliably
extract the peripheral densities at $b > 2 \, \textrm{fm}$ it is essential to use form factor 
parametrizations with correct analytic properties, as the qualitative behavior in the 
$b \rightarrow \infty$ limit is governed by the position of the singularities in the complex $t$-plane.
Such parametrizations are provided by dispersive fits to the form factor data as described in
Refs.~\cite{Hohler:1976ax,Belushkin:2006qa,Lorenz:2012tm}, and the resulting peripheral densities
are studied in Ref.~\cite{Miller:2011du}. In these dispersive fits the isovector spectral functions 
on the two-pion cut are taken as a theoretical input, and the peripheral densities obtained from 
the fits just re-express the theoretical input in a different form. The comparison with our
results therefore comes down to a theory-vs.-theory comparison of the spectral functions
near threshold (or the densities at $b > 2 \, \textrm{fm}$) as performed in 
Fig.~\ref{Fig:Improvement_ImF_V}a and c. To improve the situation one should perform dispersive fits to the 
spacelike form factor data with flexible parametrizations of the isovector spectral functions
at $4 M_\pi^2 < t < 1\, \textrm{GeV}^2$ (e.g.\ with variable strength near the two-pion threshold, 
and with variable height of the $\rho$ peak), in order to test to what extent the spectral functions
in this region could be constrained by spacelike form factor data. Such studies could answer the 
question how accurately peripheral nucleon structure could be extracted from present and future 
spacelike form factor data.

The $\Sigma^0$-$\Lambda$ transition form factor enters in the Dalitz decay $\Sigma^0 \rightarrow 
\Lambda e^+e^-$ at timelike momentum transfers $4 m_e^2 < t < (m_{\Sigma^0} - m_\Lambda)^2 = 0.006 \, 
\textrm{GeV}^2$. Precise measurements may be able to determine a combination of the slopes of the 
magnetic and electric transition form factors at $t = 0$ (magnetic and electric radii). 
The results could be compared with dispersive calculations of the transition form factors using the 
spectral functions computed in $\chi$EFT and dispersion theory, cf.~Sec.~\ref{Subsec:Improvement} 
and Fig.~\ref{Fig:ImFV-Octet} in this work and Ref.~\cite{Granados:2017cib}. 
In the context of transverse densities the slopes of the baryon form factors determine the 
$b^2$-weighted integrals of the densities, $dF_i^{B}/dt (t = 0) = \frac{1}{4}
\int d^2 b \, b^2 \, \rho_i^B (b) \; (i = 1, 2)$, 
cf.\ Eq.~(\ref{Eq:rho_def}). Because of the weighting with $b^2$ the integrals 
receive large contributions from distances $b > 1\, \textrm{fm}$ and provide sensitive tests 
of the peripheral densities.

The electromagnetic form factors of the octet baryons are also being studied in lattice 
QCD \cite{Lin:2008mr,Shanahan:2014uka,Shanahan:2014cga}. If such 
calculations could determine the transverse densities at distances $b > 1\, \textrm{fm}$, they
could be compared directly with the peripheral densities estimated in our $\chi$EFT approach.
In fact, comparing (or matching) lattice QCD and $\chi$EFT results at the level of the transverse 
densities may be very natural, as there should exist a region of ``intermediate'' distances 
$b \sim 1\, \textrm{fm}$ where the densities can be computed reliably in either approach.
This possibility deserves further study.

The combination of $\chi$EFT and $t$-channel unitarity (dispersive improvement) represents 
a general method that could be applied also to the baryon form factors of other operators 
with a two-pion cut, such as the scalar current and the energy-momentum tensor. The basic
idea and formulas of Sec.~\ref{Subsec:Improvement} remain the same, only the $\pi\pi \rightarrow B\bar B$ 
partial-wave amplitude and the pion form factor are replaced by the ones in the partial waves
excited by the other operators \cite{InPreparation}. The scalar and tensor operators also have 
transverse density representations that can be related to the baryons' partonic structure.

The studies reported here could also be extended to the form factors and transverse densities 
of the decuplet baryons \cite{Decuplet}; in the present calculation they appear only as intermediate states 
in chiral processes modifying the octet baryon matrix elements. The decuplet baryons are
unstable particles with strong decays and finite width. Their electromagnetic form factors 
can be defined in the context of $S$-matrix theory, as the residue at the poles in the 
$\pi N \rightarrow \pi N$ amplitude, and calculated consistently in the framework of
relativistic $\chi$EFT \cite{Ledwig:2011cx,Ledwig:2010ya}. The transverse densities associated 
with the $\Delta$ form factors are of special interest because new angular structures appear in 
the spin-3/2 states [cf.\ Eq.~(\ref{j_plus_rho}) for the spin-1/2 states]. 
They are also of relevance for exploring the large-$N_c$ limit of QCD, 
where the $N$ and $\Delta$ become degenerate and can be viewed as rotational states of a 
classical mean-field solution \cite{Adkins:1983ya}. The $N$-$\Delta$ transition form factors
are measured in resonance electroproduction experiments and can be described in terms of
transition densities \cite{Carlson:2007xd,Pascalutsa:2007wz}. The form factors of decuplet
baryons are studied also in lattice QCD \cite{Alexandrou:2008bn,Aubin:2008qp,Alexandrou:2009hs}.

We expect that the methods used here will lead to more predictive studies of baryon form factors in $\chi$EFT, and at the same time extend its range of applicability to higher energies. 



\appendix 
\section{Validation of EFT calculations}
\label{App:Validation}
To validate the $\chi$EFT calculations we have computed the octet baryon form factors themselves 
from the full set of $\mathcal{O}(\epsilon^3)$ diagrams, even though only the $t$-channel cut diagrams 
of Fig.~\ref{Fig:LoopsOctet}a, b and c are needed for our study of peripheral densities.
Our results for the nucleon form factors fully reproduce those of Ref.~\cite{Ledwig:2011cx} when 
reduced to the $SU(2)$ flavor case,
which is done by setting the contribution of the kaon loops to zero. The $SU(2)$ coupling constants 
$g_A$ and $h_A$ correspond to the $SU(3)$ constants $D+F$ and
$-2\sqrt{2}\mathcal{C}$, respectively. We have also verified current conservation (electromagnetic 
gauge invariance) with the $\mathcal{O}(\epsilon^3)$ results for the current matrix element.

The $\chi$EFT calculations were independently performed with two different methods:
the analytic calculations were first done with the help of FORM \cite{Vermaseren:2000nd,Kuipers:2012rf}, 
and then separately with FeynCalc~\cite{Mertig:1990an,Shtabovenko:2016sxi}. The two methods give the same
numerical results.
\section{Off-shell properties of spin-3/2 EFT}
\label{App:Delta} 
%
%
\begin{figure}[t]
\begin{tabular}{l}
\epsfig{file=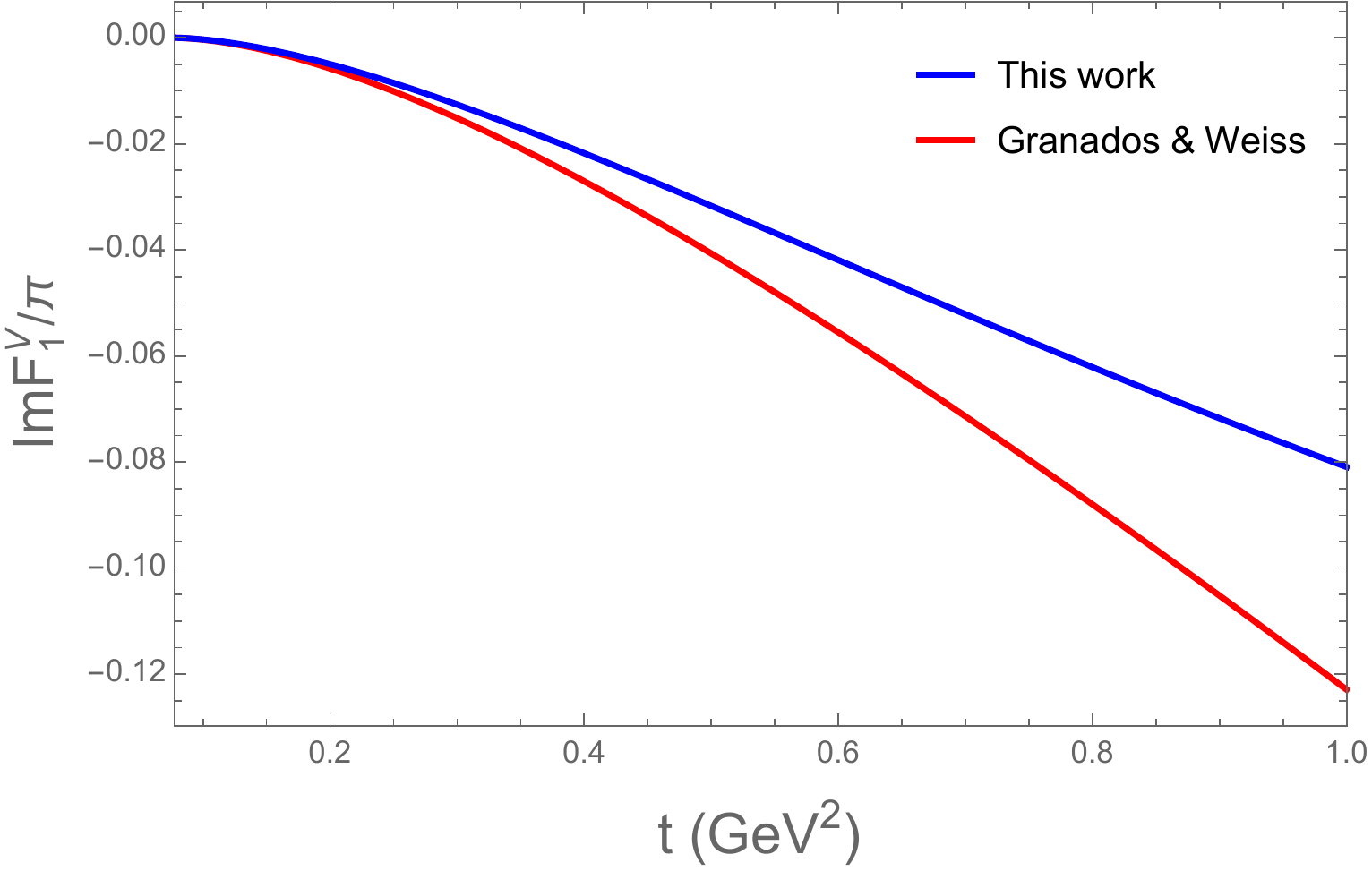,width=.45\textwidth,angle=0} 
\\[-3ex]
{\footnotesize (a)}
\\[2ex]
\epsfig{file=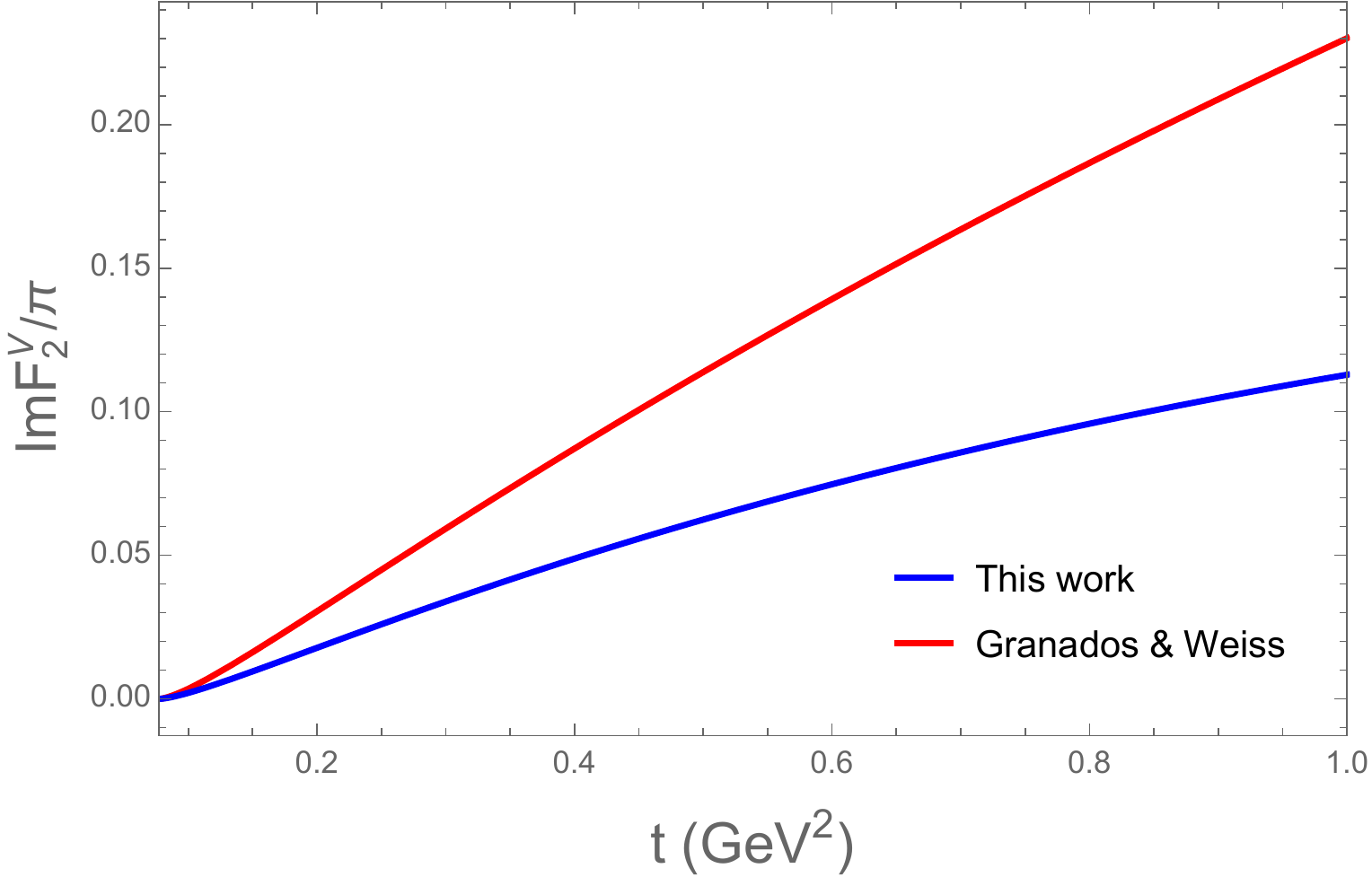,width=.45\textwidth,angle=0}
\\[-3ex]
{\footnotesize (b)}
\end{tabular}
\caption[]{\small Comparison of the $\Delta$ contribution to the nucleon isovector 
spectral functions $\textrm{Im} F_i^V(t) (i = 1, 2)$ obtained in 
Refs.~\cite{Granados:2013moa,Granados:2016jjl} and the present work. 
The curves show the total contribution resulting from the $\Delta$ triangle graph,
Fig.~\ref{Fig:LoopsOctet}c, which includes the ``intermediate $\Delta$'' and ``contact''
terms described in Refs.~\cite{Granados:2013moa,Granados:2016jjl} and in the text.
\label{Fig:Off-shell-effects}}
\end{figure} 
The $\chi$EFT used in the present study includes spin-3/2 decuplet baryons as dynamical
degrees of freedom. In this appendix we comment on issues related to the off-shell properties
of the EFT with spin-3/2 fields and compare our results to earlier calculations of 
transverse densities including the $\Delta$ isobar \cite{Granados:2013moa,Granados:2016jjl}.

The formulation of a relativistic quantum field theory with spin-3/2 fermions is inherently
not unique. Relativistic invariance necessitates working with ``redundant'' fields containing 
spin-1/2 and spin-3/2 degrees of freedom, and the projection on physical spin-3/2 degrees
of freedom can be performed unambiguously only on the mass shell, resulting in an ambiguity in 
the definition of the propagator and the vertices off the 
mass-shell \cite{Pascalutsa:1999zz,Pascalutsa:2000kd}. The ambiguity can be
formalized as an invariance under point transformations \cite{Hemmert:1997ye,Pascalutsa:2006up}.
In the context of EFT it is contained
in the overall reparametrization invariance --- the freedom to redefine the field variables, 
changing the off-shell behavior of the propagators and vertices, but leaving the on-shell 
amplitudes invariant. On-shell amplitudes calculated in EFTs with different choices of fields 
agree in the parametric order to which they were calculated but can differ by higher-order 
terms. The latter can give rise to numerical differences if the EFT
expressions are used for numerical approximation.

In Refs.~\cite{Granados:2013moa,Granados:2016jjl} the spectral functions of the nucleon form 
factors were computed in a relativistic $\chi$EFT with spin-3/2 $\Delta$ fields, using a 
Rarita-Schwinger propagator for the $\Delta$, and the minimal form of the $\pi N \Delta$ vertex
without explicit off-shell terms. The Feynman integrals obtained from the triangle diagrams 
with intermediate $\Delta$ were separated into two parts: 
\begin{itemize}
\item[(i)] An ``intermediate $\Delta$'' 
term, which results from the singularity of the integrand at the $\Delta$ pole and can be 
evaluated by contour integration with the residue at the $\Delta$ pole. This part of
the spectral functions corresponds to the triple-cut Feynman diagram in which both the 
exchanged pions and the intermediate $\Delta$ are on mass-shell. It contains the subthreshold 
singularity at $t < 4 \, M_\pi^2 - (m_\Delta^2 - m_N^2 + M_\pi^2)^2/M_\pi^2$
(or the left-hand cut of the $\pi\pi \rightarrow N\bar N$ amplitude), as represented by
the arctangent function in Eqs.~(4.21) and (4.23) of Ref.~\cite{Granados:2013moa}, as well 
as certain polynomial terms in $t$. 
\item[(ii)] A ``contact'' term, which results from the parts of the integrand that are 
non-singular at the $\Delta$ pole. This term leads to an integral of the same form as that 
resulting from the diagram with a $\pi\pi NN$ contact term in the Lagrangian, and 
can be combined with the latter. 
\end{itemize}
The separation is performed unambiguously within the reduction procedure described in 
Refs.~\cite{Granados:2013moa,Granados:2016jjl}. It is physically natural because (a) the 
intermediate $\Delta$ term does not dependent on the off-shell behavior of the $\chi$EFT, 
and the off-shell ambiguity is localized entirely in the contact term; (b) the intermediate 
$\Delta$ term can be represented in light-front time-ordered form, as an overlap integral 
of light-front wave functions describing the transition of the external $N$ to a 
$\pi \Delta$ intermediate state and back \cite{Granados:2016jjl}; (c) the scaling behavior 
of the intermediate $\Delta$ and contact terms in the large-$N_c$ limit matches that of the 
corresponding terms in contributions with intermediate $N$ states.

In the present work we use a relativistic $\chi$EFT with spin-3/2 $\Delta$ fields with
consistent $\pi N\Delta$ vertices, which are constructed to be invariant under point 
transformations by making use of the Heisenberg equations of motion of the fields.
This choice of vertices corresponds to a different off-shell behavior from that used in 
Refs.~\cite{Granados:2013moa,Granados:2016jjl}. It is instructive to compare the results
for the spectral functions obtained with the two schemes. We find that:
(a) The intermediate $\Delta$ terms of the spectral functions (as defined above
and in Refs.~\cite{Granados:2013moa,Granados:2016jjl}) are identical in the two schemes.
This confirms explicitly that these terms are independent of the off-shell behavior
of the $\chi$EFT. (b) The contact terms obtained in the two schemes are identical
at $\mathcal{O}(\epsilon^3)$, which is the accuracy of the present calculation, but differ
in higher-order contributions. In $\textrm{Im} \, F_1$ the differences appear
at $\mathcal{O}(\epsilon^5)$ (two orders higher than the nominal order), while in 
$\textrm{Im} \, F_2$ they appear at $\mathcal{O}(\epsilon^4)$ (one order higher than 
the nominal order). This shows that the off-shell ambiguity manifests
itself in the contact terms at the expected order. 

In Fig.~\ref{Fig:Off-shell-effects} we compare the numerical results for the total 
$\Delta$ contribution to the nucleon spectral functions (including both the 
intermediate $\Delta$ and contact terms) obtained in Ref.~\cite{Granados:2013moa}
and the present calculation. One sees that the results have similar behavior
in the near-threshold region but show significant differences at higher masses:
$\sim 20\%$ difference in $\textrm{Im} \, F_1$ at $t = 0.6\, \textrm{GeV}^2$,
and $\sim 50\%$ in $\textrm{Im} \, F_2$. The pattern is consistent with the 
parametric orders of the differences (see above). As noted above, these differences are
entirely due to the different contact terms in the spectral functions and could be 
compensated by adding explicit $\pi\pi NN$ contact terms to the Lagrangian. 
Since the setup of Ref.~\cite{Granados:2013moa} uses an arbitrary off-shell 
behavior, while the present formulation uses consistent vertices and can be 
extended to higher-order calculations, 
we consider the results of the present calculation more significant and use 
them in our study.

Altogether, the comparison with Ref.~\cite{Granados:2013moa} shows that the $\mathcal{O}(\epsilon^3)$
results for the spectral functions are independent of the off-shell behavior of the $\chi$EFT, as it should be. Differences between schemes appear only in higher-order
contributions and could be compensated by adding explicit $\pi\pi NN$ contact terms to 
the Lagrangian. 
\section{Isoscalar vector meson couplings}
\label{App:Vector}
In this appendix we estimate the couplings of the isoscalar vector mesons ($\omega, \phi$) 
to the $SU(3)$ octet baryons and describe the parameters in our model for the isoscalar spectral 
function, Eq.~(\ref{Eq:isoscalar_poles}). Our treatment follows the approach of Ref.~\cite{Dover:1985ba}
and makes use of $SU(3)$ symmetry, certain assumptions about the $F/D$ ratios, and empirical
meson-nucleon couplings. The coupling of the vector mesons to the octet baryons
is described by the phenomenological Lagrangian density \cite{Machleidt:2000ge}
\begin{eqnarray}
\mathcal{L}_{VBB} &=& - g_{VBB} \bar{B} \gamma^{\mu} B V_{\mu}
\nonumber \\
&&  - \frac{i f_{VBB}}{4m_N} \bar{B}\sigma^{\mu\nu} B (\partial_{\mu} V_{\nu} - \partial_{\nu} V_\mu) ,
\label{Eq:VBB_Lagrangian}
\end{eqnarray}
where $B$ is the baryon field, $V_\mu$ is the vector meson field, and $g_{VBB}$ and $f_{VBB}$ are the 
vector and tensor couplings. In the case of $SU(3)$ symmetry the flavor structure of the 
couplings is of the general form
\be
g_{VBB} &\rightarrow& g_F \, \mathrm{Tr} (\bar B [V_8, B]) 
\; + \; g_D \, \mathrm{Tr} (\bar B \{ V_8, B \} ) \nonumber
\\&&+ \; g_S \, V_1 \, \mathrm{Tr} (\bar B B ),
\label{EqVecCoup}
\ee
where $g_F$ and $g_D$ are the symmetric and antisymmetric $\bf{8} \times \bf{8} \rightarrow \bf{8}$ couplings,
and $g_S$ is the $\bf{8} \times \bf{8} \rightarrow \bf{1}$ coupling; a similar relation expresses $f_{VBB}$
in terms of $f_F, f_D$ and $f_S$. Here $B$ is the octet baryon field Eq.~(\ref{B_octet}),
$V_8$ is the octet vector meson field [cf.~Eq.~(\ref{phi_octet})],
\begin{equation}
V_8 =
\left( 
\begin{array}{ccc}
\frac{1}{\sqrt{2}} \rho^0 + \frac{1}{\sqrt{6}} \omega_8 & \rho^+ & K^{\ast +} \\[1ex] 
\rho^- & -\frac{1}{\sqrt{2}} \rho^0 + \frac{1}{\sqrt{6}} \omega_8 & K^{\ast 0} \\[1ex] 
K^{\ast -} & \bar K^{\ast 0} & -\frac{2}{\sqrt{6}} \omega_8
\end{array}
\right) ,
\label{Eq:V8_matrix}
\end{equation}
and $V_1 = \omega_1$ is the singlet vector meson field. Eq.~(\ref{EqVecCoup}) encodes the
constraints of $SU(3)$ symmetry on the physical $VBB$ couplings. 
Here we are interested in the couplings of the neutral isoscalar vector mesons, $\omega$ and $\phi$. 
Assuming ideal mixing, the physical $\omega$ and $\phi$ fields are related to the octet and
singlet fields by\footnote{In our convention the quark representation of the 
meson states is $|\omega\rangle = (|\bar u u\rangle + |\bar d d\rangle)/\sqrt{2}$ 
and $|\phi\rangle = -|\bar s s\rangle$. The $\phi$ coupling to the electromagnetic current
is positive, while the $\phi pp$ vector coupling for standard parameters is negative (see below).}
\begin{align}
& \omega= \frac{1}{\sqrt{3}} \omega_8 + \sqrt{\frac{2}{3}} \omega_1,\\
& \phi= \sqrt{\frac{2}{3}} \omega_8 - \frac{1}{\sqrt{3}} \omega_1.
\label{ideal}
\end{align}
The couplings of the physical mesons to the baryons can then be determined by 
expressing the Lagrangian Eq.~(\ref{Eq:VBB_Lagrangian}) in terms of the physical fields.

To determine the values of the $SU(3)$ couplings we use empirical information on the couplings 
of the non-strange vector mesons to the nucleon ($\rho^0, \omega$), and theoretical assumptions
about the $F/D$ ratios. The $\rho^0 NN$ and $\omega NN$ couplings are related to the $SU(3)$
couplings as
\be
g_{\rho^0 pp}  &=& \frac{1}{\sqrt{2}} (g_F + g_D) ,
\label{cond1}
\\
g_{\omega pp} &=& \frac{3 g_F - g_D}{3 \sqrt{2}} + \sqrt{\frac{2}{3}} g_S ,
\\
f_{\rho^0 pp}  &=& \frac{1}{\sqrt{2}} (f_F + f_D)
\\
f_{\omega pp} &=& \frac{3 f_F - f_D}{3 \sqrt{2}} + \sqrt{\frac{2}{3}} f_S
\ee
Assuming that the baryons' electromagnetic couplings are dominated by their coupling to 
vector mesons (vector meson dominance), the ratios of the baryon electric charges and magnetic moments 
(or their $SU(6)$ relations) provide two conditions on the ratio of $F$ and $D$ couplings
\cite{Dover:1985ba},
\be
\frac{g_F}{g_F + g_D} &=& 1,
\label{cond5}
\\
\frac{g_F + f_F}{g_F + f_F + g_D + f_D} &=& \frac{2}{5} .
\label{cond6}
\ee
Equations~(\ref{cond1})--(\ref{cond6}) allow us to calculate the values of the $SU(3)$ 
couplings in terms of the empirical $\rho^0 pp$ and $\omega pp$ vector and tensor couplings.
Using the couplings determined by the meson exchange model of the nucleon-nucleon 
interaction \cite{Machleidt:2000ge}, 
\beq
\left.
\begin{array}{rcrrcr}
g_{\rho^0 pp} &=& 3.25, & \hspace{1em} f_{\rho^0 pp} &=& 19.8, 
\\
g_{\omega pp} &=& 15.9, & f_{\omega pp} &=& 0,
\end{array}
\hspace{1em}
\right\}
\label{coupling_machleidt}
\eeq
we obtain
\begin{equation}
\left.
\begin{array}{rcrrcr}
g_F &=& 4.6, & \hspace{1em}
f_F &=& 8.4,
\\
g_D &=& 0, & 
f_D &=& 19.6,
\\
g_S &=& 15.5, & f_S &=& -1.6 .
\end{array}
\hspace{1em}
\right\}
\label{vm_su3_couplings}
\end{equation}

The coupling of the vector mesons to the electromagnetic current is summarized by the 
current-field identity 
\begin{align}
J_{\mu} = \sum_{V = \rho,\omega, \phi} \frac{M_V^2}{F_V} V_\mu .
\label{Eq:Current_field_identity}
\end{align}
The constants $F_V$ parametrize the coupling strength and can be inferred 
from the $e^+ e^-$ decay widths of the vector mesons, 
$\Gamma (V \rightarrow e^+e^-) = (\alpha M_V /3)(e/F_V)^2$,
where $\alpha = e^2/(4 \pi) \approx 1/137$ is the fine structure constant.
For the $\omega$ we obtain $F_\omega = 16.7$ and 
\beq
M_\omega^2/F_\omega = 0.037 ,
\label{em_omega}
\eeq 
using $M_\omega = 783\, \textrm{MeV}$ and the width from Ref.~\cite{Nakamura:2010zzi}.
The $\phi$ meson coupling is then fixed by assuming that the ratio of $\omega$ and
$\phi$ electromagnetic couplings follow their valence quark content [this assumption 
was used already in the vector dominance relations Eqs.~(\ref{cond5}) and (\ref{cond6})],
\beq
(M_\phi^2/F_\phi ) : (M_\omega^2/ F_\omega ) \; = \; \sqrt{2} : 1 ,
\label{em_phi}
\eeq
which gives $M_\phi^2/F_\phi = 0.052$. Note that $F_V$ is positive in our convention.
The parameters in the vector meson pole parametrization of the isoscalar spectral
function Eq.~(\ref{Eq:isoscalar_poles}) ($V = \omega, \phi$) are then calculated from the 
vector meson-baryon couplings and the electromagnetic couplings as
\begin{equation}
a_1^{VBB} = \frac{g_{VBB} M^2_{V}}{F_{V}} , \hspace{2em}
a_2^{VBB} = \frac{f_{VBB} M^2_{V}}{F_{V}} .
\label{a12_res}
\end{equation}
The results obtained for the various octet baryon states are shown in 
Table~\ref{Table:VBB-coefficients-2}.
%
%
{\tiny
\begin{table*}[t]
\centering
\begin{tabular}{|c|r|r||r|r|}
\cline{1-5} 
\hline
 \multicolumn{5}{|c|}{Couplings [GeV$^2$]} \\
\hline 
$B$ & $a_1^{\omega BB}$ & $a_1^{\phi BB}$ & $a_2^{\omega BB}$ & $a_2^{\phi BB}$  \\
         \hline
 $p, n$                                   & $0.58$ & $-0.23$ & $0$    & $0.14$     \\
 $\Lambda$                           & $0.46$ &$-0.46$ &$-0.39$    & $-0.63$ \\
  $\Sigma^\pm, \Sigma^0$    & $0.46$ & $-0.46$& $0.29$    &  $0.73$    \\
    $\Lambda - \Sigma^0$     & $0$  & $0$ & $0$     & $0$       \\
   $\Xi^0, \Xi^-$                      & $0.35$ & $-0.70$ & $-0.44$& $-0.73$ \\
\hline 
\end{tabular}
\caption[]{\small Parameters of the vector meson pole parametrization of the isoscalar 
octet baryon form factors, Eq.~(\ref{a12_res}), obtained from $SU(3)$ symmetry and
$F/D$ ratio models, Eqs.~(\ref{cond1})--(\ref{cond6}), and empirical $\rho pp$ and 
$\omega pp$ couplings. This set uses the $\rho pp$ and $\omega pp$ couplings from the meson 
exchange parametrization of the $NN$ interaction of Ref.~\cite{Machleidt:2000ge},
Eq.~(\ref{coupling_machleidt}). The couplings of the vector mesons to the electromagnetic 
current are from Eqs.~(\ref{em_omega}) and (\ref{em_phi}).
\label{Table:VBB-coefficients-2}}
\end{table*}
}
%
%
{\tiny
\begin{table*}[t]
\centering
\begin{tabular}{|c|r|r||r|r|}
\cline{1-5} 
\hline
 \multicolumn{5}{|c|}{Couplings [GeV$^2$]} \\
\hline 
 $B$                                       & $a_1^{\omega BB}$ & $a_1^{\phi BB}$ & $a_2^{\omega BB}$ & $a_2^{\phi BB}$  \\
         \hline
 $p, n$                                   & $(0.61,0.85)$ & $(-0.49,0.26)$ & $(-0.13,0.38)$    & $(-0.23, 0.28)$     \\
 $\Lambda$                           & $(0.49, 0.73)$ &$(-0.73,0.49)$ &$(-0.52,-0.01)$    & $(-1.01,-0.49)$ \\
  $\Sigma^\pm, \Sigma^0$    & $(0.49, 0.73)$ & $(-0.73,0.49)$& $(0.16, 0.67)$    &  $(0.35, 0.86)$    \\
    $\Lambda - \Sigma^0$     & $0$                 & $0$                   & $0$                   & $0$       \\
   $\Xi^0, \Xi^-$                      & $(0.37,0.61)$ & $(-0.97,-0.73)$ & $(-0.57, -0.06)$& $ (-1.11, -0.59) $ \\
\hline 
\end{tabular}
\caption[]{\small Parameters of the vector meson pole parametrization of the isoscalar 
octet baryon form factors, Eq.~(\ref{a12_res}), obtained from $SU(3)$ symmetry and
models of the $F/D$ ratios, Eqs.~(\ref{cond1})--(\ref{cond6}), and empirical $\rho pp$ and 
$\omega pp$ couplings (cf.~Table~\ref{Table:VBB-coefficients-2}). This set uses the 
range of $\omega pp$ couplings extracted from the fit to the nucleon form factor
data of Ref.~\cite{Belushkin:2006qa}, Eq.~(\ref{omega_coupling_belushkin}).
\label{Table:VBB-coefficients-2-var}}
\end{table*}
}

With the $g_{\rho^0 pp}$ and $g_{\omega pp}$ couplings of Ref.~\cite{Machleidt:2000ge} (see above)
the present model gives a small negative value of $g_{\phi NN}$, as was obtained in earlier 
studies; see Ref.~\cite{Dover:1985ba} and references therein. We note that 
the present model could also accommodate the ``quark model'' value $g_{\phi NN} = 0$, which 
would be obtained with $g_{\omega pp} = 3 g_{\rho^0 pp}$; see Ref.~\cite{Dover:1985ba} for 
discussion. We emphasize that the $\phi$ meson couplings to the strange octet baryons are generated
by $SU(3)$ symmetry regardless of the precise value of $g_{\phi NN}$, and that
we are not aiming for an accurate description of the strange form factors of the nucleon
in the present study.

To estimate the uncertainty of our parametrization of the isoscalar form factors we replace 
the fixed $g_{\omega pp}$ and $f_{\omega pp}$ values of Ref.~\cite{Machleidt:2000ge} by the 
range of values extracted from a fit to the isoscalar nucleon form factor data \cite{Belushkin:2006qa},
\beq
g_{\omega pp} = 16.7\ldots 23.1, \hspace{2em}
f_{\omega pp} = -3.6\ldots +10.3,
\label{omega_coupling_belushkin}
\eeq
and follow the variation of
the results over the allowed range. The couplings obtained in this way are shown in 
Table~\ref{Table:VBB-coefficients-2-var}.
\section*{Acknowledgments}
We thank Ina Lorenz for providing us with the parametrization of the pion form factor, and Ulf-G.~Mei\ss ner for discussions and support during the early stages of this project.
This material is based upon work supported by the U.S.~Department of Energy, 
Office of Science, Office of Nuclear Physics under contract DE-AC05-06OR23177.
We acknowledge partial financial support from the Deutsche
Forschungsgemeinschaft (Sino-German CRC 110), MINECO (Spain) and the ERDF (European Commission) grants No. FPA2013-40483-P, FIS2014-51948-C2-2-P and SEV-2014-0398. This work was also  supported by the Generalitat Valenciana under Contract PROMETEOII/2014/0068.




%
%
\newpage
%
\end{document}